\documentclass[leqno, letterpaper, 11pt]{article}

\usepackage{amsmath}
\usepackage{amsfonts}
\usepackage{amsthm}

\usepackage[pdftex]{graphicx,color} 

\usepackage{verbatim}
\usepackage{caption}

\newcommand{\bR}{{\mathbb R}}
\newcommand{\bZ}{{\mathbb Z}}

\newcommand{\bT}{{\mathbb T}}

\newcommand{\cD}{{\mathcal D}}

\newcommand{\cN}{{\mathcal N}}

\begin{document}
\pagestyle{empty}
\setlength{\textheight}{22cm}

\makeatletter
     \def\fps@figure{htbp}
\makeatother

\renewcommand{\textfraction}{0.1}
\renewcommand{\floatpagefraction}{0.66}

\captionsetup{aboveskip=1pt}
\captionsetup{labelfont=bf}
\captionsetup{labelsep=period}

\begin{center}
{\bf MODELING SNOW CRYSTAL GROWTH III:\\
three-dimensional snowfakes}
\end{center}

\begin{center}

{\sc Janko Gravner}\\
{\rm Mathematics Department}\\
{\rm University of California}\\
{\rm Davis, CA 95616}\\
{\rm e-mail: \tt gravner{@}math.ucdavis.edu}
\end{center}
\begin{center}

{\sc David Griffeath}\\
{\rm Department of Mathematics}\\
{\rm University of Wisconsin}\\
{\rm Madison, WI 53706}\\
{\rm e-mail: \tt griffeat{@}math.wisc.edu}
\end{center}

\vspace{0.1cm}

\begin{center}

{(Preliminary version, November 2007)}

\end{center}

\vspace{1cm} 

\noindent{\bf Abstract}
We introduce a three-dimensional, computationally feasible, mesoscopic model
for snow crystal growth, based on diffusion of vapor, anisotropic attachment, 
and a semi-liquid boundary layer. Several case studies are presented that faithfully 
emulate a wide variety of physical snowflakes.

\vspace{4cm}

\noindent 2000 {\it Mathematics Subject Classification\/}. Primary 82C24. Secondary
35R35, 60K35.
\vskip0.3cm

\noindent {\it Keywords\/}: Coupled lattice map, crystal growth, diffusion-limited aggregation, Stefan problem.

\vspace{0.3cm}

\noindent {\bf Acknowledgments}.
We extend our continuing appreciation and gratitude to Ken Libbrecht for sharing with us his 
unmatched collection of snowflake photographs and his extensive research on snowflake physics.
We also thank Antoine Clappier for introducing us to ray-tracing. 

\vspace{0.3cm}

\noindent {\bf Support}. JG was partially supported by NSF grant DMS--0204376 and the
Republic of Slovenia's Ministry of Science program P1--285. DG was partially supported
by NSF grant DMS--0204018.

\newpage

\addtocounter{page}{-1}

\pagestyle{headings}
\thispagestyle{empty}




\section{Introduction}

In this paper we exhibit some virtual snowflakes, 
or {\it snowfakes\/}, generated by a natural, fully three-dimensional algorithm for snow 
crystal evolution. The present study extends our earlier work
on growth and deposition \cite{GG1, GG2, GG3}, and other previous
efforts in this direction \cite{Pac, Rei}. The key features of our model are {\it diffusion\/}
of vapor,  anisotropic {\it attachment\/} of water molecules, and a narrow
{\it semi-liquid layer\/} at the boundary. All three ingredients
seems to be essential for faithful emulation of the morphology observed in nature.
The algorithm assumes a mesoscopic (micron) scale
of basic units for the ice crystal and water vapor, which eliminates inherent
randomness in the diffusion and the attachment mechanism. This brings
the process within reach of realistic simulation; by contrast,
any three-dimensional approach based on microscopic dynamics is completely beyond the scope of present computing technology. We refer the reader to 
\cite{GG3} for a brief history of snow crystal observation and modeling, background on our approach in a two-dimensional setting, and many references to the literature. See also \cite{NR} for another 
attempt at spatial mesoscopic modeling.

There are many papers and books, for a variety of audiences,
dealing with snowflake photography and classification, the underlying
physics, or some combination thereof, so we will not offer a
comprehensive review here.
Excellent introductions to the subject include {\it the\/} classic book by Nakaya
\cite{Nak}, early empirical studies and classification schemes \cite{BH} and \cite{ML}, 
and more recent papers and books by K.~Libbrecht \cite{Lib1, Lib2, Lib3, Lib4, Lib5, LR}. 
Among research papers that attempt to decipher the three-dimensional aspects of snow crystals, 
the standout reference is \cite{TEWF}; also worth mentioning are \cite{Iwa}, \cite{NK} and \cite{Nel}. 
The single most convenient resource for comparison of our simulations to physical crystals is Libbrecht's field guide \cite{Lib6}.

\vskip0.2cm
\hskip-0.5cm
\begin{minipage}{8cm}
\vskip-0.15cm
\includegraphics[trim=0.0cm 0cm 0.0cm 0.1cm, clip, height=1.5in]{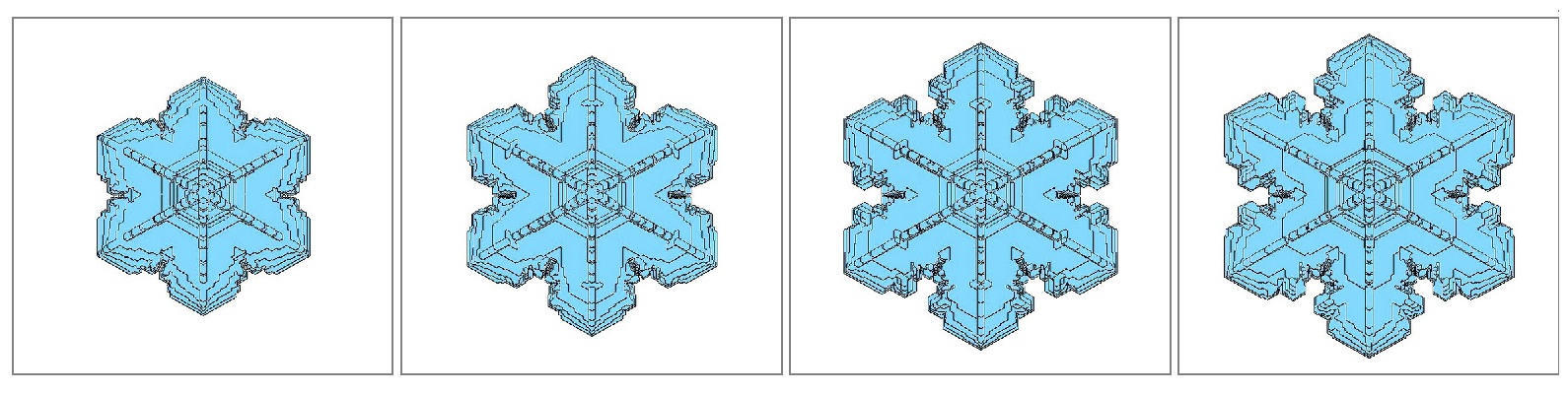}
\end{minipage}
\vskip-0.25cm 
\begin{minipage}[b]{8cm}
\includegraphics[trim=0cm 0cm 0cm 0cm, clip, height=2.25in, angle=0]{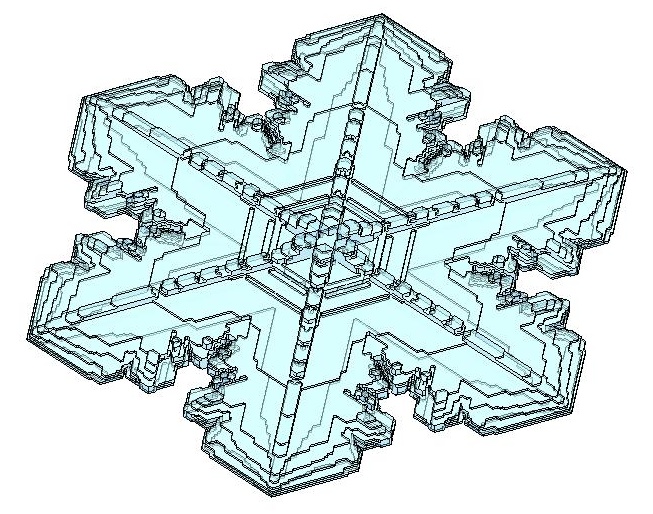}
\end{minipage}
\hskip0.4cm
\begin{minipage}[b]{8cm}
\includegraphics[trim=0cm 0cm 0cm 0cm, clip, height=1.8in, angle=0]{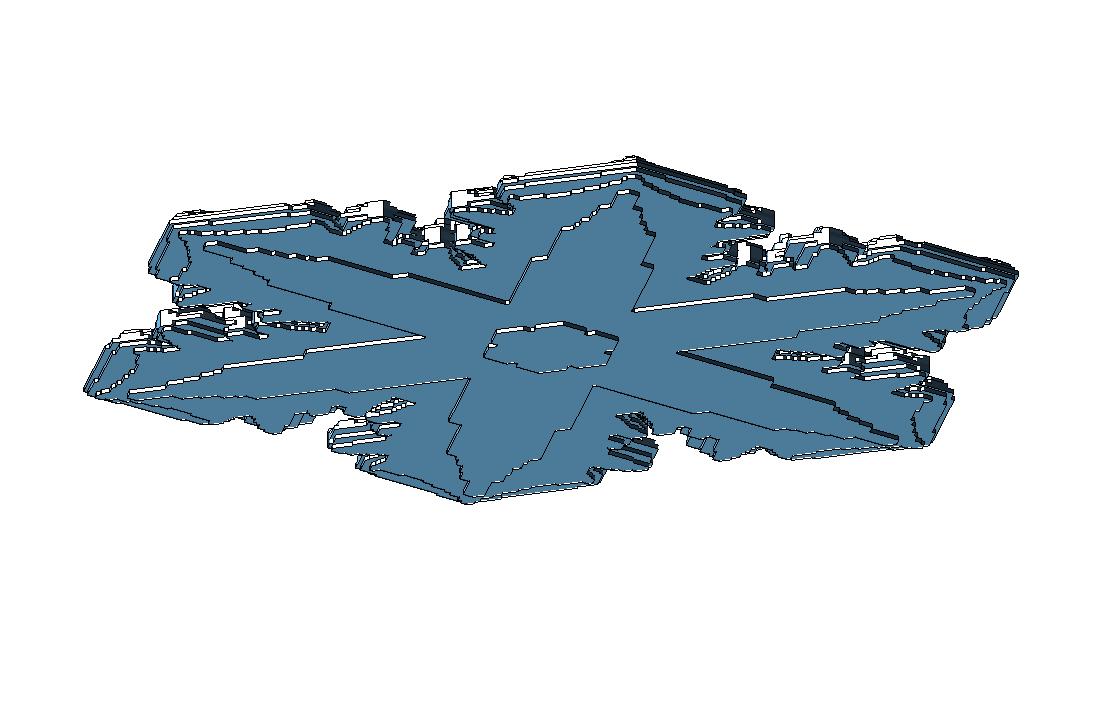}
\end{minipage}
\vskip-.3cm

{\bf Fig.~1.} Tip instability and oblique top ({\it left\/}) and bottom ({\it right\/}) views of the final crystal.

As a preview of the capabilities of our model, let us illustrate the crystal 
tip instability and initiation of side branching studied in the laboratory by Gonda and Nakahara \cite{GN}. 
A sequence of four still frames from their paper was reproduced in \cite{GG3} so we will not show it here. 
But Fig.~1 depicts the top view of a corresponding snowfake at four different times (12, 15, 18, and 21 thousand), 
and oblique views of the crystal's top and bottom at the final time. 
The parameters are:  
$\beta_{01}=2.8$, $\beta_{10}=\beta_{20}=2.2$, $\beta_{11}=\beta_{21}=1.6$, 
$\beta_{30}=\beta_{31}=1$, $\kappa\equiv .005$, $\mu_{10}=\mu_{20}=.001$, $\mu=.0001$ otherwise, $\phi=.01$, 
and $\rho=.12$. Their role, and that of the initial state, will be described in Section 2. 
Similarity between the real and 
simulated sequences is striking: in both instances a defect arises at a characteristic distance from 
the crystal tip, becomes more pronounced, and later gives rise to a side branch with its own ridge structure 
similar to that of the main branch. Note also that our snowfake has its ridges and most 
of its markings on the top side; the bottom is almost featureless. This is due to a small downward 
drift in our model, an aspect we will discuss later in more detail. The direction of the 
drift represents the motion of the crystal in the opposite direction --- we prefer 
upward motion because interesting features then appear on top, although this would obviously 
correspond to the bottom of a falling snowflake.  We should also note that 
the drift value means that, during its evolution, our simulated crystal moved for about 200 space units, 
which is comparable to the diameter it reached. This is typical of drift values that
erase features on one side without otherwise significantly changing the morphology. Our model 
thus predicts that a significantly larger range of motion during growth is not possible 
for most interesting physical snow crystals, such as dendrites or plates. Another example of
our algorithm's potential to make 
new predictions about basic aspects of snow crystal growth is the location of 
markings. From micrographs, 
it is almost impossible to tell 
whether these are on the top, bottom, or inside a given physical specimen, 
so little attention has been paid to this issue to date. We have gathered a considerable 
amount of evidence that inside markings are quite common (cf.~Sections 7, 8 and 9). 

Our account will focus on seven case studies that reproduce many  
features commonly observed in actual snowflakes: ridges, ribs, 
flumes and other ``hieroglyphs,'' formation of side branches, 
emergence of sandwich plates, hollow columns, hollow prism  
facets, and so forth. We also explore dependence on the 
density of vapor, and the aforementioned effect of drift, and inhibition of side branches by the semi-liquid layer. 
Varying meteorological conditions during growth are considered very important \cite{Lib6} so we include several 
examples, such as plates with dendritic tips and capped 
columns, that are believed to arise due to sudden changes in the weather. However, 
we will encounter snowfakes that grew in a homogeneous environment but give 
the impression that they did not. We will occasionally address dependence of the 
final crystal on its early development, and conclude with a few eccentric examples that may be 
too brittle to occur in nature. These typically arise near a phase boundary, 
when the dominant direction of growth is precarious. A complete collection of snowfakes 
from our case studies (with some additional information, such as simulation array sizes), 
and a slide show are available for download from:
\centerline{\tt http://psoup.math.wisc.edu/Snowfakes.htm}

The first order of business, in the next section, is to describe the snowfake algorithm in detail.
Four subsequent sections discuss computer implementation and visualization tools, mathematical foundations, 
parameter tuning, and extensions of the model. 
The remainder of the paper is then devoted to the case studies.

\section{The algorithm for three-dimensional snow crystal growth}

Our basic assumptions are as follows:
\begin{itemize}
\item[A1.]  The mesoscopic (micron-scale) building blocks are (appropriately scaled)
            translates of the {\it fundamental prism\/}, which has
            hexagonal base of side length $1/\sqrt 3$ and height $1$;
\item[A2.]  In its early stages of growth, from microscopic to mesoscopic, the crystal 
            forms a hexagonal prism, and then it maintains this simple polyhedral shape until 
            it reaches the size of a few microns across.
\item[A3.]  Diffusion outside the growing crystal is isotropic 
            except possibly for a small drift in the $\bZ$-direction;
\item[A4.]  Crystallization and attachment rates depend on the direction and
            local convexity at the boundary;
\item[A5.]  There is a melting rate at the boundary, creating a quasi-liquid layer.
\end{itemize}

\noindent Note that the side (rectangular) faces of the fundamental
prism are commonly referred to as {\it prism\/} faces, while the top and
bottom (hexagonal) ones are called {\it basal\/} faces.

The lattice for our model is $\bT\times \bZ$, where $\bT$ is the planar triangular lattice (see Fig.~2). 
This is not precisely the crystalline lattice of hexagonal ice {\it Ih\/}, which is obtained 
by removing certain edges and sites from $\bT\times \bZ$, and then applying a periodic deformation \cite{NR}, but
we are constructing a mesoscopic model that should obscure such fine details. Therefore,
each $x\in \bT\times \bZ$ has 8 neighbors, 6 in the $\bT$-direction and 2 in the $\bZ$-direction.

At each discrete time $t=0,1,2,\dots$
and with each site $x\in \bT\times \bZ$, we associate a Boolean variable and two varieties of mass:
the state of the system at time $t$ at site $x$ is
$\xi_t(x)=(a_t(x), b_t(x), d_t(x))$ where the attachment flag
$$
a_t(x)=
\begin{cases}
1 &\quad\text{if $x$ belongs to the crystal at time $t$},\\
0 &\quad\text{otherwise;}
\end{cases}
$$
and
$$
\begin{aligned}
&b_t(x)=\text{the {\it boundary mass\/} at $x$ at time $t$}\quad &&\text{({\it frozen\/} if $a_t(x)=1$, {\it quasi-liquid\/} if $a_t(x)=0$)\/}, \\
&d_t(x)=\text{the {\it diffusive mass\/} at $x$ at time $t$}\quad&&\text{({\it vapor\/})}.
\end{aligned}
$$
Our dynamics assumes that the diffusive and the quasi-liquid mass both change to ice when
the site joins the crystal, and stay in that state thereafter.
The two types of mass can coexist on the boundary of the snowfake, 
but only boundary mass persists inside the snowfake while only diffusive mass occurs outside and away from the boundary.

The initial state will consist
of frozen mass 1 at each site of some finite set, on which also $a_0\equiv 1$,
with $a_0$ and $b_0 \equiv 0$ and $d_0\equiv\rho$ everywhere else. In keeping with assumption (A2), 
the most natural choice for this finite set, a singleton at the origin, often does not work well, 
as its $\bZ$-direction neighbors see 7 neighbors off the crystal's boundary. 
This means that it is common, even for low $\rho$, that the dynamics immediately triggers 
a rapid expansion in the $\bZ$-direction. To prevent this singularity, our canonical initial state consists of 
a hexagon of radius 2 and thickness 1, consisting of 20 sites. Other non-symmetric initial states will be discussed later.

\newcommand\orig{\text{\bf 0}}

Let us now describe the update rule of our snowflake simulator, which performs steps 
({\it i\/})--({\it iv\/}) below in order every discrete time unit.  
The reader should observe that total mass is conserved by each step, and hence by the dynamics as a whole. 

Write $\cN_x^{\bT}=\{x\}\cup\{y: y$ is a neighbor of $x$ in the $\bT$-direction$\}$,
$\cN_x^{\bZ}=\{x\}\cup\{y: y$ is a neighbor of $x$ in the $\bZ$-direction$\}$ for the
$\bT$-{\it neighborhood\/} and $\bZ$-{\it neighborhood\/} of $x$, respectively. We also let
$\cN_x= \cN_x^{\bT}\cup\cN_x^{\bZ}$,
and set
$$
\begin{aligned}
&A_t=\{x: a_t(x)=1\}=\text{the snowfake at time $t$};\\
&\partial A_t=\{x\notin A_t: a_t(y)=1\text{ for some }y\in \cN_x\}=
\text{the boundary of the snowfake at time $t$};\\
&{\bar A}_t=A_t\cup \partial A_t.
\end{aligned}
$$
The complement of a set $A$ is denoted by $A^c$. 
Also, we use $^\circ$ (degree) and $'$ (prime) notation to denote amounts of
mass before and after a step or substep is completed. If there is more than one intermediate
step, we use double primes.
This is necessary since some mass allocations may change more than
once during a single cycle of the steps. At the end of each cycle the time $t$  advances to
$t+1$.

\vskip0.25cm

\noindent{\bf Steps of the update rule:}

\noindent{\it i\/. Diffusion}

Diffusive mass evolves on $A_t^c$ in two, or possibly three, substeps.
The first substep is by discrete diffusion with uniform weight $\frac 17$
on the center site and each of its $\bT$-neighbors.
Reflecting boundary conditions are used at the edge of the crystal. In other words,
for $x\in {\bar A}_t^c$,
\begin{equation}
d_t'(x)=\frac 17\sum_{y\in \cN_x^{\bT}}d_t^\circ(y).
\tag{1a}
\end{equation}
The second substep does the same in the $\bZ$-direction:
\begin{equation}
d_t''(x)=\frac 47 d_t'(x)+\frac 3{14}\sum_{y\in \cN_x^{\bZ},y\ne x}d_t'(y).
\tag{1b}
\end{equation}
For $x\in\partial A_t$ any term in the sum in (1a) (resp.~(1b)) corresponding to  $y\in A_t$
is replaced by $d_t^\circ(x)$ (resp, $d_t'(x)$).

The reason for the weights in (1b) is as follows. Imagine we tessellate $\bR^3$ with translates of the fundamental prism and scale the lattice $\bT\times \bZ$ so that the lattice
points are in the centers of these prisms. The ``bonds'' in the top left frame of Fig.~2 
thus all have unit length and we eventually visualize the
crystal by drawing prisms that are centered about sites of $A_t$. Rule (1b)
ensure that diffusion on the scaled lattice is isotropic, in agreement with assumption A2.

As mentioned in the Introduction, there is also good reason to consider the more general case of diffusion with drift in the $\bZ$-direction, corresponding to downward (or upward) motion of
the snowflake. The third diffusion substep is thus:
\begin{equation}
d_t'''(x)= (1-\phi\cdot(1-a_t(x-e_3))\cdot d_t''(x)+\phi\cdot(1-a_t(x+e_3))\cdot
d_t''(x+e_3),
\tag{1c}
\end{equation}
where $e_3=(0,0,1)$ is the third basis vector.
Parameter $\phi$ measures the strength of the drift, and needs to be
small for the dynamics to remain diffusion-limited.

\vskip0.25cm

\noindent{\it ii\/. Freezing}

Assume that $x\in \partial A_t$, and denote
\begin{equation}
n_t^{\bT}(x)=\#\{y\in \cN_x^{\bT}: a_t^\circ(y)=1\}\wedge 3,
\quad n_t^{\bZ}(x)=\#\{y\in \cN_x^{\bZ}: a_t^\circ(y)=1\}\wedge 1.
\tag {2a}
\end{equation}
Proportion $1-\kappa(n_t^{\bT}(x), n_t^{\bZ}(x))$ of the diffusive mass at $x$
becomes boundary mass. That is,
\begin{equation}
\begin{aligned}
&b_t'(x)=b_t^\circ(x)+(1-\kappa(n_t^{\bT}(x), n_t^{\bZ}(x)))d_t^\circ(x), \\
&d_t'(x)=\kappa(n_t^{\bT}(x), n_t^{\bZ}(x)) d_t^\circ(x).\\
\end{aligned}
\tag{2b}
\end{equation}
The seven parameters $\kappa(i,j)$, $i\in \{0,1\}$,
$j\in \{0,1,2,3\}$, $i+j>0$, constitute one of the ingredients that
emulate the dynamics of the quasi-liquid layer at the boundary of the crystal.
The other ingredient, $\mu$, appears in step {\it iv\/} below.
We assume that $\kappa$ decreases in each coordinate since ``more concave corners'' at the boundary $\partial A_t$, i.e., those with more neighbors in $A_t$, should
catch diffusing particles more easily.

\vskip0.25cm

\noindent{\it iii\/. Attachment}

Assume again that $x\in \partial A_t$ and define the neighborhood counts as in (2a).
Then $x$ needs boundary mass at least $\beta(n_t^{\bT}(x), n_t^{\bZ}(x))$ to
join the crystal:
\begin{equation}
\text{
If $b_t^\circ(x)\ge \beta(n_t^{\bT}(x), n_t^{\bZ}(x))$, then $a_t'(x)=1$.}
\tag{3}
\end{equation}
Again, we have seven parameters $\beta(i,j)$, $i\in \{0,1\}$,
$j\in \{0,1,2,3\}$, $i+j>0$, and the assignment only makes physical sense
if $\beta$ decreases in each coordinate.

In addition, we assume that
$a_t'(x)=1$ automatically whenever $n_t^{\bT}(x)\ge 4$ and $n_t^{\bZ}(x)\ge 1$.
This last rule fills holes and makes the surface of the crystal smoother, without 
altering essential features of the dynamics.

At sites $x$ for which $a_t'(x)=1$, the diffusive mass becomes boundary mass:
$b_t'(x)=b_t^\circ(x)+d_t^\circ(x)$, $d_t'(x)=0$. 
Attachment is permanent, and there are no further dynamics at attached sites.
Thus we do not model sublimation, although it may play
a significant role in the last stages of snow crystal evolution
(cf.~p.~27 of \cite{Lib6}). 

\vskip0.25cm

\noindent{\it iv\/. Melting}

Proportion $\mu(n_t^{\bT}(x), n_t^{\bZ}(x))$ of the boundary mass at each
boundary site becomes diffusive mass. Thus, for $x\in\partial A_t$,
\begin{equation}
\begin{aligned}
&b_t'(x)=(1-\mu(n_t^{\bT}(x), n_t^{\bZ}(x)))b_t^\circ(x),\\
&d_t'(x)=d_t^\circ(x)+
\mu(n_t^{\bT}(x), n_t^{\bZ}(x)) b_t^\circ(x).
\end{aligned}
\tag{4}
\end{equation}
Again, $\mu$ is decreasing in each coordinate.

\vskip0.25cm

Fig.~2 summarizes our model in three frames. At the upper left is a portion of the 
underlying lattice $\bT\times \bZ$. The central site represented as a 
larger black ball has its neighborhood indicated in black, 
and a translate of the fundamental prism is centered at that site.
In the upper right detail, blue translates of the fundamental prism are drawn around 
each site of a small crystal. Seven boundary
sites are depicted in red and each is labeled by its boundary
configuration. For example, the ``21'' site has 2 horizontal
($\bT$-) neighbors and 1 vertical ($\bZ$-) neighbor, and consequently
needs boundary mass $\beta_{21}$ to join the crystal. Finally, the lower panel shows a 
flowchart for the algorithm. There are three epochs in the life of a site. Away from the crystal's
boundary, it only exchanges diffusive mass $d_t$ with its neighbors. 
Once the crystal grows to reach the site's neighborhood, two additional effects, 
melting and freezing, promote exchange between diffusive mass $d_t$ and boundary mass $b_t$. 
Final changes occur once boundary mass exceeds the threshold $\beta$ (which depends on the neighborhood configuration): 
the site attaches and the two types of mass merge into $b_t$.

\vskip-0.1cm

\hskip-0.5cm
\begin{minipage}{8cm}
 
\resizebox{16cm}{!}{\begin{picture}(0,0)%
\includegraphics{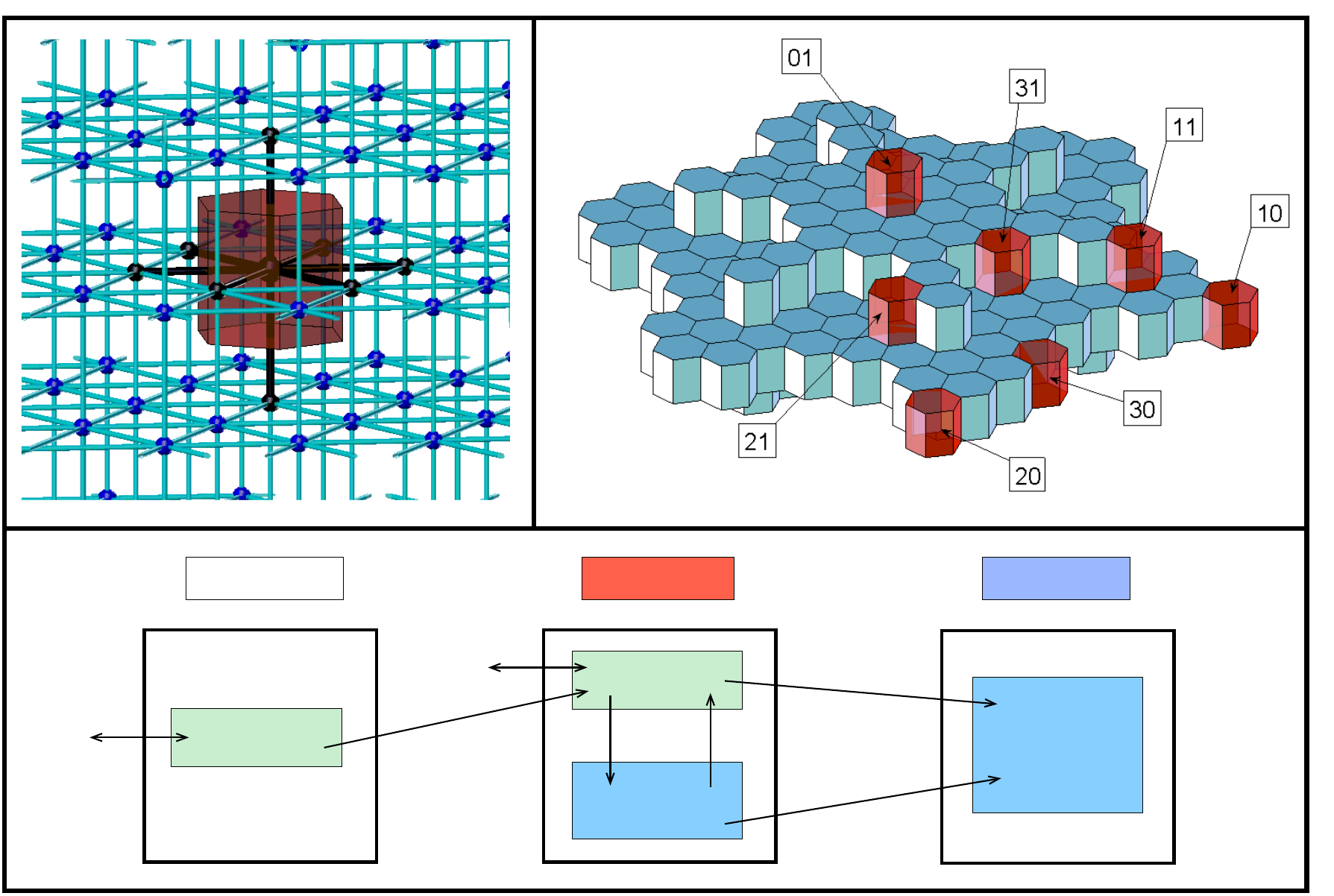}%
\end{picture}%
\setlength{\unitlength}{2693sp}%
\begingroup\makeatletter\ifx\SetFigFont\undefined%
\gdef\SetFigFont#1#2#3#4#5{%
  \reset@font\fontsize{#1}{#2pt}%
  \fontfamily{#3}\fontseries{#4}\fontshape{#5}%
  \selectfont}%
\fi\endgroup%
\begin{picture}(12554,8434)(542,-8700)
\put(6672,-6763){\makebox(0,0)[lb]{\smash{{\SetFigFont{10}{12.0}{\familydefault}{\mddefault}{\updefault}{\color[rgb]{0,0,0}$d_t$}%
}}}}
\put(6672,-7915){\makebox(0,0)[lb]{\smash{{\SetFigFont{10}{12.0}{\familydefault}{\mddefault}{\updefault}{\color[rgb]{0,0,0}$b_t$}%
}}}}
\put(5196,-6440){\makebox(0,0)[lb]{\smash{{\SetFigFont{7}{8.4}{\familydefault}{\mddefault}{\updefault}{\color[rgb]{0,0,0}diffusion}%
}}}}
\put(2891,-7304){\makebox(0,0)[lb]{\smash{{\SetFigFont{10}{12.0}{\familydefault}{\mddefault}{\updefault}{\color[rgb]{0,0,0}$d_t$}%
}}}}
\put(1437,-7101){\makebox(0,0)[lb]{\smash{{\SetFigFont{7}{8.4}{\familydefault}{\mddefault}{\updefault}{\color[rgb]{0,0,0}diffusion}%
}}}}
\put(10451,-7336){\makebox(0,0)[lb]{\smash{{\SetFigFont{10}{12.0}{\familydefault}{\mddefault}{\updefault}{\color[rgb]{0,0,0}$b_t$}%
}}}}
\put(10122,-5766){\makebox(0,0)[lb]{\smash{{\SetFigFont{8}{9.6}{\familydefault}{\mddefault}{\updefault}{\color[rgb]{0,0,0}attached}%
}}}}
\put(6346,-5776){\makebox(0,0)[lb]{\smash{{\SetFigFont{8}{9.6}{\familydefault}{\mddefault}{\updefault}{\color[rgb]{0,0,0}boundary}%
}}}}
\put(8146,-7306){\makebox(0,0)[lb]{\smash{{\SetFigFont{7}{8.4}{\familydefault}{\mddefault}{\updefault}{\color[rgb]{0,0,0}when $b_t\ge\beta$}%
}}}}
\put(6121,-7261){\makebox(0,0)[lb]{\smash{{\SetFigFont{7}{8.4}{\familydefault}{\mddefault}{\updefault}{\color[rgb]{0,0,0}$\kappa$}%
}}}}
\put(2431,-5776){\makebox(0,0)[lb]{\smash{{\SetFigFont{8}{9.6}{\familydefault}{\mddefault}{\updefault}{\color[rgb]{0,0,0}non-boundary}%
}}}}
\put(7066,-7261){\makebox(0,0)[lb]{\smash{{\SetFigFont{7}{8.4}{\familydefault}{\mddefault}{\updefault}{\color[rgb]{0,0,0}$\mu$}%
}}}}
\put(4231,-7486){\makebox(0,0)[lb]{\smash{{\SetFigFont{7}{8.4}{\familydefault}{\mddefault}{\updefault}{\color[rgb]{0,0,0}when in contact }%
}}}}
\put(4231,-7711){\makebox(0,0)[lb]{\smash{{\SetFigFont{7}{8.4}{\familydefault}{\mddefault}{\updefault}{\color[rgb]{0,0,0}with the crystal}%
}}}}
\end{picture}%
}
 
\end{minipage}

\vskip-0.1cm

{\bf Fig.~2.} The stacked triangular lattice $\bT\times \bZ$ ({\it top left\/}), 
coding of boundary configurations ({\it top right\/}), and a flowchart for the growth algorithm ({\it bottom\/}).

\vskip0.5cm

\section{Notes on computation and visualization}

Following the same strategy as for our previous two-dimensional model \cite{GG3},
the dynamics actually run on the cubic lattice $\bZ^3$, which can be mapped onto 
$\bT^2\times \bZ$. Our basic computational engine 
is written in C, but MATLAB is used for mapping and visualization. 
As mentioned previously, the snowfakes are depicted by drawing
visible boundaries of translates of the fundamental prism centered on sites of
$A_t$. Since this straightforward procedure makes jagged vertical
boundaries, we apply a smoothing algorithm at the boundary that enlarges the crystal 
by no more than one mesoscopic unit. (This algorithm is {\it not\/} applied 
to the small snowfake in Fig.~2.) MATLAB's {\tt patch} routine renders the faces. 
For better results we then emphasize edges using the {\tt line} routine.

MATLAB's visualization tools certainly provide adequate representations
for detailed investigation of the resulting crystals. They do not, however,
give a satisfactory comparison with the best snowflake photographs \cite{LR,Lib5,Lib6}, 
typically taken from directly above the (predominantly two-dimensional)
crystal, which is in turn illuminated from  below. This viewpoint can be effectively 
simulated by ray-tracing, as implemented here by the POV-Ray software \cite{POV}. 
Our program automatically outputs a file with a triangulation of the crystal's boundary,
which is then used by the {\tt mesh2} command in POV-Ray.

We would like to point out that both the algorithm and
visualization procedures require considerable computing power  
and memory. At present (fall 2007), our simulations are very time consuming, 
barely feasible on commercial personal computers. (In fact, an adaptive resolution 
algorithm is necessary to make the boundary 
descriptions manageable.) Progress in studying snowfakes is therefore 
quite slow, precluding systematic classification of the dynamics. Our goal has been to find 
representative examples that seem to replicate physical snow crystals and thereby 
shed light on their evolution.

For computational efficiency, if the diffusion step is isotropic one can exploit symmetry by taking the finite lattice to be a discrete hexagonal prism with patched wrap edge conditions. When $\phi=0$ and the initial state has complete symmetry, it thereby
suffices to compute the dynamics on $\frac 1{24}$ of the whole space. 
There are two good reasons for giving up complete symmetry of the rule. First, the
initial state may not be symmetric, and second, the diffusion may have a
drift. For computational efficiency, we only give up reflectional symmetry around the $xy$-plane 
(recall that the drift is only in the $\bZ$-direction), allowing the initial state
to depend on the $z$ coordinate, but retaining its hexagonal symmetry in the $x$ and $y$
coordinates. This increases the space and time demands of the fully symmetric program by a factor of 2.

The program stops automatically when the density at the edge of the lattice falls below 
a given proportion of the initial density (typically $2\rho/3$ or $\rho/2$), 
or when the crystal gets too close to the edge (snowfake radius greater than 80\% the radius of the system).

\section{Connection to pde, and size of the parameter space} 

Mathematically, our algorithm is a discrete space and time version 
of a {\it free boundary\/}, or {\it Stefan\/}, problem 
\cite{Lib2, Lib3, Lib4}. This is a partial differential equation 
(pde) in which the crystal is represented by a growing 
set $A_t$ and the density (i.e., supersaturation) of vapor
outside it as $u=u(x,t)$. Then $u$ is 0 on  the boundary $\partial A_t$, 
and satisfies the diffusion equation outside the crystal
\begin{equation}
\frac {\partial u}{\partial t}=\Delta u,\quad x\in A_t^c.
\tag {1.1}
\end{equation}
The velocity of the boundary at a point 
$x\in\partial A_t$ with outside normal $\nu$ is given by a 
function 
\begin{equation}
w\left( \frac{\partial \rho}{\partial \nu}, \nu\right).
\tag{1.2}
\end{equation}
Considering the slow growth of $A_t$, diffusion
equation (1.1) may be simplified to its equilibrium counterpart
$\Delta u=0$ \cite{Lib2, Lib3, Lib4}, which makes this into 
an anisotropic version of the {\it Hele-Shaw\/} problem. 

Presumably under diffusion scaling, in which space 
is scaled by $\epsilon$, time by $\epsilon^{-2}$, and $\epsilon\to 0$, 
the density field and the occupied set in our model converge to 
a solution of the Stefan problem. We hope to provide rigorous justification 
for this connection, and identification of the limit $w$ in 
terms of model parameters, in future work.

The boundary velocity function $w=w(\lambda,\nu)$ is defined for $\lambda\ge 0$ 
and three-dimensional unit vectors $\nu\in S^2$. In order to develop a 
rigorous mathematical theory, the most convenient assumptions are that $w$
is continuous in both variables,  
nondecreasing in $\lambda$, and satisfies $w(\lambda,\nu)\le C\lambda$ for 
some constant $C$ independent of $\lambda$ and $\nu$. Under these
conditions, the non-isotropic Stefan problem (1.1--1.2) has a unique viscosity 
solution at all times $t\ge 0$, starting from any smooth initial crystal. This is 
proved in \cite{Kim} for the isotropic case (when $w$ is constant); 
assuming the listed properties of $w$, the proof extends to our general setting. 
We should note, however, it has long been known that the crystal's boundary 
will not remain smooth \cite{SB}. Indeed, this will be no mystery once we present 
our simulations, which feature a considerable variety of singularities and instabilities. 
Presumably these make direct numerical computation with the 
pde very challenging, explaining why numerical pde-based models for 
snow crystal growth have not been satisfactory (cf.~\cite{Sch}). 
For further mathematical theory and references, we refer the reader to \cite{Kim, CK}. 

For the model studied here, $w(\lambda, u)$ will be linear in $\lambda$, 
since the attachment and melting rates are independent of the vapor density. 
This may not always be the case; in fact, some of the 
literature even considers the possibility that $w$ is 
non-monotone in $\lambda$ \cite{Lib3, GG3}. Analysis of such cases would 
present new theoretical challenges, and from simulations of our 
3d model it appears that nonmonotonicity is not needed for observed phenomena 
in nature. Monotone nonlinearity, arising from monotone density dependent rates,  
is harder to dismiss and worth further investigation -- for instance,  
it is possible that $w$ vanishes for very small $\lambda$. 

Once we accept that our scheme approximates the  
viscosity solution of (1.1--1.2), the macroscopic evolution of the 
crystal is uniquely determined by its initial state and 
the velocity function $w$. In turn, $w$ is determined by very few physical parameters,
perhaps just two: temperature and atmospheric pressure \cite{Lib2, Lib3, Lib4}. Therefore,
possible evolutions from a fixed seed comprise a three-dimensional manifold
(its coordinates being the supersaturation level, temperature, and pressure)
in an infinite-dimensional space of possible velocities $w$. Much of the ongoing snow crystal 
research constitutes an attempt to understand the structure of this manifold,
a daunting task since the underlying (perhaps quantum) attachment physics is very poorly understood,
controlled homogeneous environments are hard to design, and crystal evolution is difficult to record.  
Our model does not have these problems. Instead, its
main weakness is the number of free parameters that need to be tuned to approximate
$w$ at a particular temperature and pressure. It helps that our parameters have
intuitive meaning, but finding a particular realistic snowfake involves approximating 
an a priori infinite-dimensional object $w$ by one of finite but high dimensionality. 
The challenge is compounded by very incomplete information -- all that is typically observable 
in nature is the final crystal, which may have been subjected to numerous changes in conditions
and orientation during growth, as well as sublimation and perhaps even artifacts of
the recording process. It is thus no surprise that our parameter selection is
an arduous and imprecise task.

In the next section we will describe some ad hoc rules that we have used to generate our case 
studies, but the issue of parameter selection is in dire need of further investigation. 
What we can say is that the best examples are quite sensitive to perturbations in $w$. Thus they require
good approximations and a large number of judicious parameter choices. 
In addition, the dependence on the initial seed is often quite dramatic. 
These observations underscore both the marvel and the fragility of natural snowflakes.

At the same time, we wish to emphasize the conceptual simplicity of our model. 
The large parameter space is a consequence of geometry rather than an excessive number of modeling ingredients. Apart from the two scalar parameters -- density $\rho$ and drift $\phi$ -- we have only 
three vector parameters --- attachment threshold $\beta$, freezing rate 
$1-\kappa$, and melting rate $\mu$ --- whose high dimensionality 
arises from the many possible boundary arrangements.  
The parameter set can be reduced, but 
{\it some\/} tuning will always be necessary, as illustrated by the ``random'' 
crystal in Fig.~3. This was obtained by choosing $\kappa\equiv .1$, 
$\mu\equiv .001$, $\rho=.1$, $\phi=0$, and all $\beta$'s equal to 1 
except $\beta_{01}=1.73$ and $\beta_{10}=\beta_{20}=1.34$. These values are in a sensible 
neighborhood of the parameter space, but the last two attachment rates were selected by chance. 
The result has some physically reasonable features, but one immediately notices an excessive 
density of branches and inordinately high ridges. 

\begin{center}
{\includegraphics[trim=0cm 0cm 0cm 0cm, clip, height=5cm]{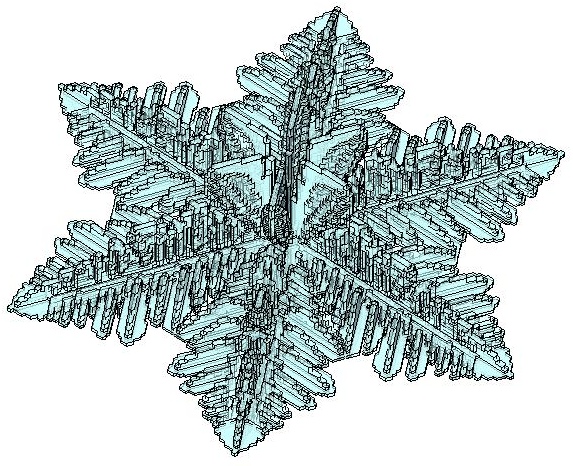}}
\end{center}
\vskip-0.5cm

{\bf Fig.~3.} A ``failed'' snowfake.
\vskip1cm

\section{Effective choice of parameters for simulations}

While optimal choices of parameters requires considerable guesswork, 
there are a few guidelines we have developed. Some come from mathematical 
arguments, others from experimentation; both are described in this section. 

Our simulator represents diffusion by discrete averaging in time $t$, which 
is also discrete. The bulk effects of this operation expand at the rate $\sqrt t$, although 
the extreme radius of its influence (or {\it light cone\/}) grows linearly in $t$. If the initial 
density $\rho$ of our discrete vapor field is too large, then the  
crystal may expand in some direction as fast as the light cone, or perhaps fall behind 
it by $\mathcal O(\sqrt t)$.  We call parameter sets leading to this behavior the 
{\it Packard regime\/}; it is clearly not 
physical, as it depends on the discrete nature of the averaging.
However, systems of this sort are able to generate fractal plates reminiscent of Packard 
snowflakes \cite{Pac, GG3} and exhibit one variety of faceting (cf.~\cite{NR}).
In our simulations we systematically avoid the Packard regime 
by keeping the density low. For the extremal points of our snowfakes not to expand 
at light speed, the conditions are 
$$
(1-\kappa_{01})\rho<\beta_{01},\quad (1-\kappa_{10})\rho< \beta_{10}, 
$$
as is easy to see from the description of the rule. Our densities are typically 
considerably smaller, since large densities generate expansion that is too rapid to be realistic, 
at least in its initial stages. As mentioned previously, a surprisingly important role 
is also played by the choice of initial seed.

On the other hand, it is clear that a very large melting rate will 
stop growth altogether. This happens if the flow out of the boundary mass  
exceeds the flow in just before that mass exceeds the threshold for attachment. 
A sufficient condition for continual growth in all directions is therefore 
$$
\mu_{01}\beta_{01}<(1-\kappa_{01})\rho, \quad \mu_{10}\beta_{10}<(1-\kappa_{10})\rho, 
$$ 
since the 01 and 10 boundary arrangements always have the slowest potential growth. 
In the great majority of examples we will present, parameters for the 20 and 10 arrangements 
agree. In this case, the last condition is necessary as well --- if it does not hold, then
the growth is convex-confined in the $\bT$-direction. 

Let us now describe a few rules of thumb when searching for snowfakes that emulate nature.
We commonly start with a {\it reduced\/} parameter set. Namely, we set the 
$\kappa$'s to a common value, say, $\kappa\equiv .1$. Then we select
two different $\beta$ parameters, $\beta_{01}$ and $\beta_{10}=\beta_{20}=\beta_{11}$,
with all the remaining $\beta$'s fixed to 1. The size of $\beta_{20}$ controls the
strength of the convexifying mechanism, assumed to be the same in both the $xy$ and $z$ 
directions. Indeed, if $\beta_{20}$ is large, then the crystal will remain 
a perfect hexagonal prism for a long time.
The only other parameters are the common value of all $\mu$'s and 
the vapor density $\rho$. This is a more manageable
four-parameter space that encodes four essential elements of
three-dimensional snowflake growth, each with a single tunable
parameter: diffusing supersaturation level
($\rho$), convexifying strength ($\beta_{20}$), quasi-liquid layer smoothing 
($\mu$), and preference for the $\bZ$-direction over the $\bT$-direction ($\beta_{01}/\beta_{20}$). 
This scheme is used to identify the neighborhood of a desired morphological type in phase space. 
Then parameters are perturbed for added realism.

One of the most important lessons of our two-dimensional model \cite{GG3} was
that the melting parameter $\mu$ inhibits side-branching and is therefore
important for dendrite formation. When $\mu\equiv 0$, it seems impossible to
avoid an excessive density of branches. Indeed, this role of $\mu$ 
is easily understood. Namely, $\mu$ creates a positive density at the 
boundary, due to flow out of the boundary layer. This density has the effect of 
reducing the ambient vapor density by a fixed amount, independent of location, 
and hence disproportionately affects regions of smaller density. (To a very rough 
first approximation \cite{Lib4}, the expansion speed is proportional 
to $\sqrt\rho/\sqrt t$ when $\mu\equiv 0$.)  Since there is clearly less mass between branches 
than at the tips, growth and side branching there gets stunted by increasing $\mu$. 

Realistic ``classic'' dendrites occur for a relatively narrow range of choices for $\mu$, once the
other parameters are held fixed. Typically, though, the other parameters need to be
perturbed along with $\mu$; increasing $\mu$ alone tends to erode all complex structure. 

The markings seen on snow crystal plates are sometimes called {\it hieroglyphs\/}. These often  
have fairly regular geometric forms, such as ridges, flumes, ribs, and circular shapes,  
but can also exhibit more chaotic patterns. In photomicrograph collections  
\cite{BH, LR, Lib5, Lib6} it is usually unclear whether the marks are on the outside
of the crystal or within what we call sandwich plates. In our experiments, the 
inner structures are {\it much\/} more prevalent, so we are glad to observe that they are 
abundant in nature \cite{EMP}. To obtain nice {\it outer\/} markings, 
the ratio $\beta_{01}/\beta_{20}$ needs to be sufficiently large, but there is then a 
tendency for the crystal to become too three-dimensional. 
Again, the correct choice is often rather delicate. Inner markings occur generically for small values of this ratio. 

Finally, different $\kappa$'s may appear to be a more natural mechanism
to enforce anisotropy than different $\beta$'s, as they directly 
correspond to sticking, or {\it killing\/}, of particles at the crystal's boundary. However, for this
effect to be significant, the $\kappa$'s need to be very close to 1; otherwise
the killing at the crystal boundary is too rapid to make a difference, and
then the already slow growth proceeds at an even more sluggish pace. While
less physically appealing, we view the $\beta$'s as a reasonable compromise for the sake of computational efficiency.

\vskip0.5cm

\section{Variants and extensions of the model} 

\subsection{Uniform snowfakes} 

Since attachment thresholds $\beta$ vary, the 
mass of the final crystal is not uniform. There is  
a variant of our algorithm that removes this defect with little change 
in observed morphology. Assume that there is no automatic filling of 
holes; instead, boundary mass exactly 1 is needed for 
attachment when $n_t^{\bT}(x)\ge 4$ and $n_t^{\bZ}(x)\ge 1$.
Then a uniform crystal is obtained by performing the following additional step 
just after step {\it iii} in the simulator:

\noindent{\it iii'\/. Post-attachment mass redistribution}

To redistribute any excess mass from the attached site to its unattached neighbors, 
let
$$
n_t^{c}(x)=\#\{y\in \cN_x: a_t^\circ(y)=0\}
$$
be the number of non-attached neighbors. Then, for every $x$ with $a_t^\circ(x)=0$,
$$
b_t'(x)=b_t^\circ(x)+\sum_{y:a_t^\circ(y)=1} \frac {b_t^\circ(y)-1}{n_t^c(y)}.
$$

\subsection{Simulation without symmetry}

As explained in Section 3, at the cost of a 24-fold slowdown compared to our fully symmetric model, 
implementation of the algorithm without exploiting symmetry makes it possible to study 
the evolution from arbitrary initial seeds. Such an extension is necessary in order to 
produce snowfakes corresponding to exotic forms such as triangular crystals, split stars, 
and bullets. We have conducted a few experiments along these lines with our planar model \cite{GG3}, 
but in three dimensions a simulator dramatically faster than our current one is needed. 
We have future plans to develop a suitably high-performance parallel version.

\subsection{Random dynamics} 

Our only three-dimensional snowfakes to date are deterministic, since randomness would also 
require the just discussed simulation without symmetry. 
We propose to include an additional  parameter $\epsilon$ 
representing residual noise on the mesoscopic scale, as we did in the two-dimensional setting \cite{GG3}. 
Again, $\epsilon$ would need to be quite small, say on the order $10^{-5}$. 
The random perturbation of diffusive mass from \cite{GG3} is not suitable in 3d since it is not 
physical to violate mass conservation. Instead, a small random slowdown in the 
diffusion rate is more appropriate. To this end, first denote the (linear) 
operation on the field $d_t^\circ$ in (1a--1c) by $\cD$; thus 
step {\it i\/} can be written as $d_t'''=\cD(d_t^\circ)$. 
Next, let $\xi_t(x)$, $t\ge0$, $x\in \bT\times\bZ$, be independent~random variables, equal to 
$\epsilon>0$ or 0, each with probability $1/2$. Here the field $\xi$ represents 
the proportion of particles  that refuse to diffuse at position $x$ and time $t$. 
The randomized step {\it i\/} now reads
$$
d_t'''=\cD((1-\xi_t)d_t^\circ)+\xi_t d_t^\circ=\cD(d_t^\circ)+\xi_t d_t^\circ- \cD(\xi_td_t^\circ). 
$$
In a natural way, this represents small random temperature fluctuations in space and time. 

Similarly, one could introduce a small proportion of particles that refuse to freeze in (2b), 
or melt in (4); e.g., (2b) would be replaced by 
\begin{equation*}
\begin{aligned}
&b_t'(x)=b_t^\circ(x)+(1-\kappa(n_t^{\bT}(x), n_t^{\bZ}(x)))d_t^\circ(x)(1-\xi_t(x)), \\
&d_t'(x)=\kappa(n_t^{\bT}(x), n_t^{\bZ}(x)) d_t^\circ(x)(1-\xi_t(x))+d_t^\circ(x)\xi_t(x).\\
\end{aligned}
\end{equation*}
 
\vskip0.5cm

\section{Case study $i$ : ridges and plates}

Our prototypical snowfake has $\rho=.1$
and the canonical initial state of radius 2
and thickness 1. Fig.~4 depicts the crystal after 70000 time steps, when
its radius is about 350. Its parameters are $\beta_{01}=2.5$, $\beta_{10}=\beta_{20}=\beta_{11}=2$,
$\beta_{30}=\beta_{21}=\beta_{31}=1$, $\kappa\equiv .1$, $\mu\equiv .001$, 
and $\phi=0$.  

\vskip-0.3cm
\null\hskip-0.6cm
\begin{minipage}[b]{8cm}
\includegraphics[trim=8cm 1cm 8cm 2cm, clip, height=2.9in]{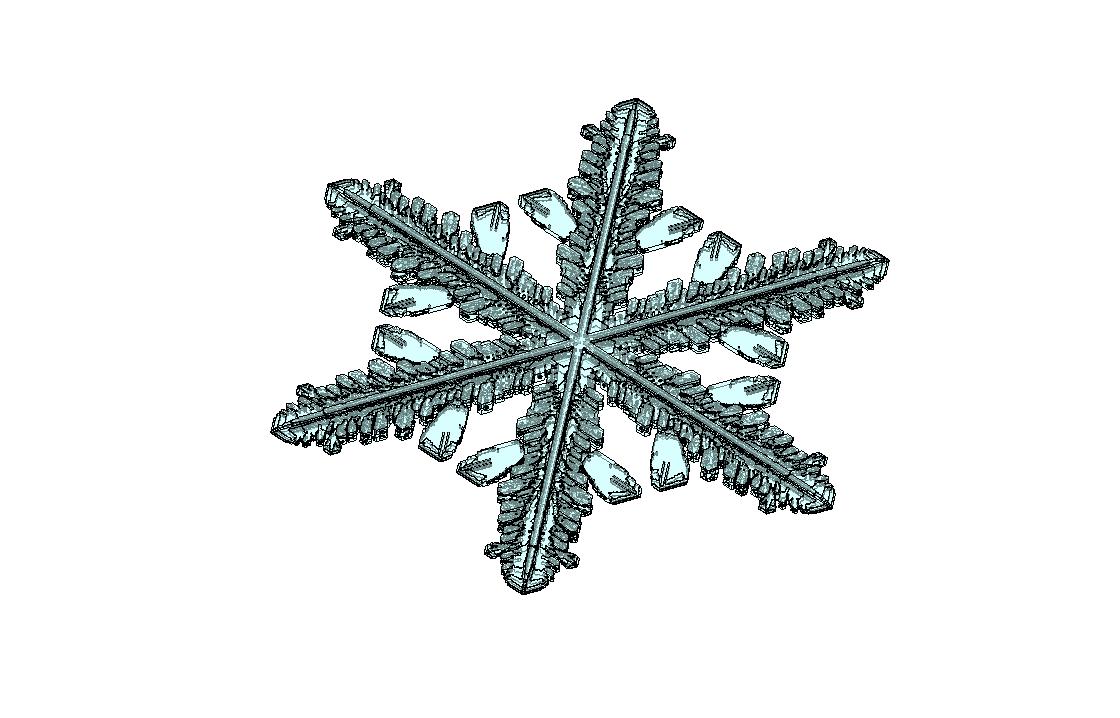}
\end{minipage}
\hskip0.5cm
\begin{minipage}[b][7cm][t]{15cm}
\vskip-0.25cm
\includegraphics[trim=1.2cm 0cm 1.2cm 0cm, clip, height=2.5in]{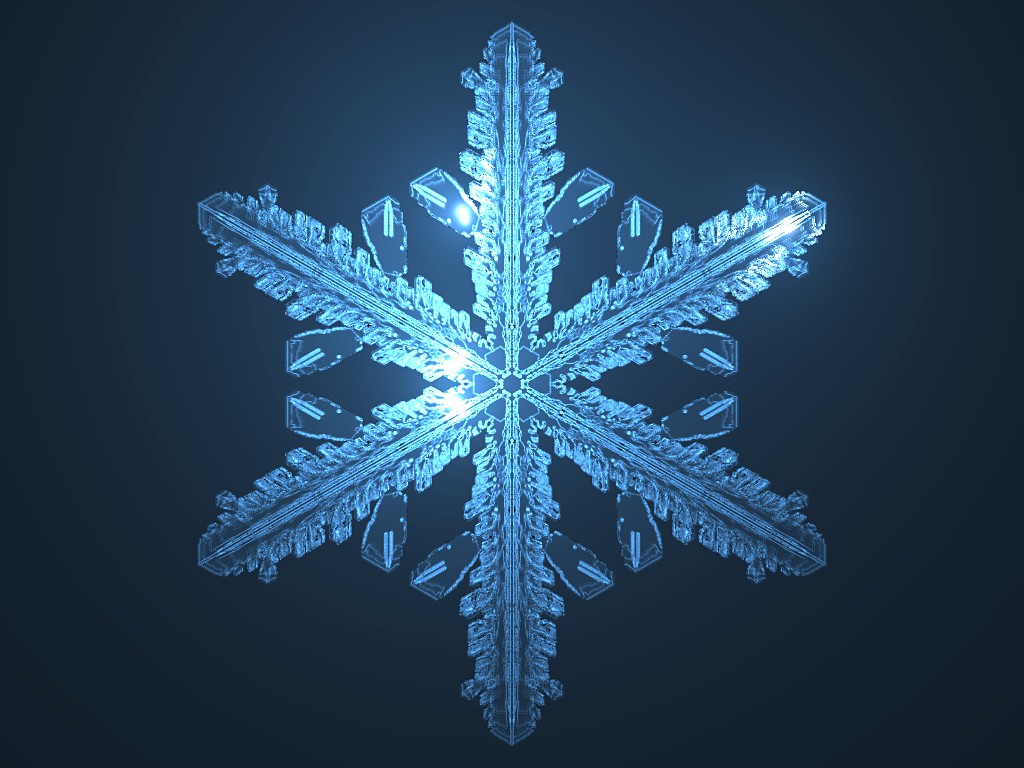}
\end{minipage}
\vskip-0.75cm
{\bf Fig.~4.} The oblique (MATLAB-rendered) and top (ray-traced) views of the crystal.
\vskip0.5cm

\vfill\eject

We invite the reader to compare the simulated crystal with some
of the photographs at \cite{Lib5} and especially with Fig.~1(h) in
\cite{TEWF}, a snowflake obtained at temperature about $-13^\circ$C.
We think of our length unit as about $1\mu$m, so even the sizes of the two
objects roughly match.
Perhaps the most striking features shared by the snowfake in Fig.~4
and physical ones are the {\it ridges\/}, elevations in the middle
of each main branch, with less pronounced
counterparts on the side branches. We begin by illustrating
how these ridges are formed and maintained. In the process we also
encounter the {\it branching instability\/}, when the initial growth of
a thin hexagonal plate is no longer viable and it gives birth to the six main
branches.

As shown in Fig.~5, ridges are formed quite early in the evolution, by mesoscopic bumps
known as {\it macrosteps\/} that are near, but not too near, the center of the plate. 
This is how the ridges grow ({\it very\/} slowly) in the vertical direction --- 
compare with times 4044 and 7099, which also feature such bumps. The ridges spread to a characteristic width, 
but sharpen to a point near the branch tip. One can also observe the
commonly observed {\it flumes\/} (called {\it grooves\/} in \cite{Lib6}) that form on both 
sides of a ridge.

\begin{center}
   {\includegraphics[trim=18cm 6cm 17cm 5cm, clip, height=3cm]{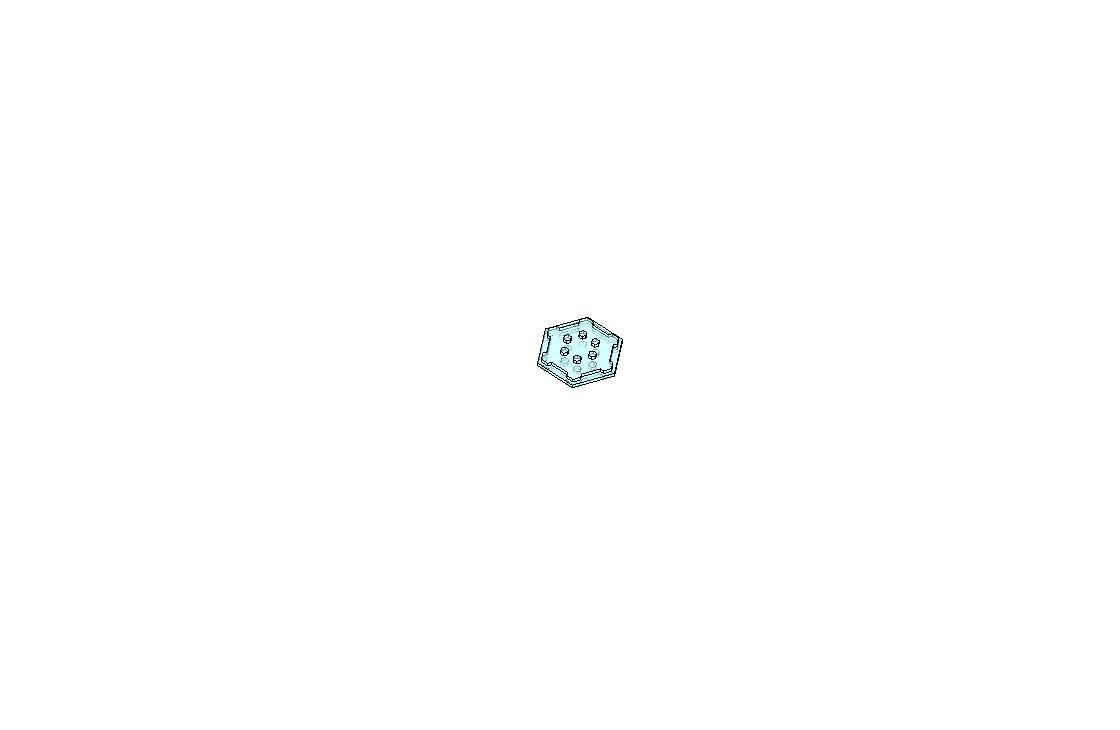}
   \includegraphics[trim=18cm 6cm 16.5cm 5cm, clip, height=3cm]{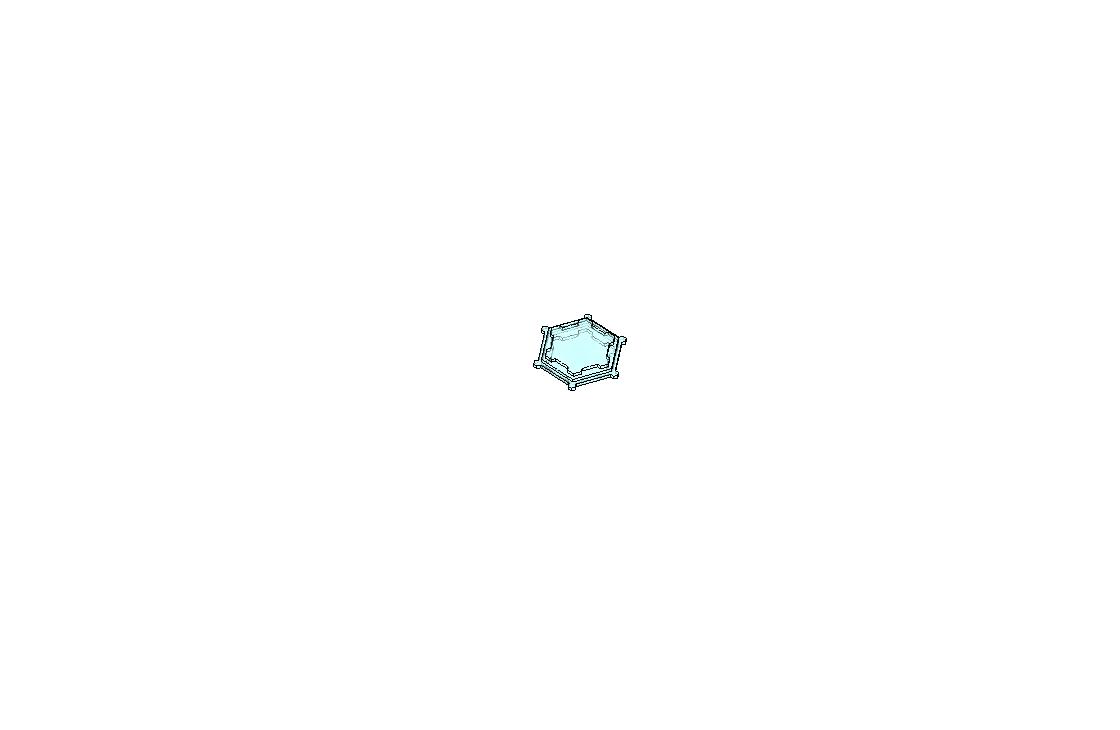}
  \includegraphics[trim=17.5cm 6cm 16cm 5cm, clip, height=3cm]{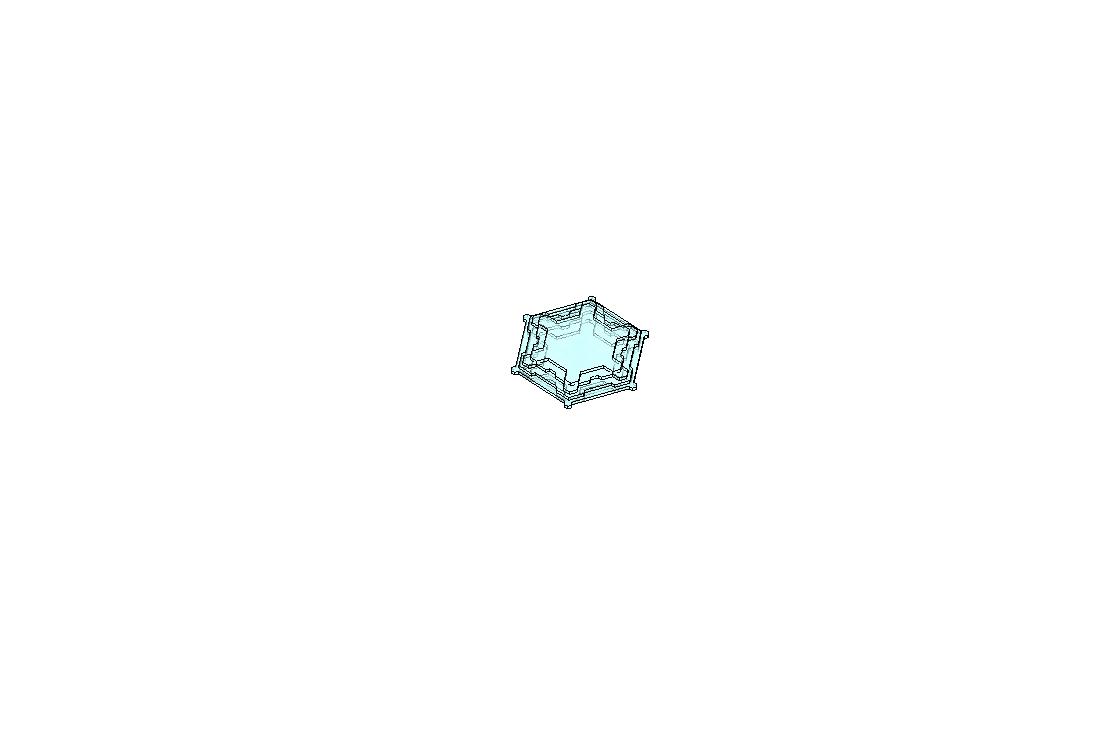}
\includegraphics[trim=16cm 6cm 14cm 5cm, clip, height=3cm]{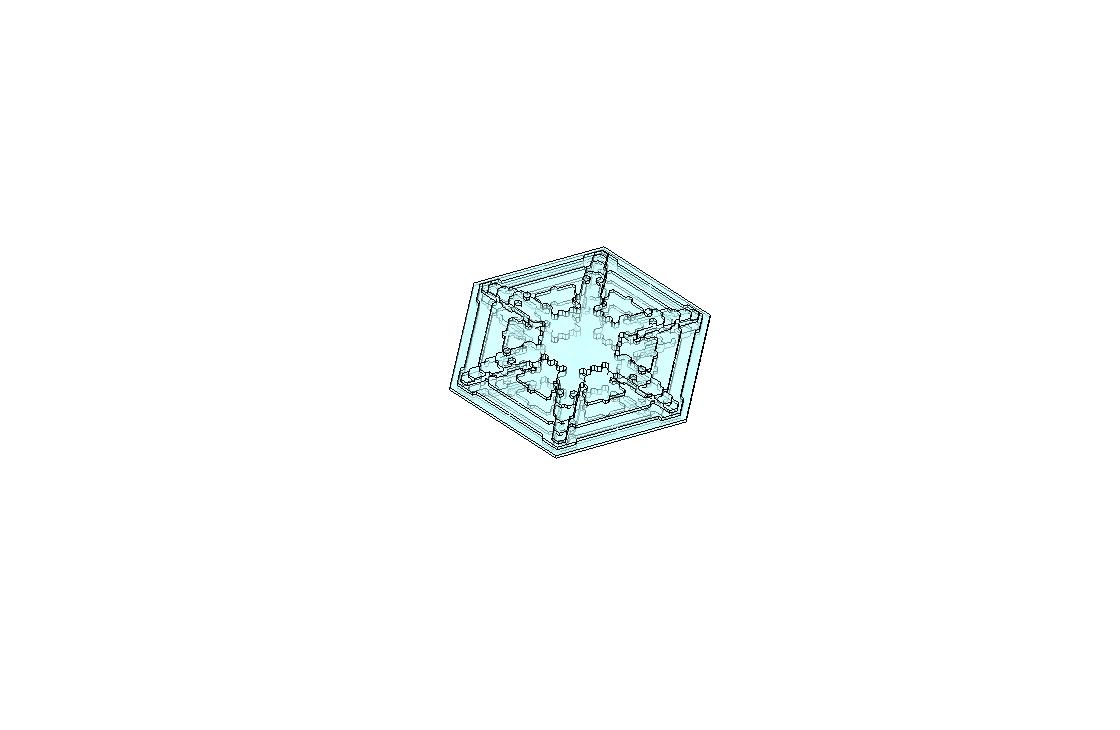}
\includegraphics[trim=14cm 6cm 13cm 5cm, clip, height=3cm]{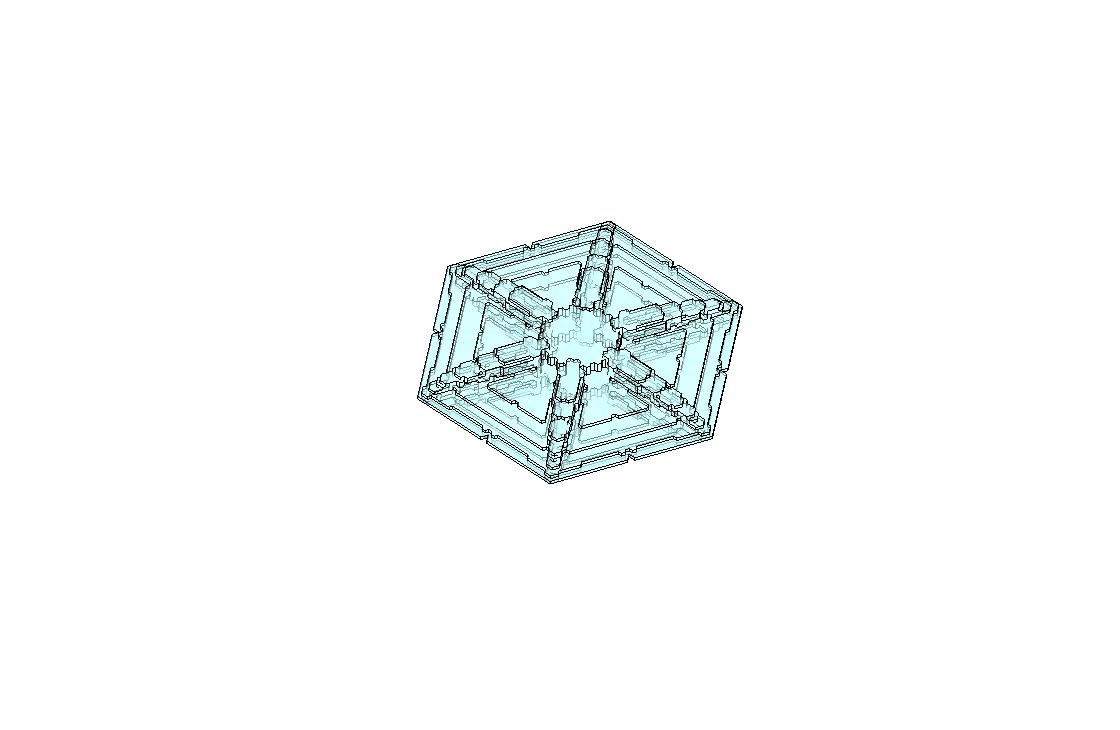}
   \includegraphics[trim=13.5cm 6cm 12cm 5cm, clip, height=3cm]{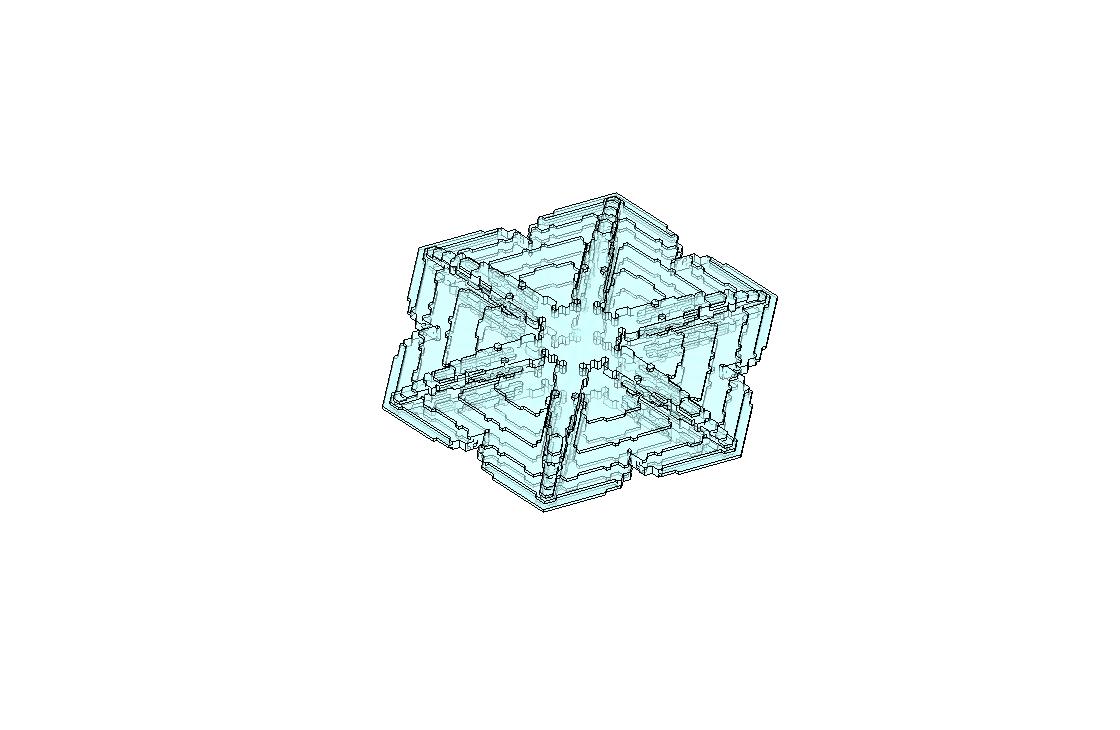}
   \includegraphics[trim=11cm 6cm 10cm 5cm, clip, height=3cm]{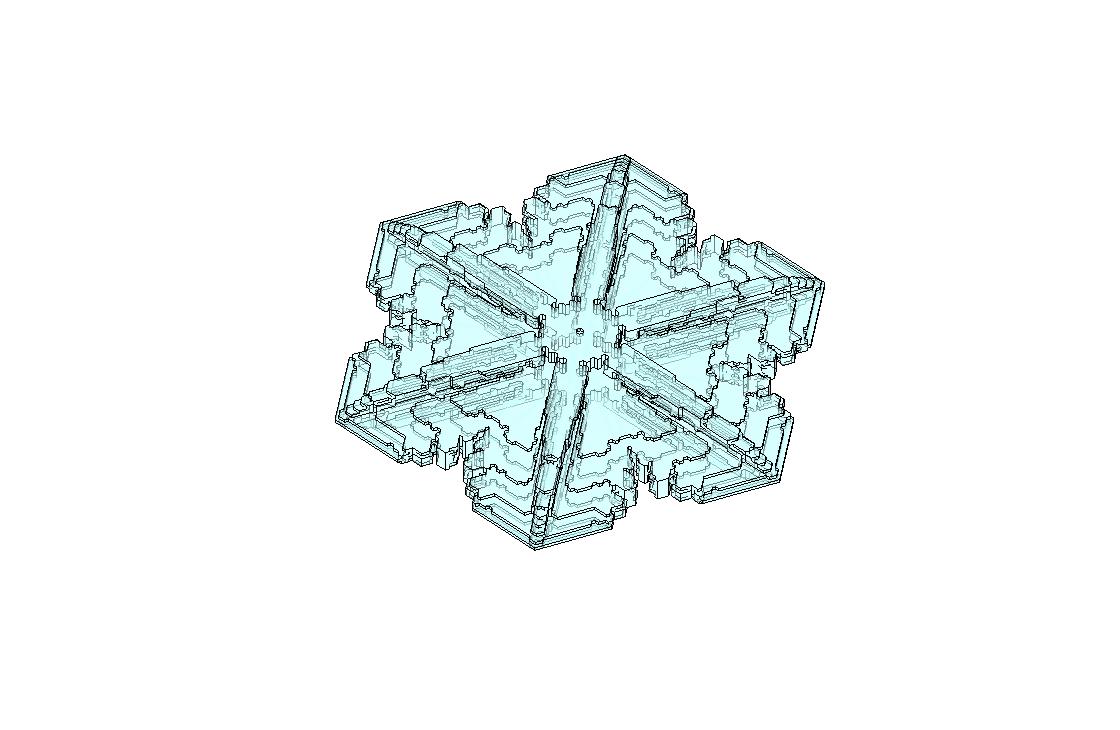}
   }
\end{center}

\vskip-0.5cm
{\bf Fig.~5.} The crystal at times 820, 863, 1600, 4044, 5500, 7099, and 9500.
\vskip0.5cm

The small indentation that emerges, due to lower vapor density,
in the middle of each prism facet at time 5500, has appeared several times before. 
However, this is the first instance when the growth is unable to repair it. 
Instead, the growth there virtually stops, while the six main arms continue to grow
and eventually produce two types of side branches: common, relatively thick
double-plated branches that we call {\it sandwich plates\/}, and more unusual thin plates with their
own ridges. The tip of each arm assumes its characteristic shape
by the final frame of Fig.~5.

It is perhaps surprising how dramatically this scenario depends 
on the initial (micron scale) state. Keeping everything else the same, we change the
initial prism to one with radius 2 and thickness 3. The previous rather complex and aesthetically
pleasing evolution is replaced by a growing double plate (Fig.~6). (Remarkably, even adding a small 
drift does not help matters much.) This dichotomy arises frequently in our model --- within a 
neighborhood of the parameter space that produces planar crystals there are
two stable attractors: one with outside ridges and the other a split plate
with ridges on the inside. As much of the literature points out, split
plates are extremely common in physical crystals (cf.~\cite{Iwa}).

\vskip-0.25cm
\begin{minipage}[b]{7cm}
\includegraphics[trim=8cm 1cm 8cm 2cm, clip, height=2.5in]{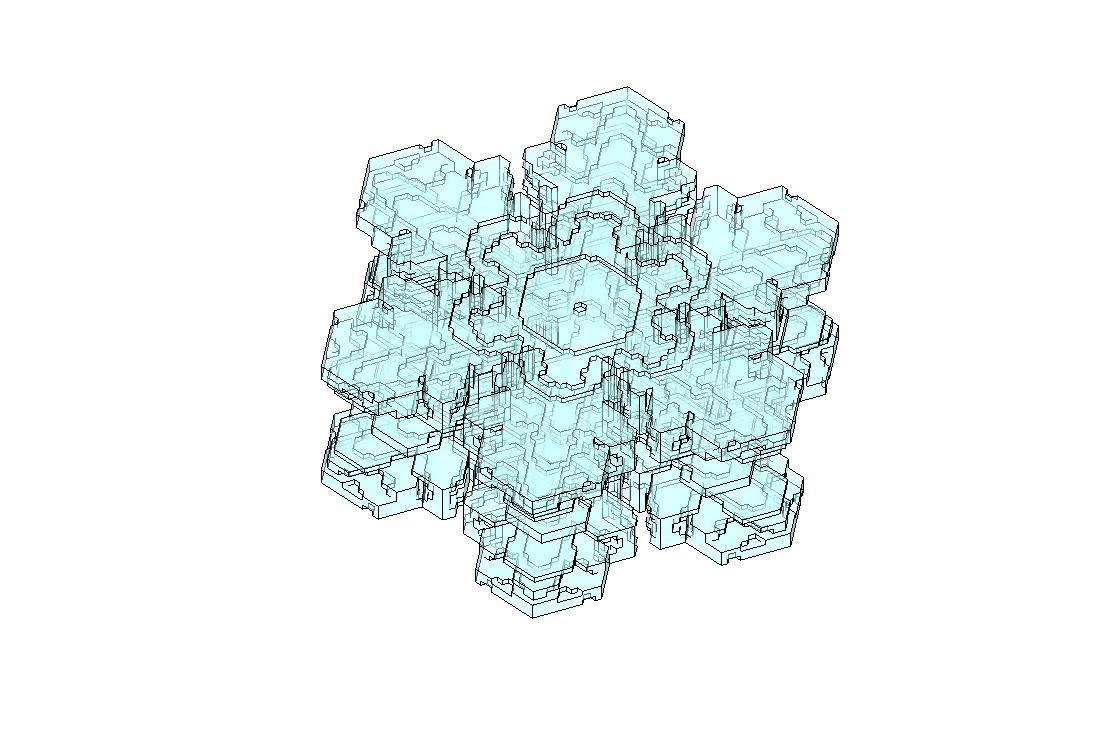}
\end{minipage}
\hskip0.05cm
\begin{minipage}[b][5cm][t]{3cm}
\includegraphics[trim=5cm 8cm 0cm 6cm, clip, height=1.0in]{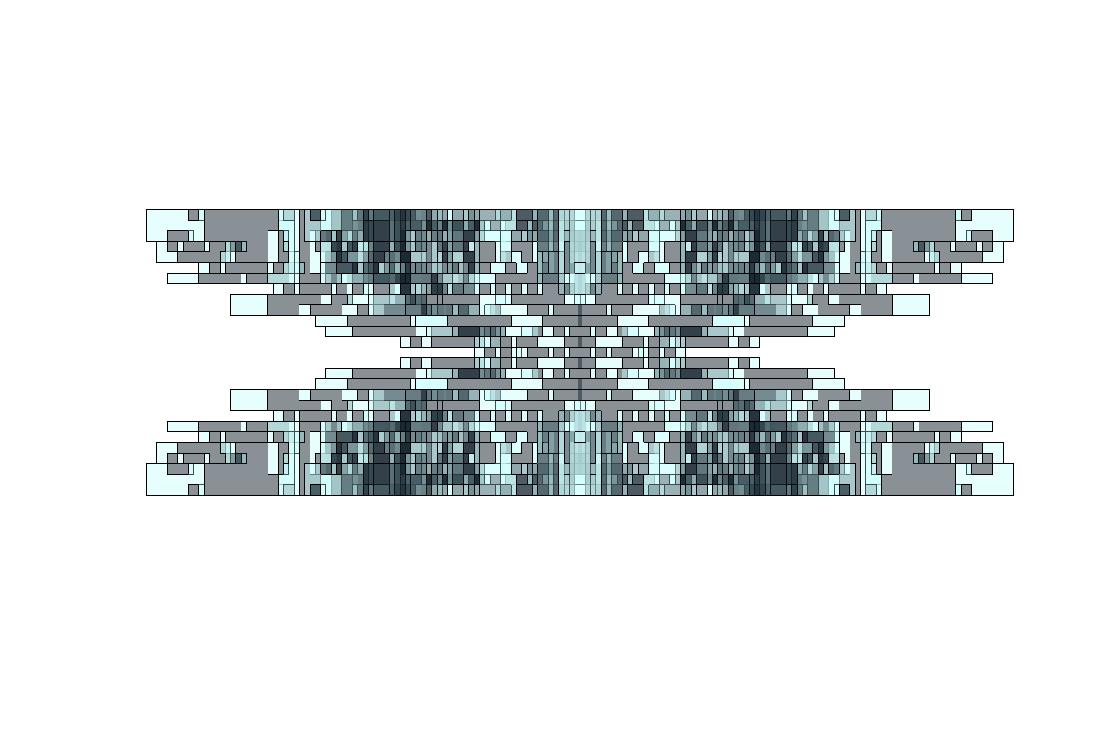}
\end{minipage}

\vskip-1cm
{\bf Fig.~6.} Oblique and side views of the crystal from a different initial
state.
\vskip0.2cm

Finally, let us experiment with changing the density $\rho$. We exhibit five
crystals, each with the canonical initial condition and all other parameters 
of the prototype unchanged, but at different
densities and different final times. Dramatically lower density does
promote faceting (\cite{Lib6, LR}), but a moderate perturbation
seems to mainly promote slower growth, without a
change in morphology.

\null\hskip-0.6cm
\begin{minipage}[b]{8cm}
\includegraphics[trim=8cm 1cm 8cm 2cm, clip, height=2.9in]{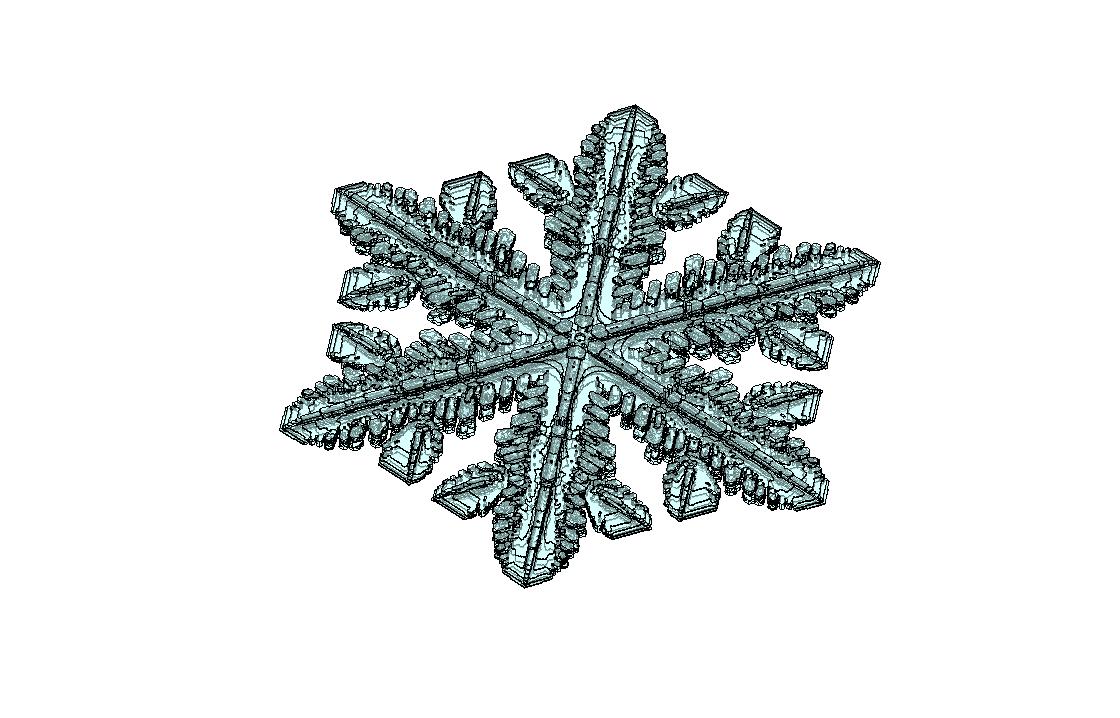}
\end{minipage}
\hskip0.5cm
\begin{minipage}[b][7cm][t]{15cm}
\vskip-0.25cm
\includegraphics[trim=1.2cm 0cm 1.2cm 0cm, clip, height=2.5in]{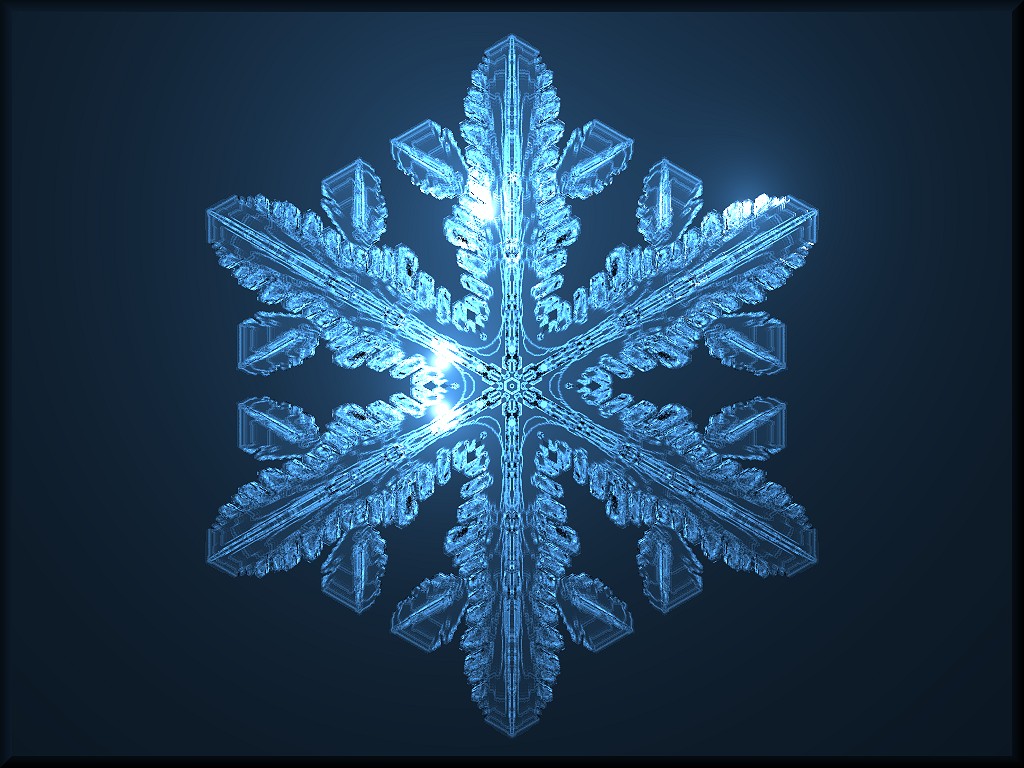}
\end{minipage}
\hfill\break
\vskip-2.25cm
\null\hskip5cm
\includegraphics[trim=14.5cm 6cm 12.5cm 4cm, clip, height=1.75in]{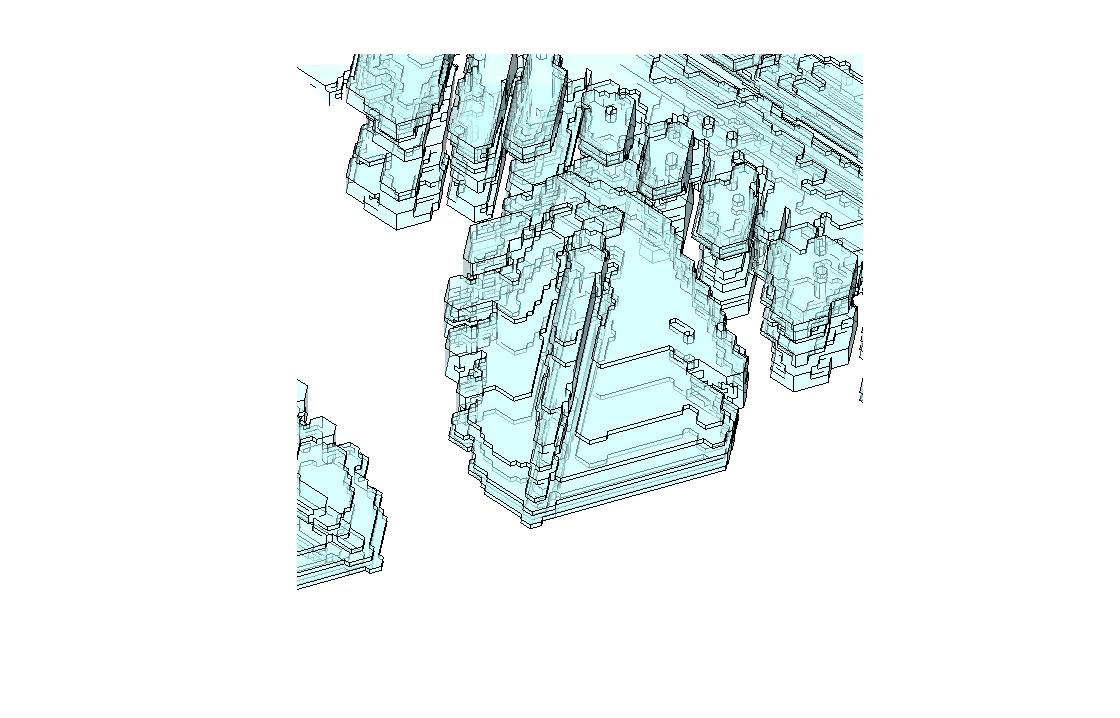}
\vskip-0.2cm
{\bf Fig.~7.} At density $\rho=.15$, the side branches have particularly well-defined
ridges.
\vskip0.5cm

\vfill\eject

\null\hskip-0.6cm
\begin{minipage}[b]{8cm}
\includegraphics[trim=8cm 1cm 8cm 2cm, clip, height=2.9in]{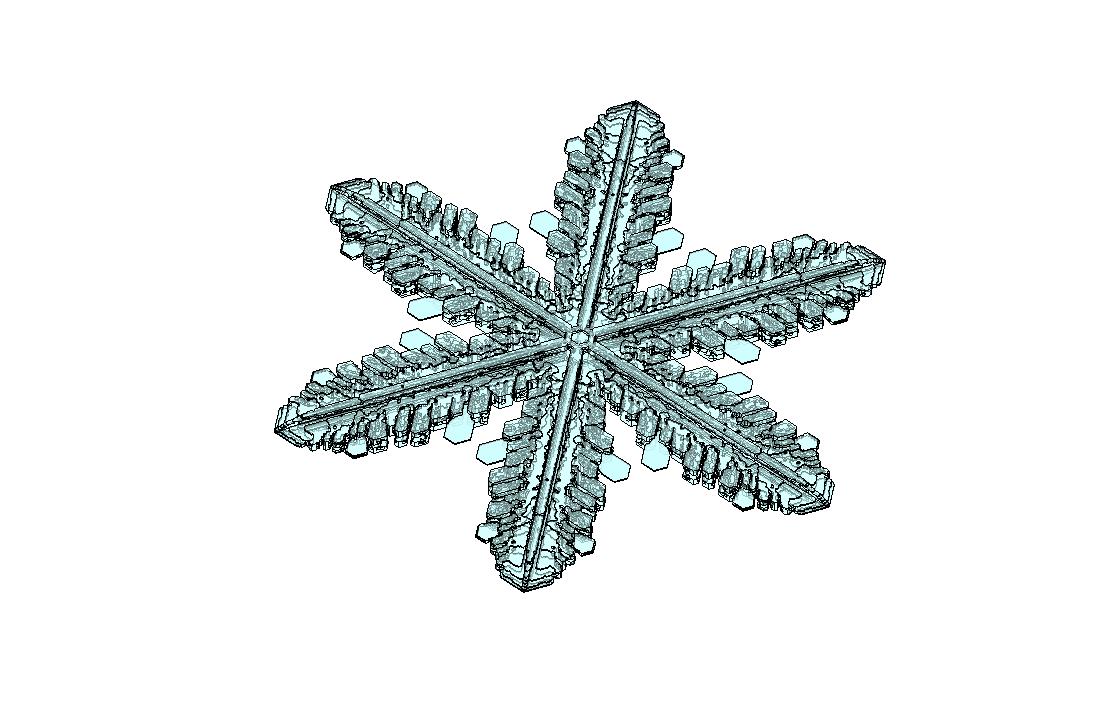}
\end{minipage}
\hskip0.5cm
\begin{minipage}[b][7cm][t]{15cm}
\vskip-0.25cm
\includegraphics[trim=0.7cm 0cm 0.7cm 0cm, clip, height=2.5in]{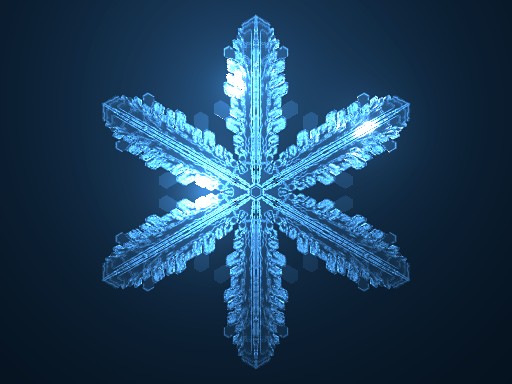}
\end{minipage}
\hfill\break
\vskip-2.25cm
\null\hskip5cm
\includegraphics[trim=14cm 2cm 11cm 6cm, clip, height=1.75in]{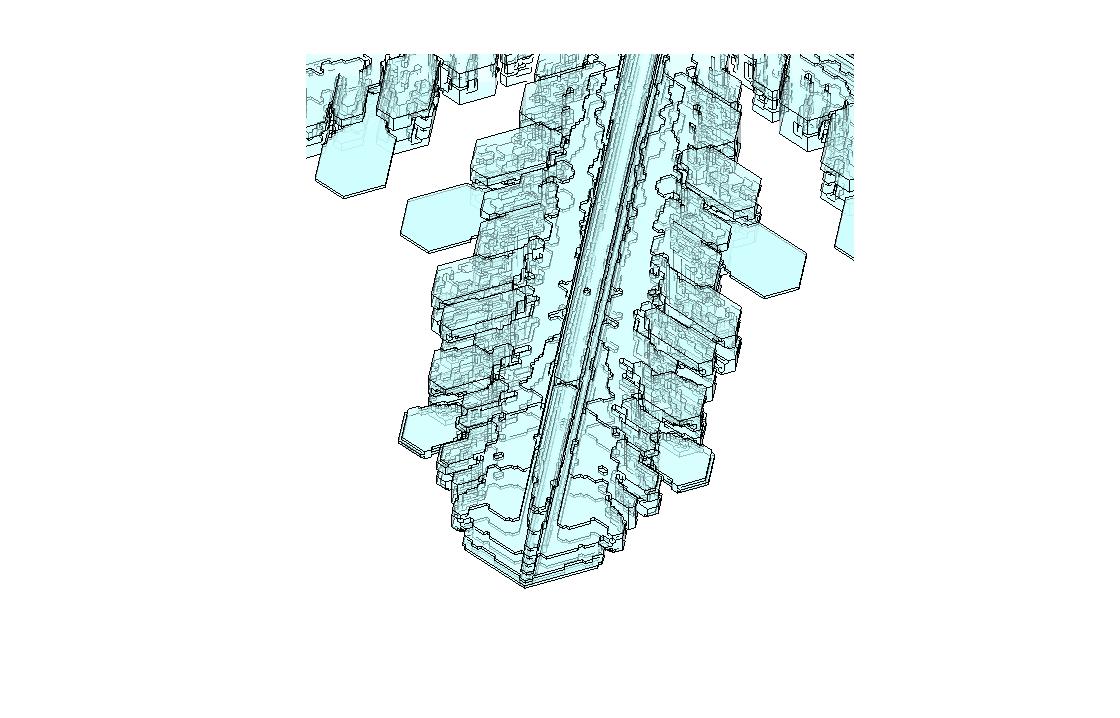}
\vskip-0.6cm
{\bf Fig.~8.} At density $\rho=.09$, the flumes are well-delineated.
\vskip1.5cm

\null\hskip-0.6cm
\begin{minipage}[b]{8cm}
\includegraphics[trim=8cm 1cm 8cm 2.0cm, clip, height=2.9in]{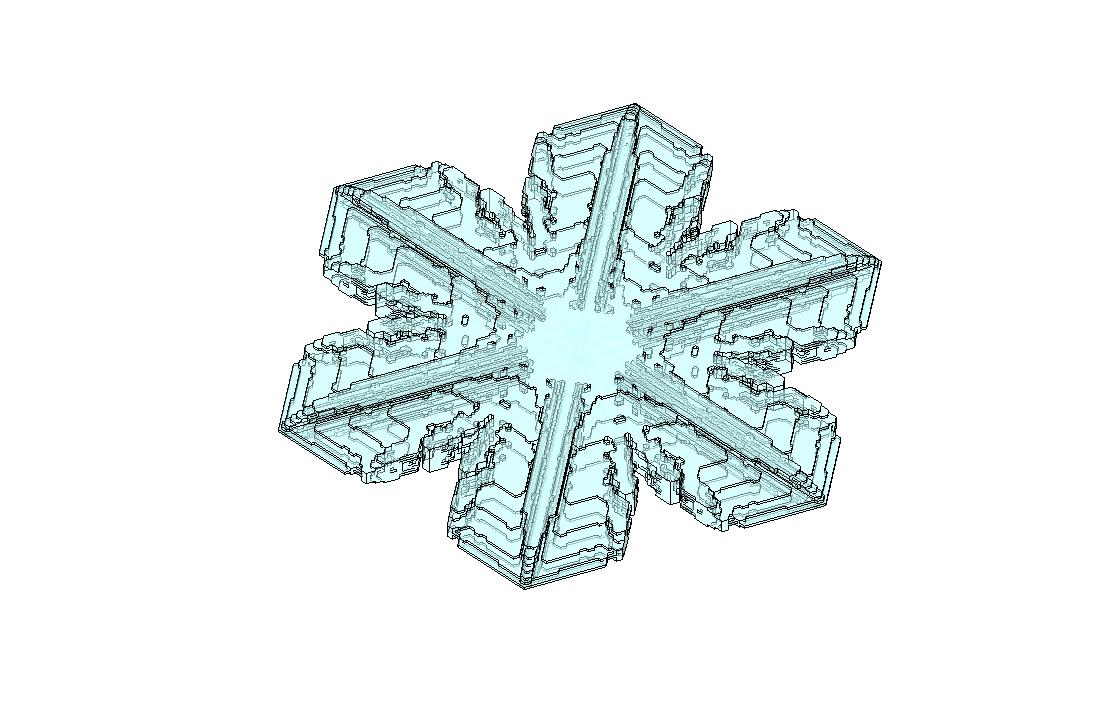}
\end{minipage}
\hskip0.5cm
\begin{minipage}[b][7cm][t]{15cm}
\vskip-0.25cm
\includegraphics[trim=0.7cm 0cm 0.7cm 0cm, clip, height=2.5in]{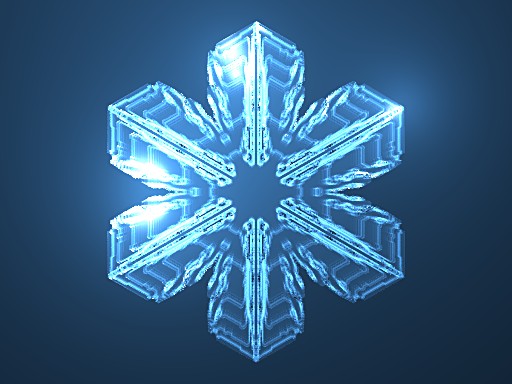}
\end{minipage}
\vskip-0.75cm
{\bf Fig.~9.} Density $\rho=.05$ results in sectored plates.
\vskip1cm

\null\hskip-0.6cm
\begin{minipage}[b]{8cm}
\includegraphics[trim=8cm 1cm 8cm 2cm, clip, height=2.9in]{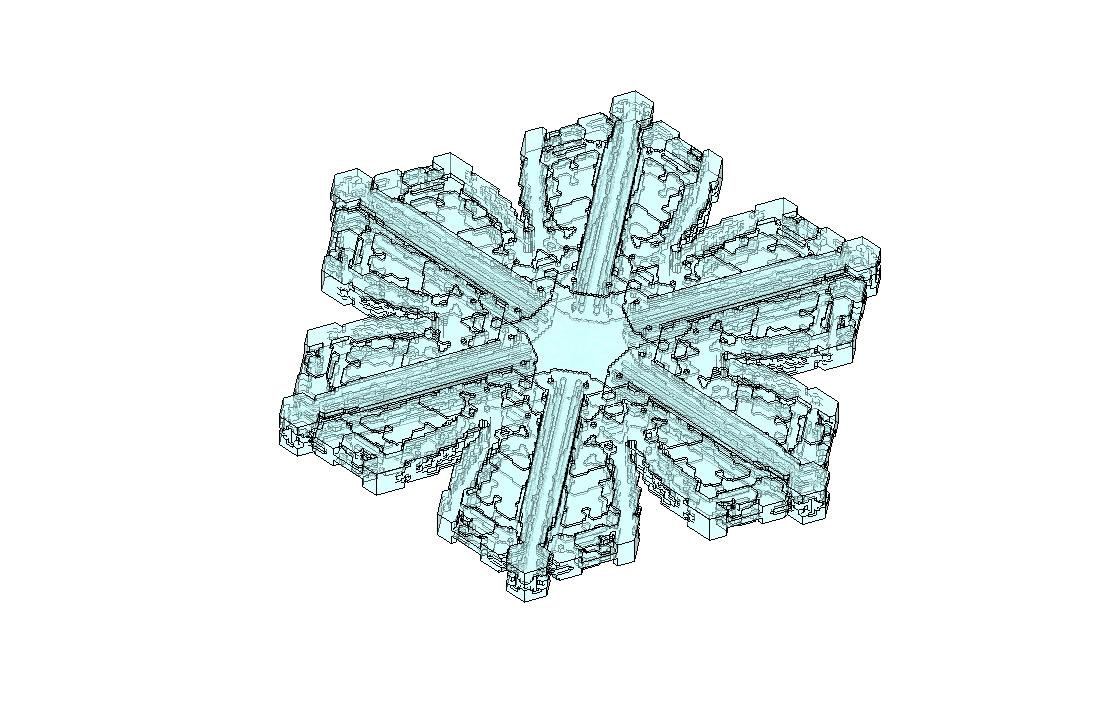}
\end{minipage}
\hskip0.5cm
\begin{minipage}[b][7cm][t]{15cm}
\vskip-0.25cm
\includegraphics[trim=0.7cm 0cm 0.7cm 0cm, clip, height=2.5in]{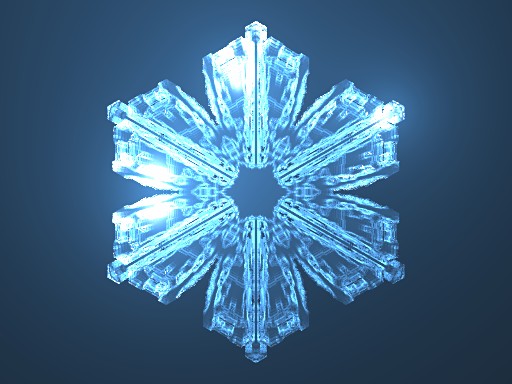}
\end{minipage}
\vskip-0.75cm
{\bf Fig.~10.} Density $\rho=.045$ results in sectored branches.
\vskip1.5cm

\null\hskip-0.6cm
\begin{minipage}[b]{8cm}
\includegraphics[trim=8cm 1cm 8cm 2cm, clip, height=2.9in]{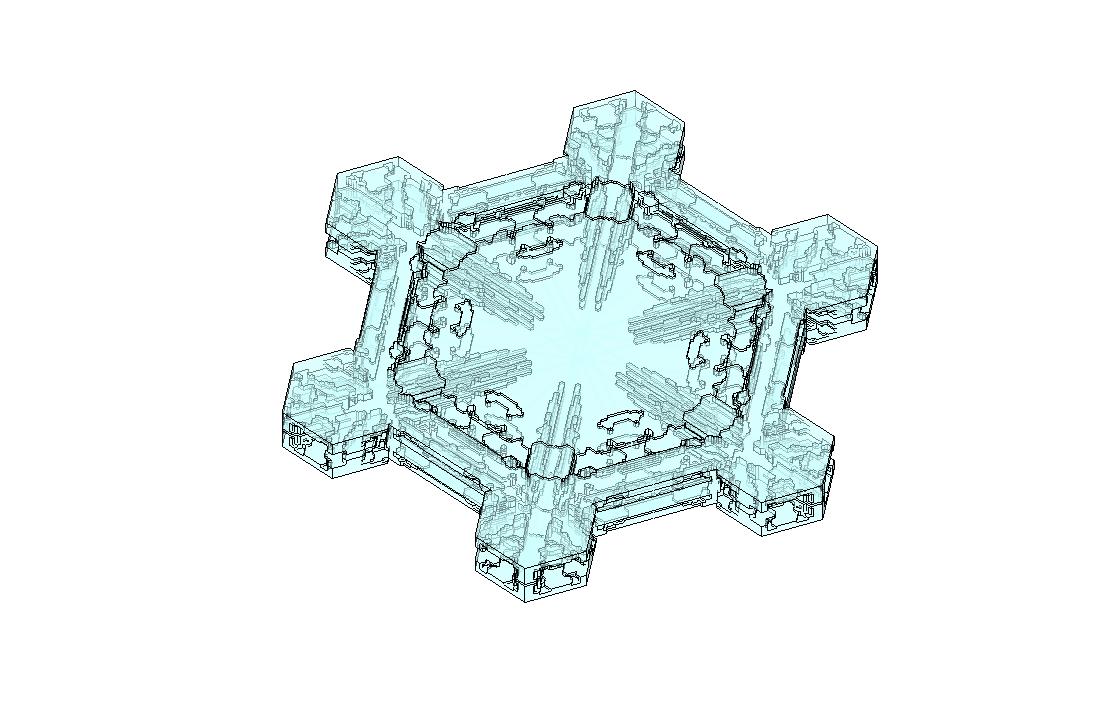}
\end{minipage}
\hskip0.5cm
\begin{minipage}[b][7cm][t]{15cm}
\vskip-0.25cm
\includegraphics[trim=0.7cm 0.0cm 0.7cm 0.0cm, clip, height=2.5in]{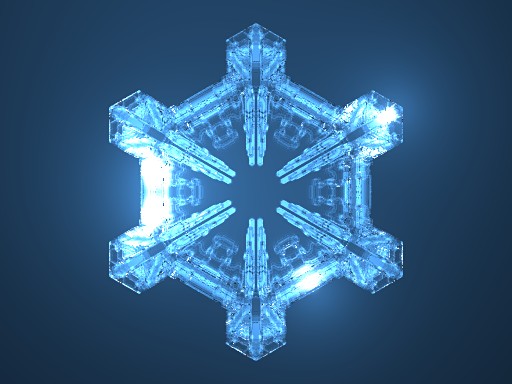}
\end{minipage}
\vskip-0.75cm
{\bf Fig.~11.} Density $\rho=.4$ results in sandwich plates with inner ridges.
\vskip.75cm

The example in Fig.~11 (pictured at time $120000$)
never undergoes the branching instability illustrated in Fig.~5,
although it does develop fairly standard ridges that persist until
about time $40000$. This is the time shown in the first frame of Fig.~12; 
subsequent frames show the evolution in time increments of 10000.
We observe that a completely different {\it sandwich instability\/} takes
place: first the tips and then the sides of the snowfake thicken and
develop sandwich plates. It is also clear from the time sequence
that this morphological change is accompanied by a significant slowdown
in growth. We should emphasize that this slowdown is {\it not\/} due to the
depletion of mass on a finite system: much larger systems
give rise to the same sandwich instability well before the edge density diminishes
significantly. Neither is this slowdown accompanied by a significant
growth in the $\bZ$-direction --- in the period depicted, the radius in the $\bZ$-direction
increases from 6 to 7, whereas the radius in the $\bT$-direction increases
from 67 to 87. Instead, much of the growth fills space between the
ridges, the remnants of which end up almost completely below the surface.

\begin{center}
{\includegraphics[trim=13cm 5cm 11cm 4cm, clip, height=2.8cm]{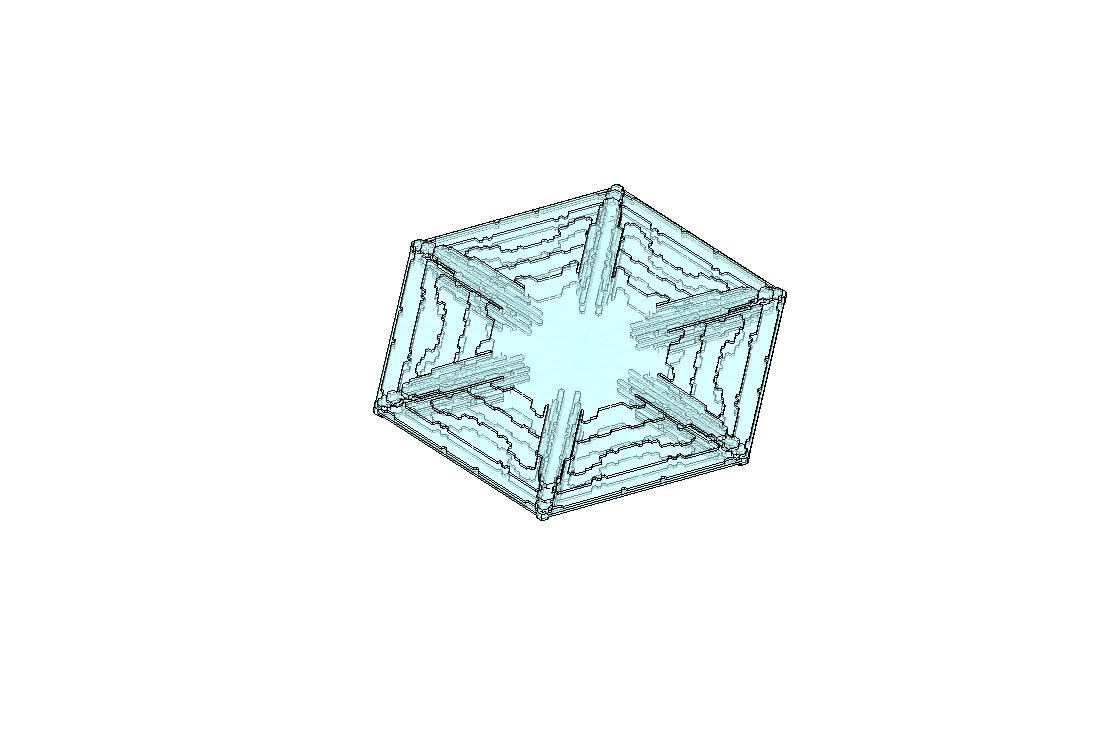}}
{\includegraphics[trim=12cm 5cm 10.5cm 4cm, clip, height=2.8cm]{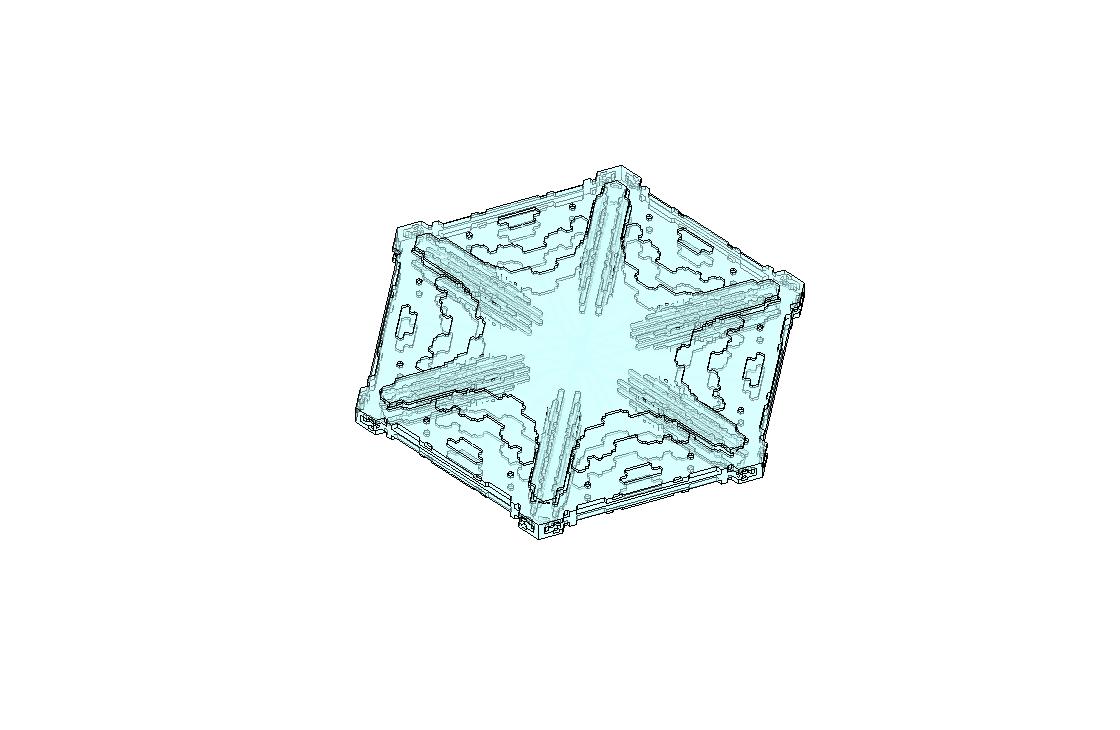} }
{\includegraphics[trim=11cm 5cm 10cm 4cm, clip, height=2.8cm]{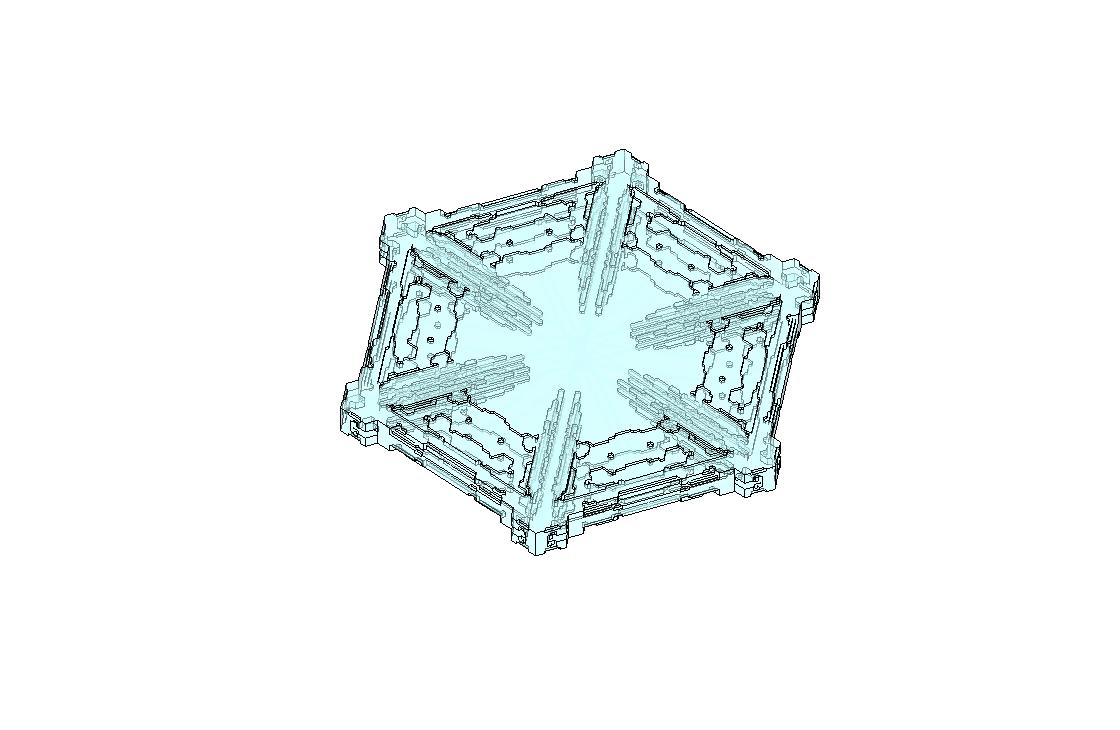}  }
{\includegraphics[trim=11cm 5cm 10cm 4cm, clip, height=2.8cm]{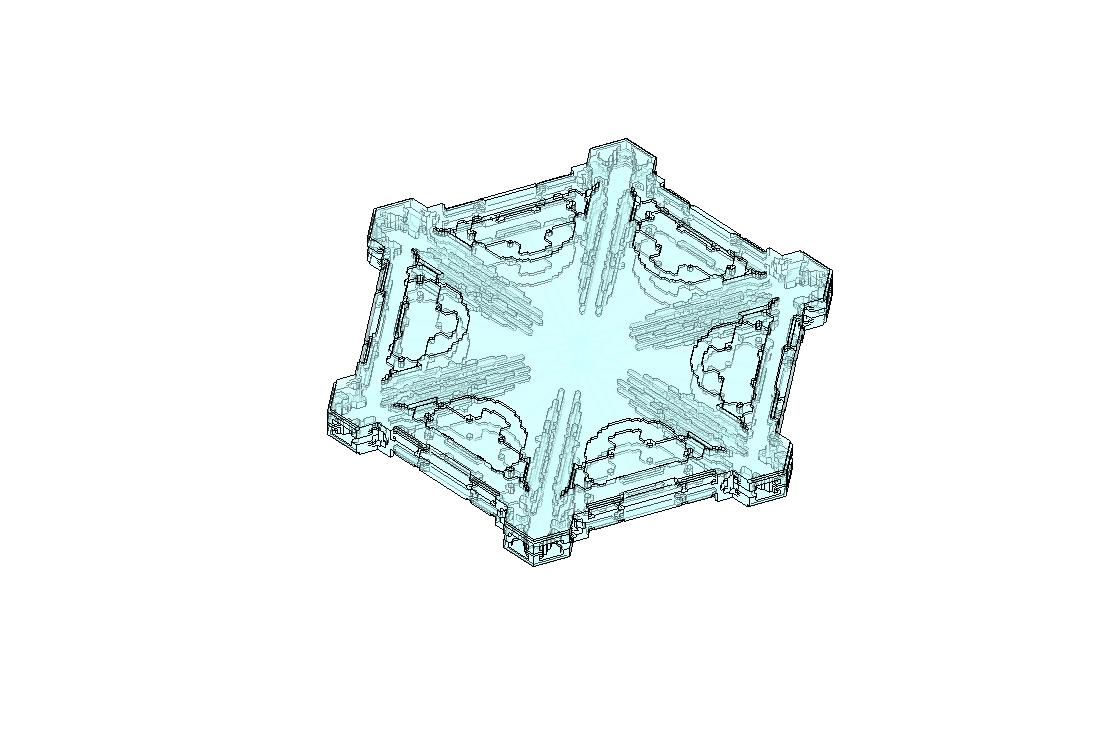}  }
{\includegraphics[trim=11cm 5cm 9.5cm 4cm, clip, height=2.8cm]{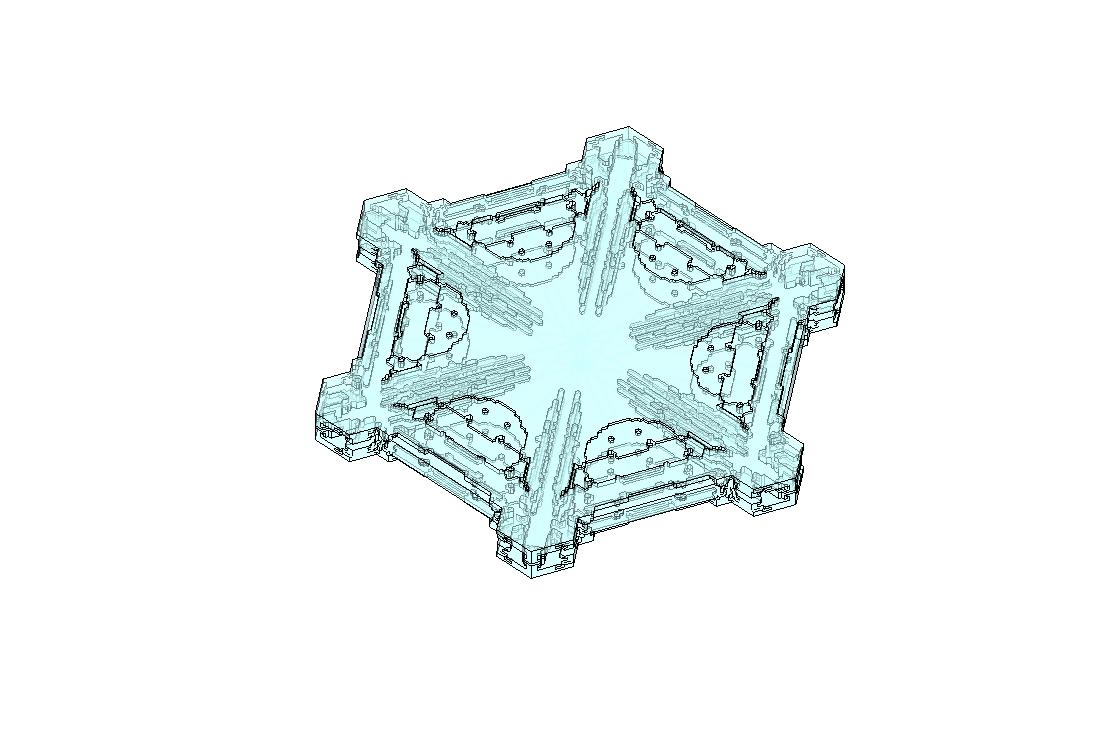}   }
\end{center}

\vskip-0.5cm
{\bf Fig.~12.} The crystal of Fig.~11 at earlier times.
\vskip0.5cm
Note that the snowfake of Fig.~10 is also experiencing the sandwich
instability at about the capture time. The difference in that case is that the 
growing crystal also experienced the branching instability earlier in its development.

\vskip1.25cm

\section{Case study $ii$ : classic dendrites}

\null\hskip-0.6cm
\begin{minipage}[b]{8cm}
\hskip0.5cm
\includegraphics[trim=8cm 1cm 8cm 2cm, clip, height=2.9in]{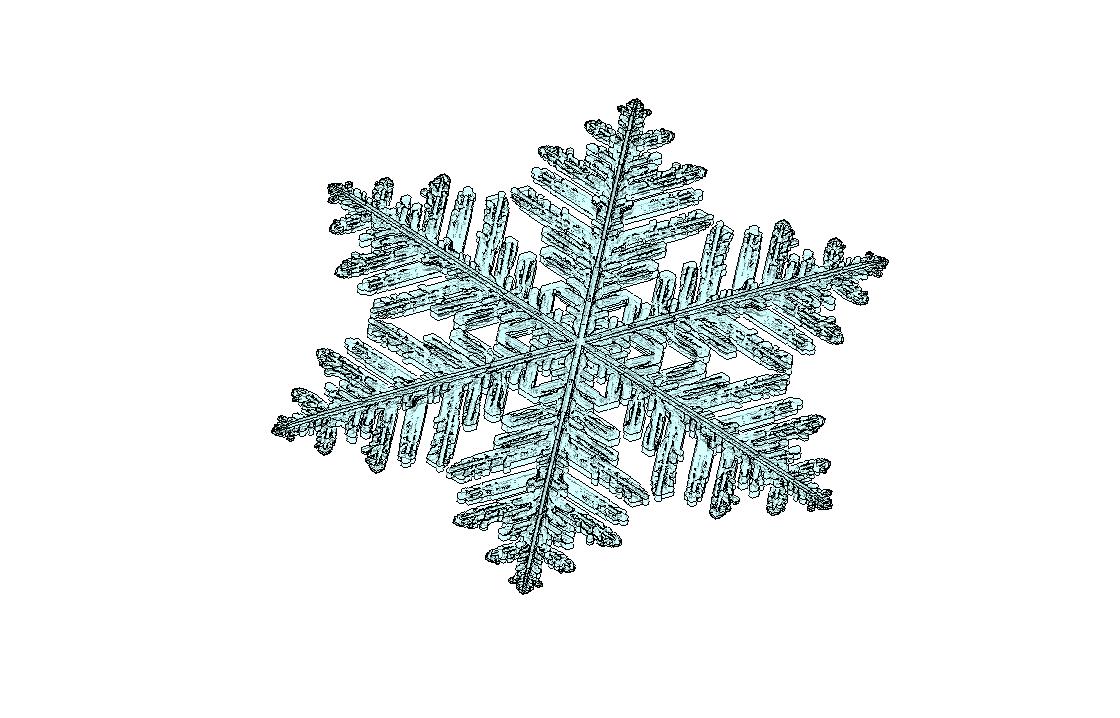}
\end{minipage}
\hskip1cm
\begin{minipage}[b][7cm][t]{15cm}
\vskip-0.25cm
\includegraphics[trim=1.4cm 0cm 1.4cm 0cm, clip, height=2.5in]{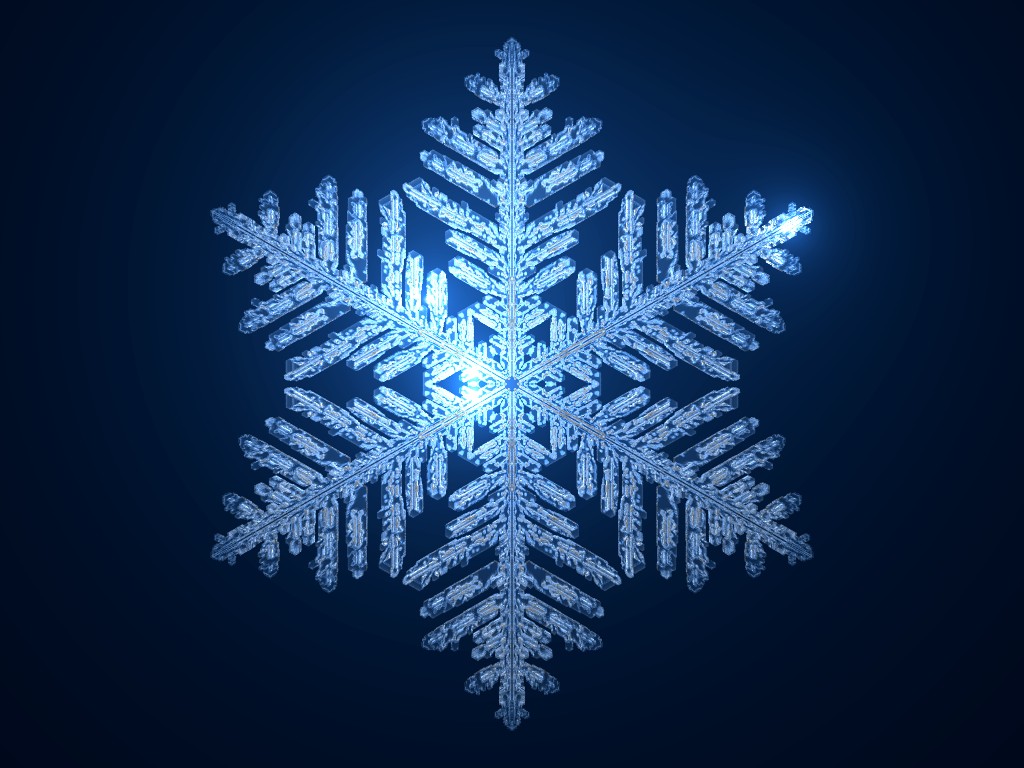}
\end{minipage}
\vskip-0.8cm
{\bf Fig.~13.}  $\rho=.105$ : a fern dendrite.
\vskip0.5cm

For this series of snowfakes, $\beta_{01}=1.6$, $\beta_{10}=\beta_{20}=1.5$, $\beta_{11}=1.4$,
$\beta_{30}=\beta_{21}=\beta_{31}=1$, $\kappa\equiv .1$, all $\mu\equiv .008$, $\phi=0$,   
and growth starts from the canonical initial state.  
We will again look at how morphology is affected by different vapor densities $\rho$. 
The simulations argue persuasively that the frequency of side branches decreases with decreasing 
$\rho$. When $\rho=.105$, the branches are so dense that the crystal is rightly 
called a fern, while the examples with $\rho=.1$ and $\rho=.095$ have the 
classic look of winter iconography. These are our largest 
crystals, with radii around 400. A more substantial decrease in $\rho$ 
eliminates any significant side branching on this scale, resulting in a simple star for $\rho=.09$. 
As should be expected from Section 7, further decrease finally produces a sandwich 
instability at the tips, resulting in thick double plates. In this instance, slow growth 
at the branch tips is accompanied by significant fattening in the $\bZ$-direction. 

\vskip1cm
\null\hskip-0.6cm
\begin{minipage}[b]{8cm}
\includegraphics[trim=8cm 1cm 8cm 2cm, clip, height=2.9in]{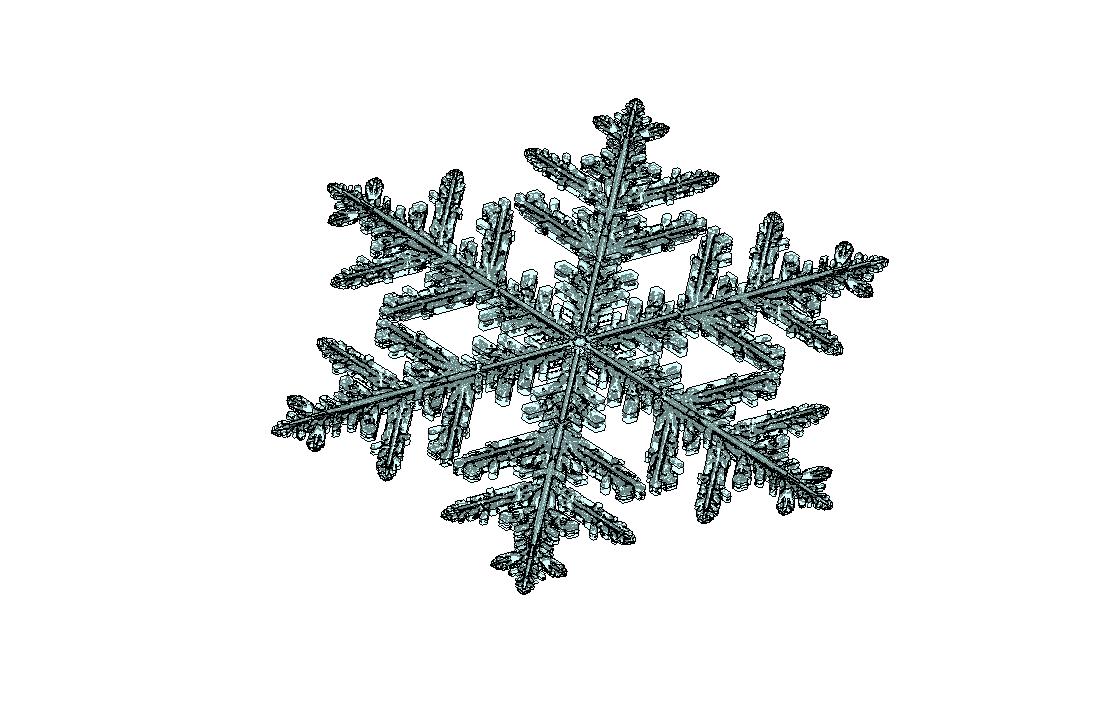}
\end{minipage}
\hskip0.5cm
\begin{minipage}[b][7cm][t]{15cm}
\vskip-0.25cm
\includegraphics[trim=1.4cm 0cm 1.4cm 0cm, clip, height=2.5in]{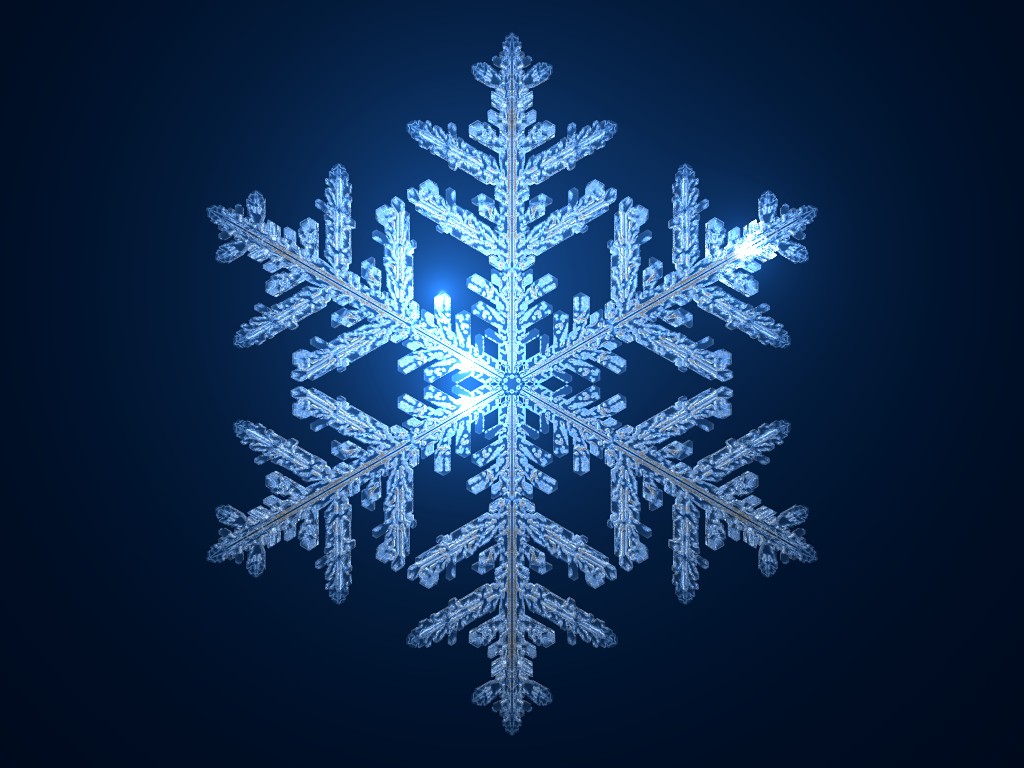}
\end{minipage}
\vskip-0.75cm
{\bf Fig.~14.}  $\rho=.1$ : a classic dendrite.
\vskip2cm

\null\hskip-0.6cm
\begin{minipage}[b]{8cm}
\includegraphics[trim=8cm 1cm 8cm 2cm, clip, height=2.9in]{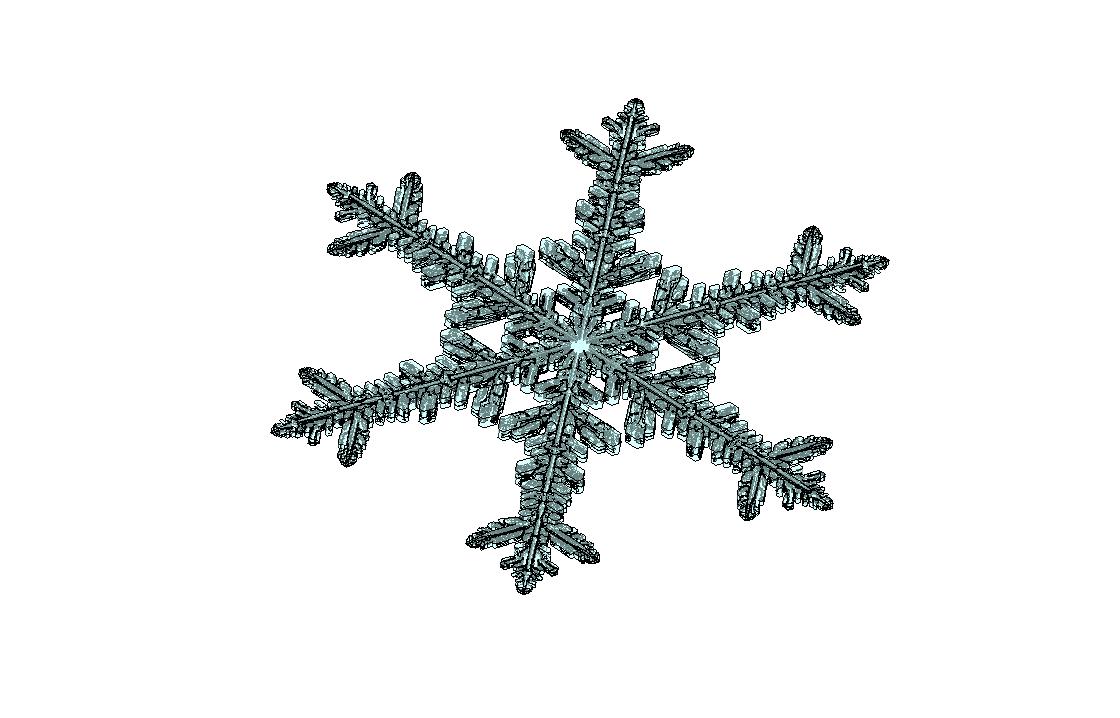}
\end{minipage}
\hskip0.5cm
\begin{minipage}[b][7cm][t]{15cm}
\vskip-0.25cm
\includegraphics[trim=1.4cm 0cm 1.4cm 0cm, clip, height=2.5in]{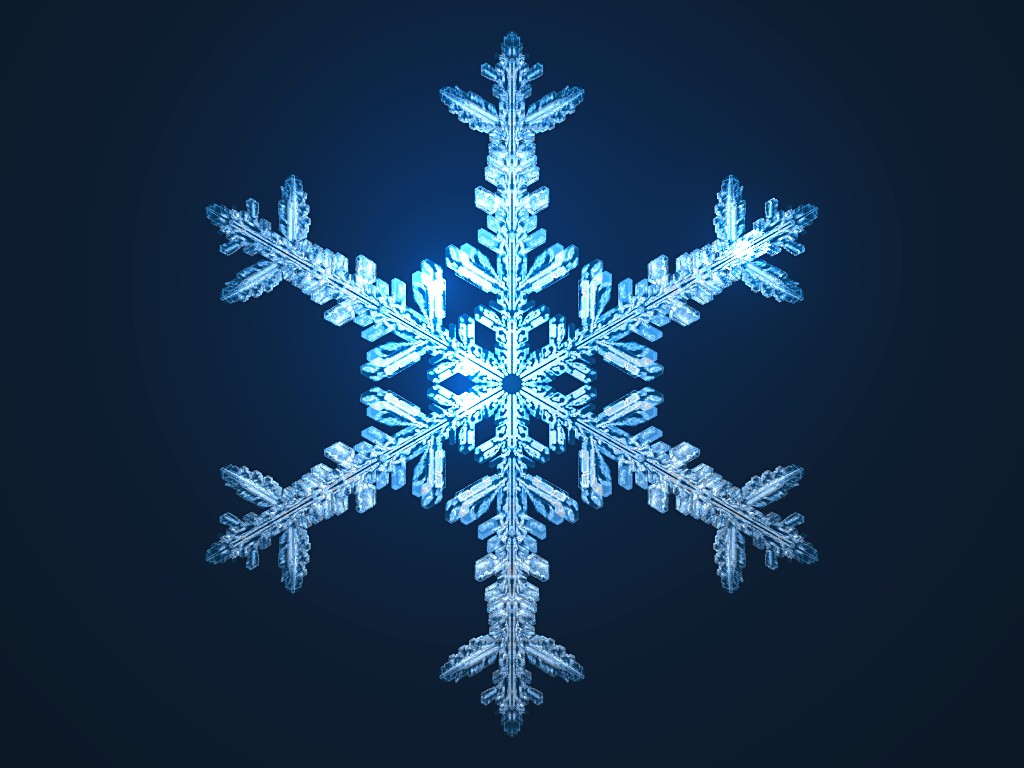}
\end{minipage}
\vskip-0.75cm
{\bf Fig.~15.}  $\rho=.095$ : fewer side branches.
\vskip0.5cm

\null\hskip-0.6cm
\begin{minipage}[b]{8cm}
\includegraphics[trim=8cm 1cm 8cm 2cm, clip, height=2.9in]{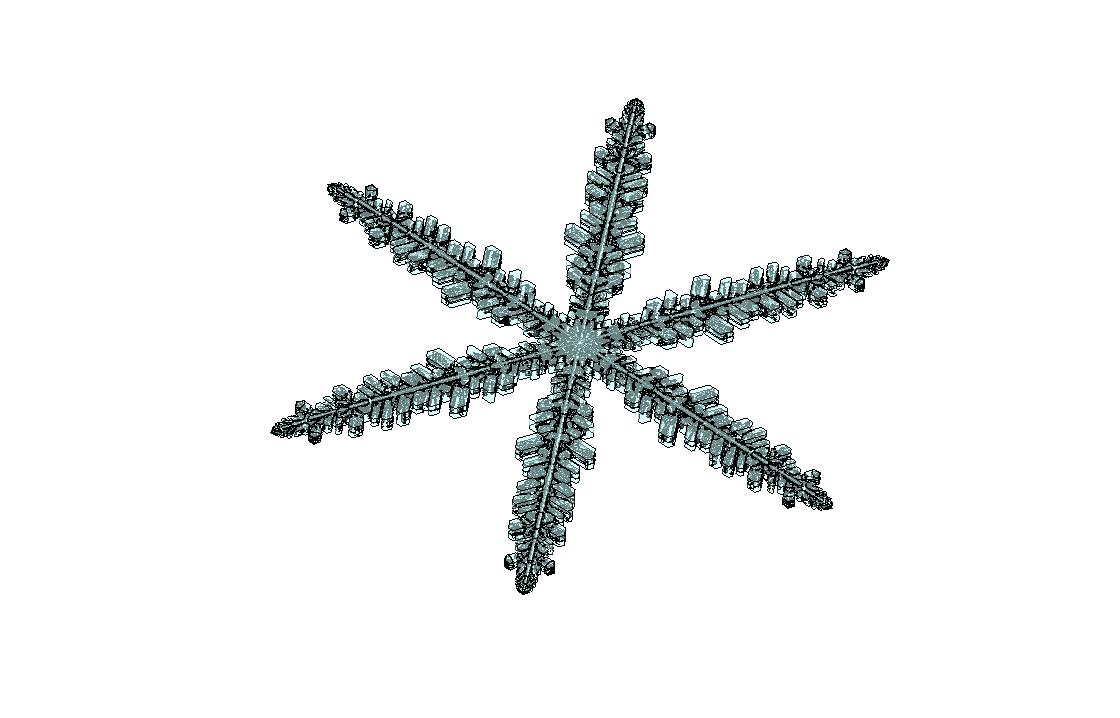}
\end{minipage}
\hskip0.5cm
\begin{minipage}[b][7cm][t]{15cm}
\vskip-0.25cm
\includegraphics[trim=1.4cm 0cm 1.4cm 0cm, clip, height=2.5in]{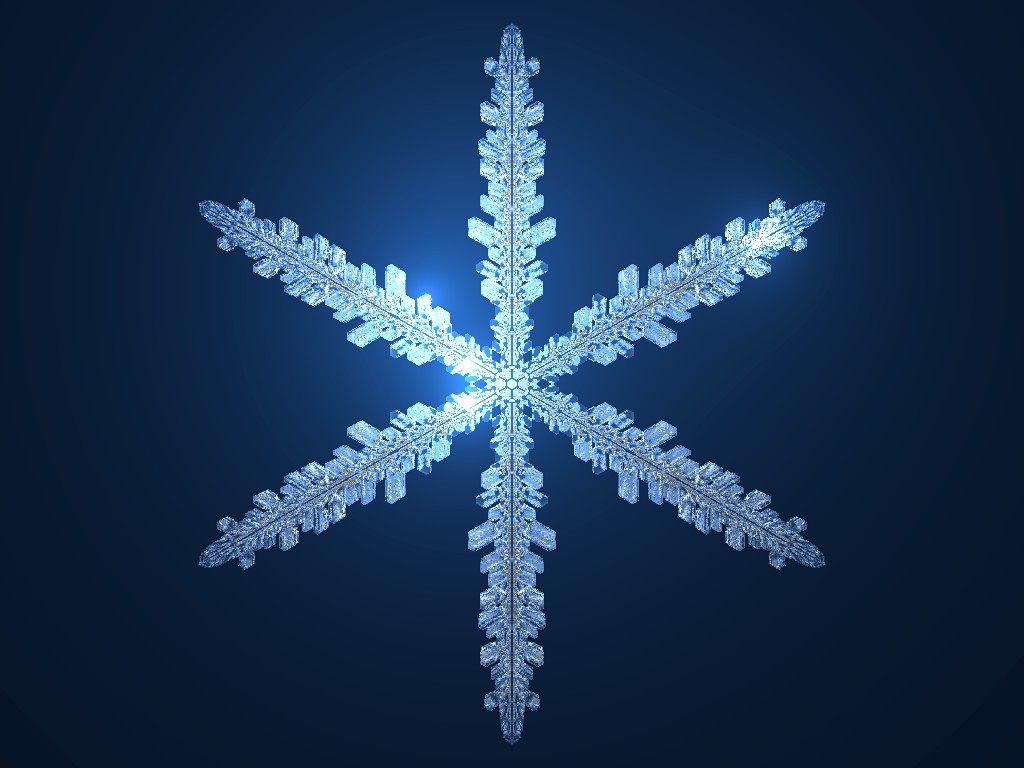}
\end{minipage}
\vskip-0.75cm
{\bf Fig.~16.}  $\rho=.09$ : no significant side branches on this scale.
\vskip0.0cm

\null\hskip-0.6cm
\begin{minipage}[b]{8cm}
\includegraphics[trim=8cm 1cm 8cm 2cm, clip, height=2.9in]{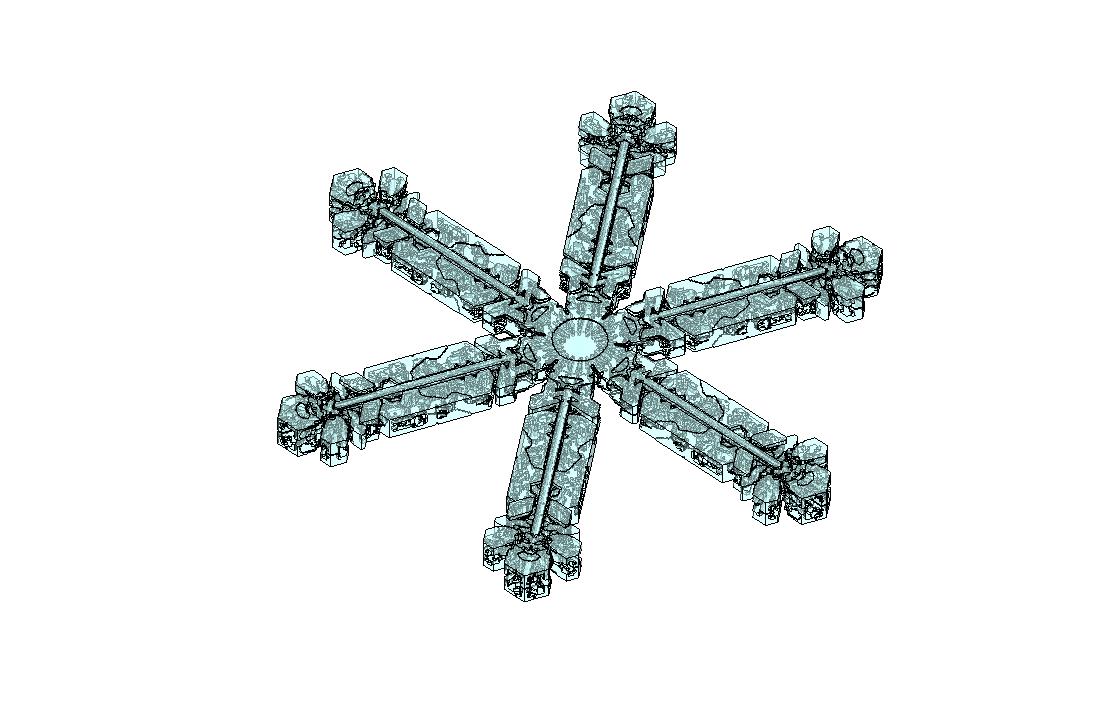}
\end{minipage}
\hskip0.5cm
\begin{minipage}[b][7cm][t]{15cm}
\vskip-0.25cm
\includegraphics[trim=1.2cm 0cm 1.2cm 0cm, clip, height=2.5in]{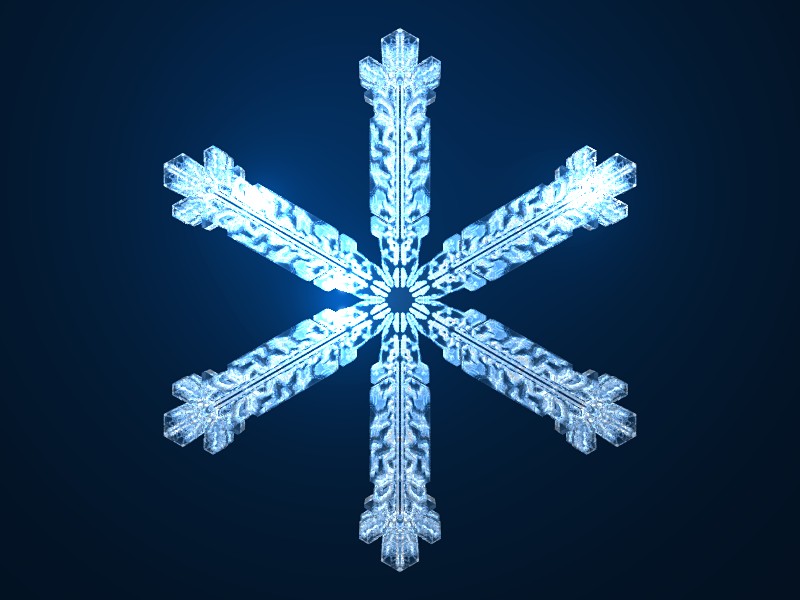}
\end{minipage}
\vskip-0.75cm
{\bf Fig.~17.}  $\rho=.082$ : the tip undergoes a sandwich instability.
\vskip0.5cm

The crystal in Fig.~17 is captured at about time 60000. The series of close-ups 
in Fig.~18 provides another illustration of the sandwich instability --- snapshots of the 
same snowfake are shown at time intervals of 1000, starting from time 37000.   

\begin{center}
{\includegraphics[trim=14cm 10cm 12cm 0cm, clip, height=2.8cm]{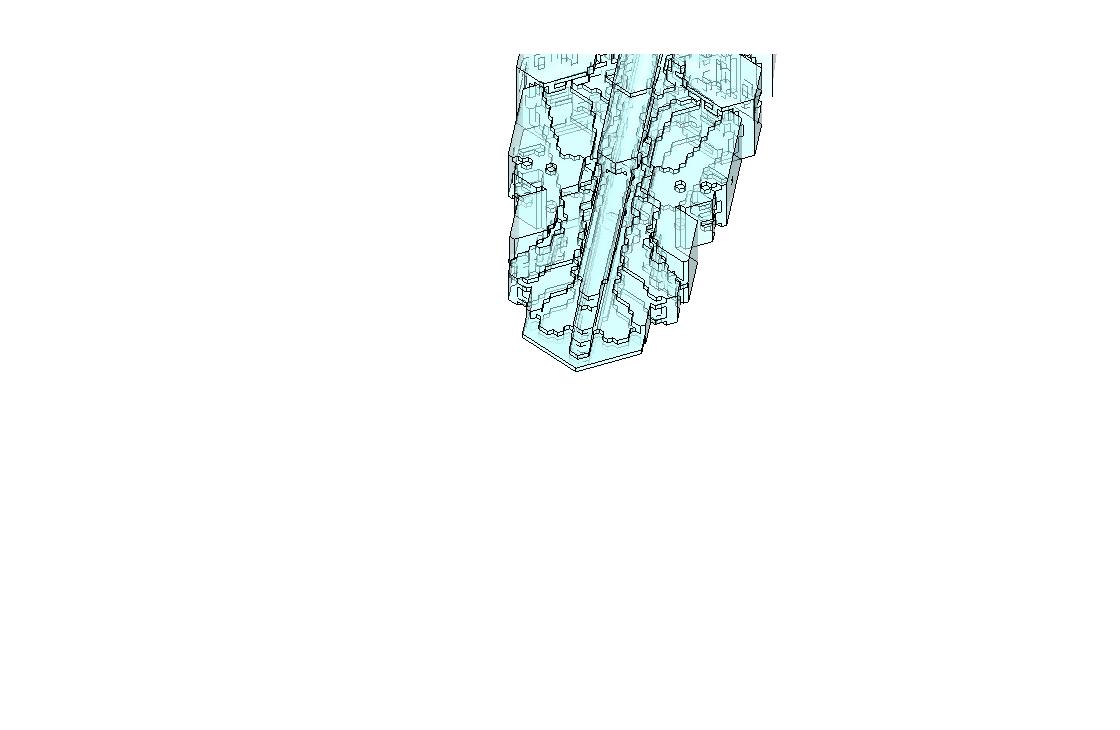}}
{\includegraphics[trim=14cm 10cm 12cm 0cm, clip, height=2.8cm]{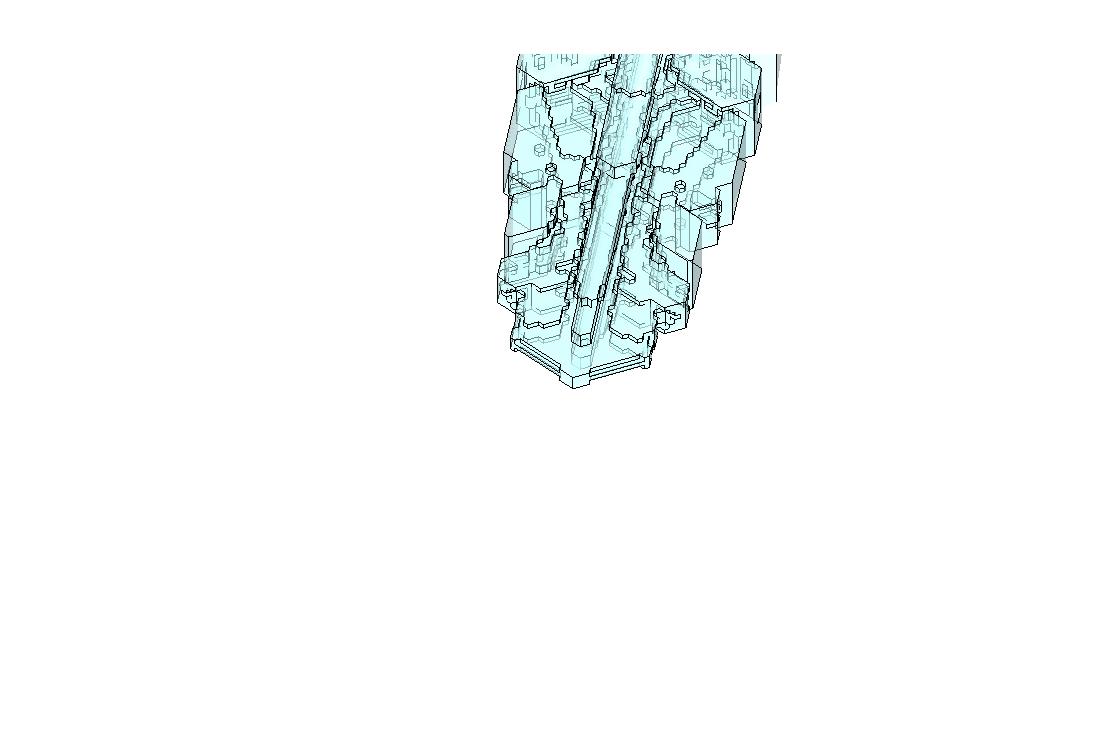}}
{\includegraphics[trim=14cm 10cm 12cm 0cm, clip, height=2.8cm]{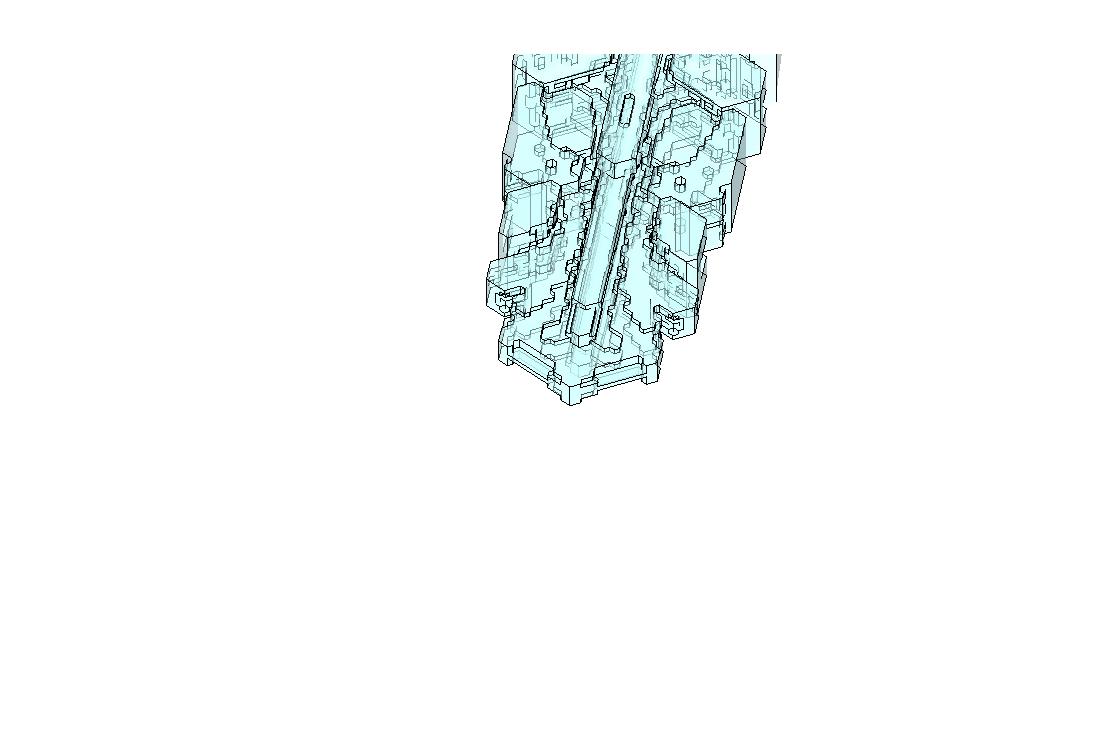}}
{\includegraphics[trim=14cm 10cm 12cm 0cm, clip, height=2.8cm]{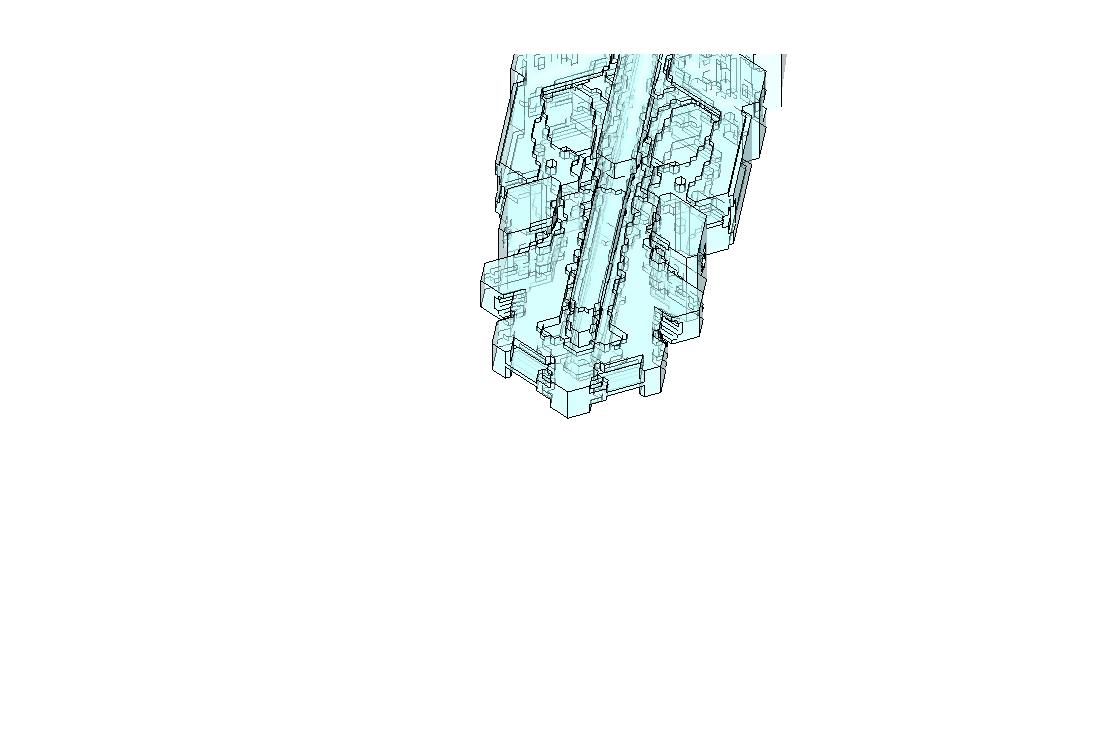}}
{\includegraphics[trim=14cm 10cm 12cm 0cm, clip, height=2.8cm]{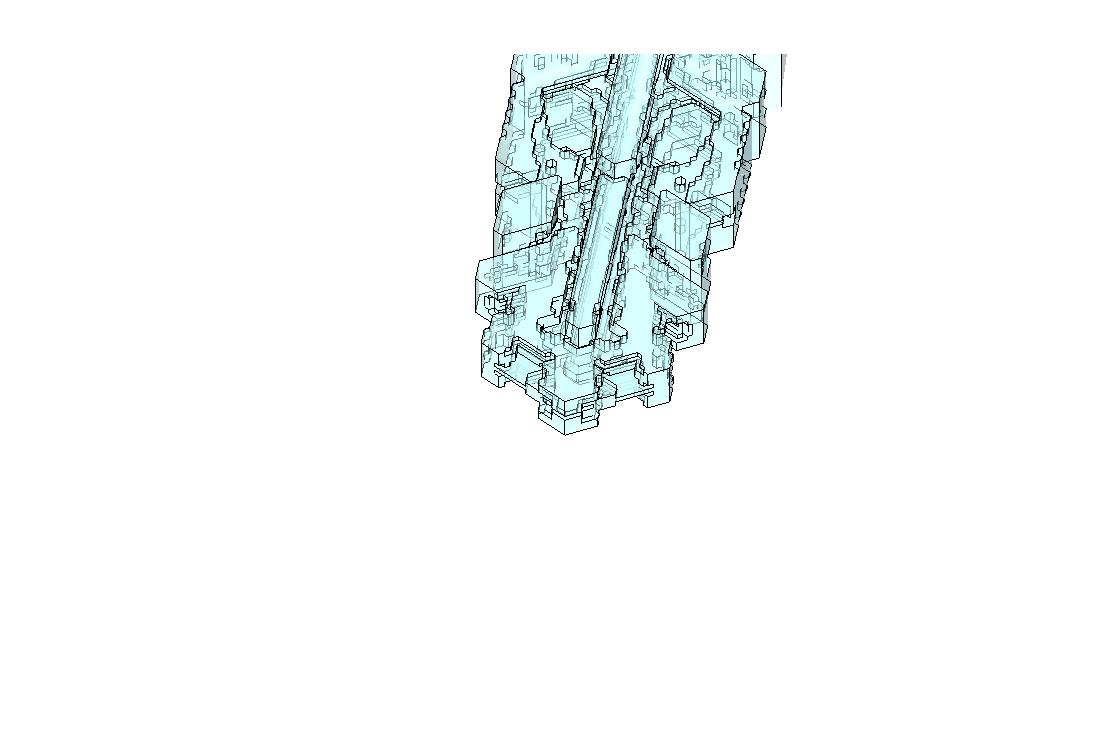}}
{\includegraphics[trim=14cm 10cm 12cm 0cm, clip, height=2.8cm]{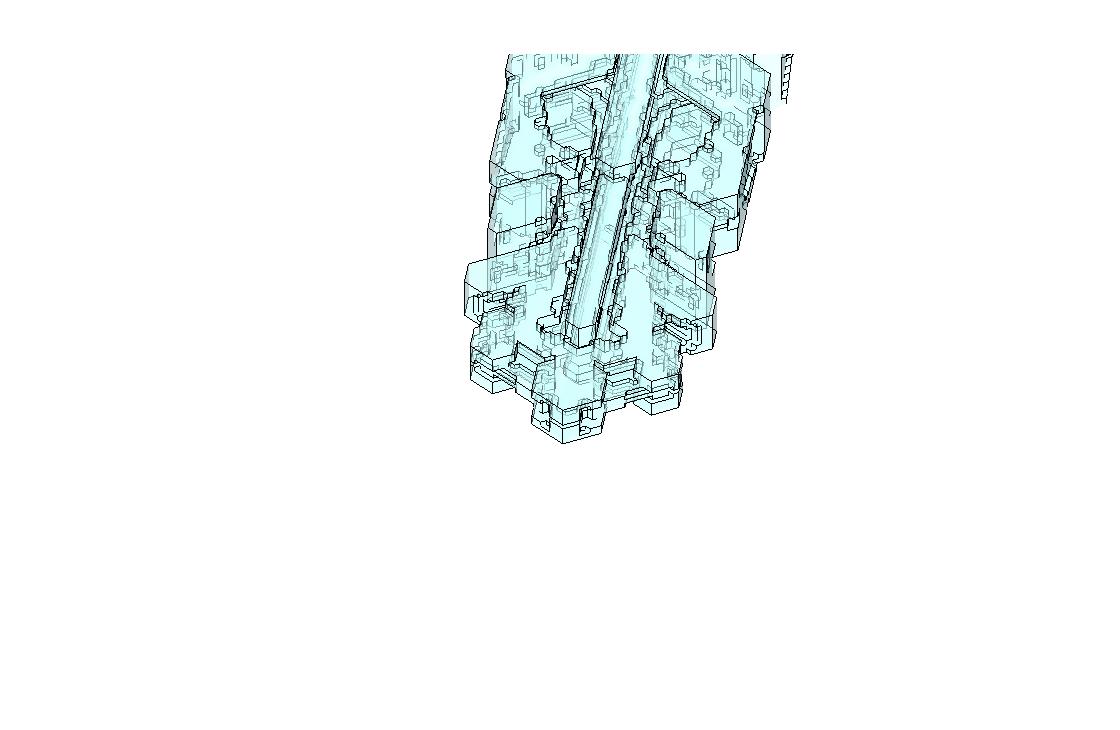}}
 
\end{center}

\vskip-0.5cm
{\bf Fig.~18.} Close-up of the sandwich instability at $\rho=.082$.
\vskip0.5cm

Our final example, with $\rho=.081$, demonstrates that a further decrease in density makes the 
crystal increasingly three-dimensional.

\null\hskip-0.6cm
\begin{minipage}[b]{8cm}
\includegraphics[trim=8cm 1cm 8cm 2cm, clip, height=2.9in]{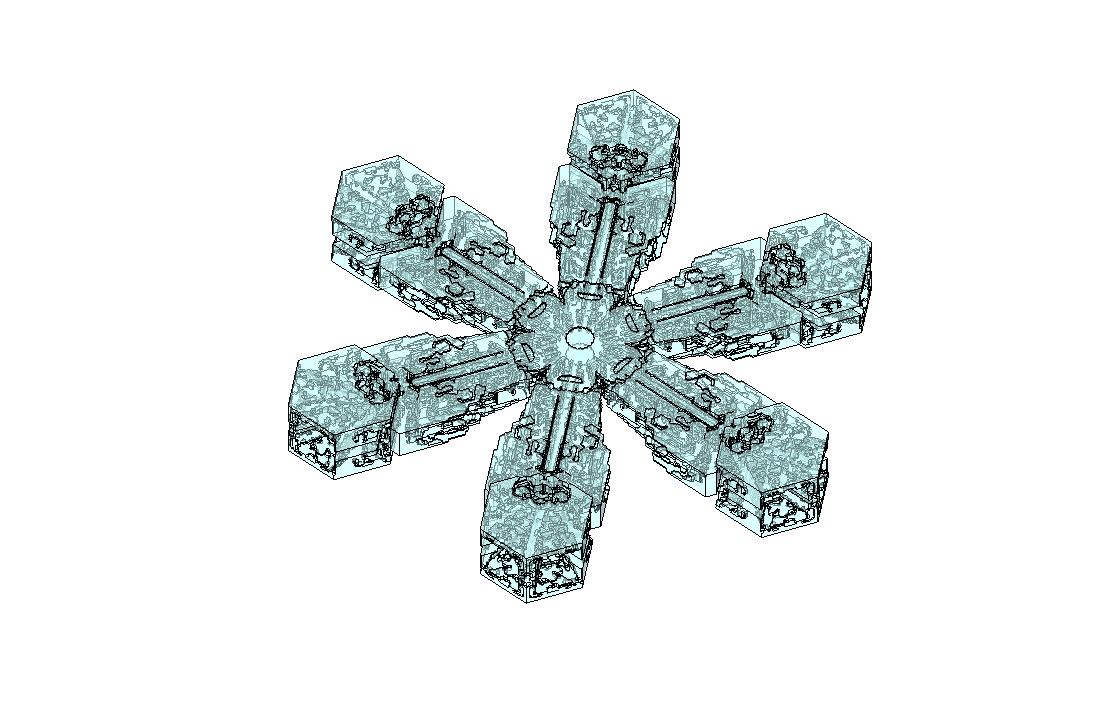}
\end{minipage}
\hskip0.5cm
\begin{minipage}[b][7cm][t]{15cm}
\vskip-0.25cm
\includegraphics[trim=0.7cm 0cm 0.7cm 0cm, clip, height=2.5in]{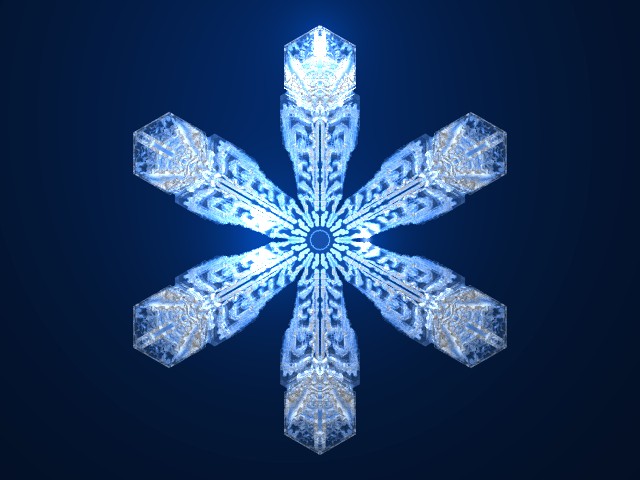}
\end{minipage}
\vskip-0.75cm
{\bf Fig.~19.}  Fattening from the tip inward at $\rho=.081$.
\vskip1cm

\section{Case study $iii$ : sandwich plates}

When growth in the $\bZ$-direction is much slower than 
in the $xy$-plane, outer ridges never develop. Instead, 
the dynamics grows a featureless prism, which, when 
sufficiently thick, undergoes a sandwich instability 
producing inner ridges. Much later the crystal experiences
the branching instability, with plate-like branches that 
bear a superficial resemblance to Packard snowflakes \cite{Pac, GG2}
during early stages.  

Throughout the evolution the external surface of the crystal has few
markings, whereas inside features include ridges and {\it ribs\/}, 
which signify gradual thinning of the plates from the center outward 
before the branching instability. 

The sole surface designs are {\it reverse shapes\/}, 
which occur when the crystal grows in the $\bZ$-direction from 
buds that arise close to the tips. These macrostep nuclei result in 
rapid growth of a single layer in the $\bT$-direction until 
this layer outlines a nearly circular hole near the crystal's 
center; the hole then proceeds to shrink much more slowly. 

We note that this observation provides a convincing explanation for the 
circular markings seen on many snow crystal photographs
\cite{Lib6, LR}. It also suggests that ribs are predominantly 
inner structures. While outer ribs could occur 
due to instabilities or changing conditions (cf.~Fig.~11), there 
is scant evidence of them in electron microscope photographs \cite{EMP},
which completely obscure inner structure. On the other hand, those photos  
reveal an abundance of sandwich plates, which appear
as the crystal centers, at the tips of the six main arms, and as side branches.

We now present two examples. Both start from the canonical seed.
In the first, depicted in Fig.~20, $\beta_{01}=6$, 
$\beta_{10}=\beta_{20}=2.5$, $\beta_{11}=2$, and the remaining 
$\beta$'s are 1. All $\kappa$'s are .1, except that $\kappa_{01}=.5$, 
$\mu\equiv .0001$, and $\rho=.08$. The final radius of the crystal
at the capture time 100000 is about 150. Note that the main ridge is
interrupted: while initially it connects the two plates
(and it has darker color in the ray-traced image as the background can be 
seen through it), it later splits and each plate has its own ridge. 
There is a suggestion of this phenomenon in real crystals (e.g., on p.~26 of \cite{Lib6}). 

\null\hskip-0.6cm
\begin{minipage}[b]{8cm}
\includegraphics[trim=8cm 1cm 8cm 2cm, clip, height=2.9in]{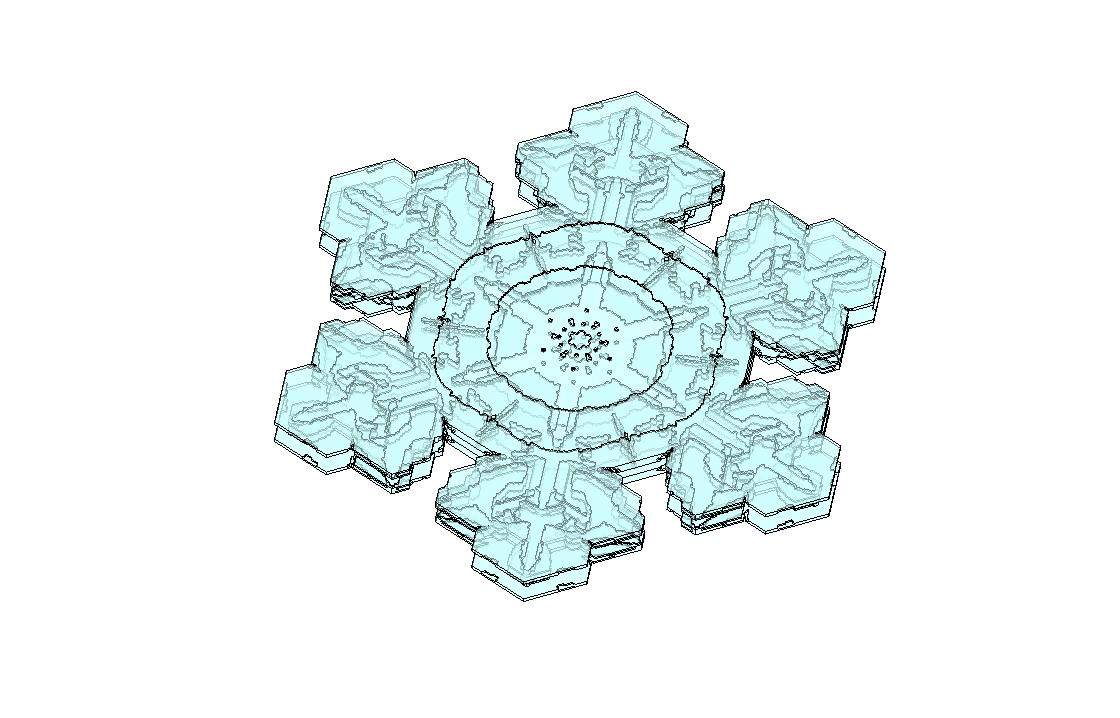}
\end{minipage}
\hskip0.5cm
\begin{minipage}[b][7cm][t]{15cm}
\vskip-0.25cm
\includegraphics[trim=0.7cm 0cm 0.7cm 0cm, clip, height=2.5in]{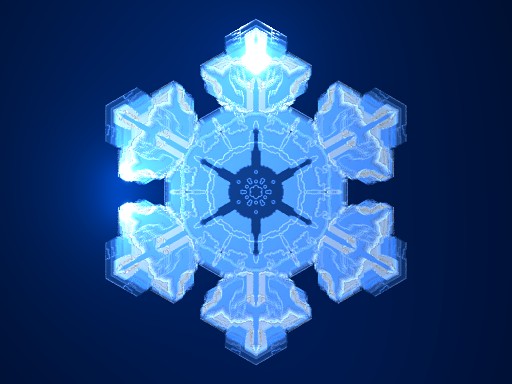}
\end{minipage}
\vskip-0.75cm
{\bf Fig.~20.} A sandwich plate.
\vskip1cm

Our second example (Figs.~21 and 22) has interrupted main ridges and a few ribs.  
The parameter set now has $\beta_{01}=6.5$, $\beta_{10}=\beta_{20}=2.7$, and $\rho=.15$. 
The remaining values are as before, and the final sizes (this one at $t=36100$)
are comparable. We provide a few intermediate stages and a detail of the 
inner structure. Observe the buds at times 25883 and
31671; also note that the outermost rib at time 19000 later disappears. 

\null\hskip-0.6cm
\begin{minipage}[b]{8cm}
\includegraphics[trim=8cm 1cm 8cm 2cm, clip, height=2.9in]{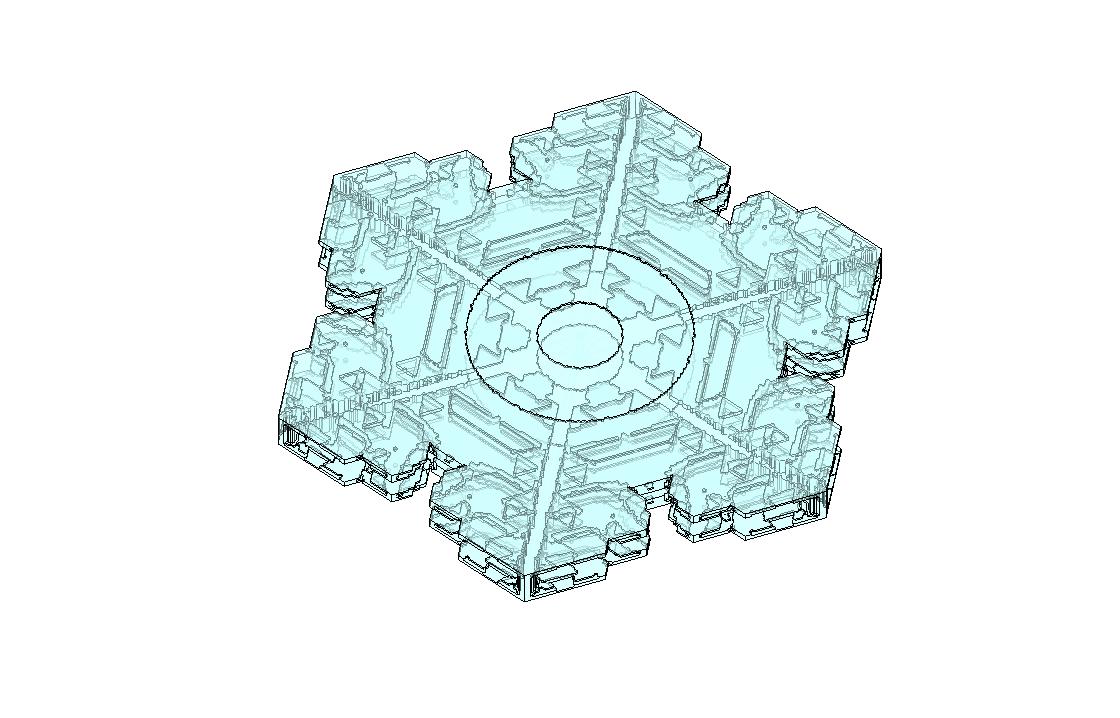}
\end{minipage}
\hskip0.5cm
\begin{minipage}[b][7cm][t]{15cm}
\vskip-0.25cm
\includegraphics[trim=0.7cm 0cm 0.7cm 0cm, clip, height=2.5in]{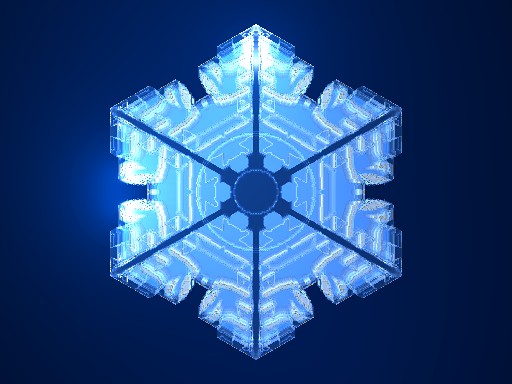}
\end{minipage}
\vskip-0.75cm
{\bf Fig.~21.} Another sandwich plate.
\vskip1cm

\begin{center}
{\includegraphics[trim=12cm 5cm 12cm 4cm, clip, height=2.5cm]{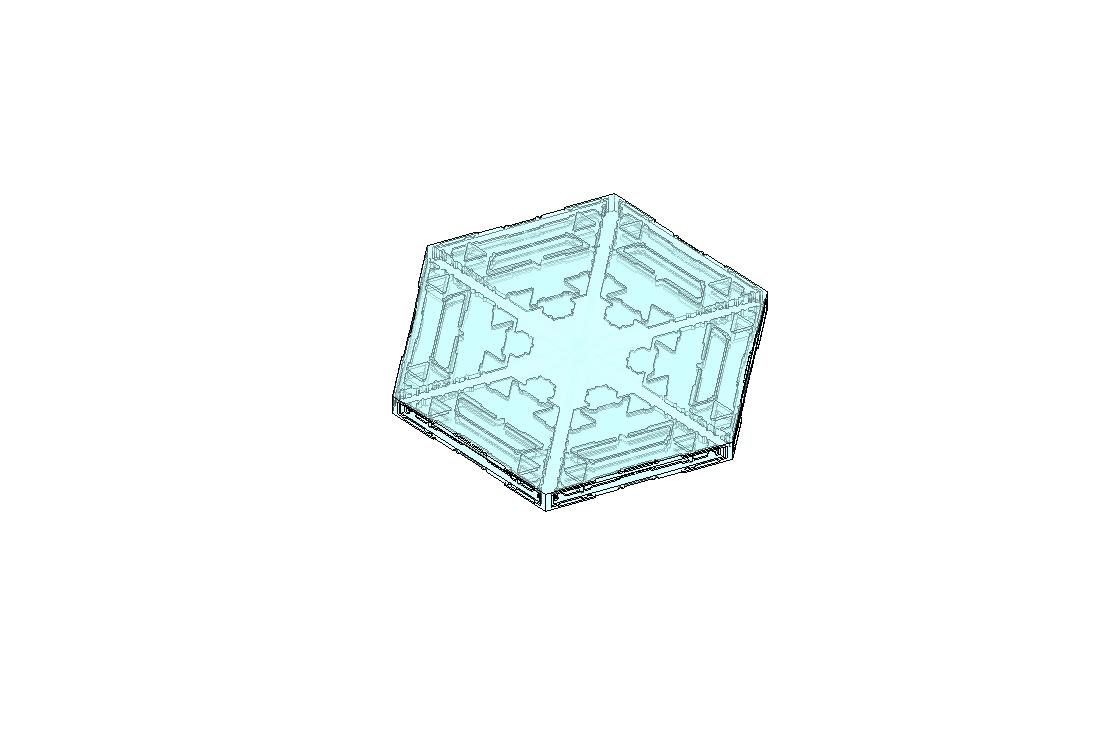}}
{\includegraphics[trim=12cm 5cm 11cm 4cm, clip, height=2.5cm]{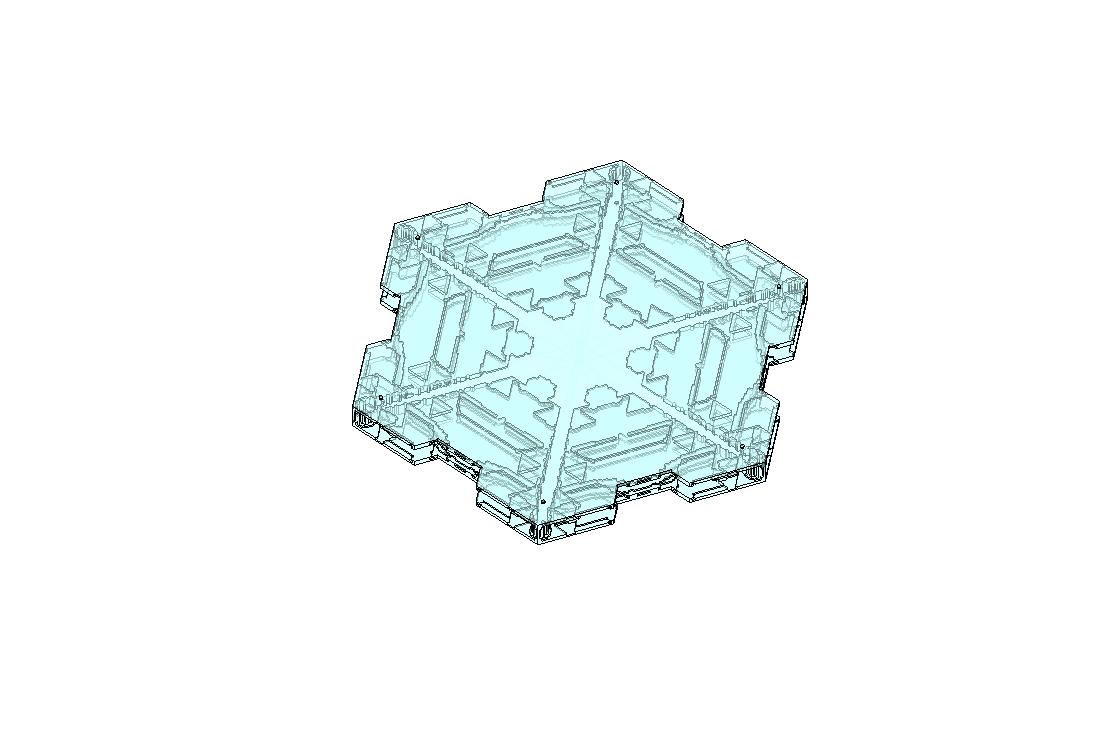}}
{\includegraphics[trim=12cm 5cm 11cm 4cm, clip, height=2.5cm]{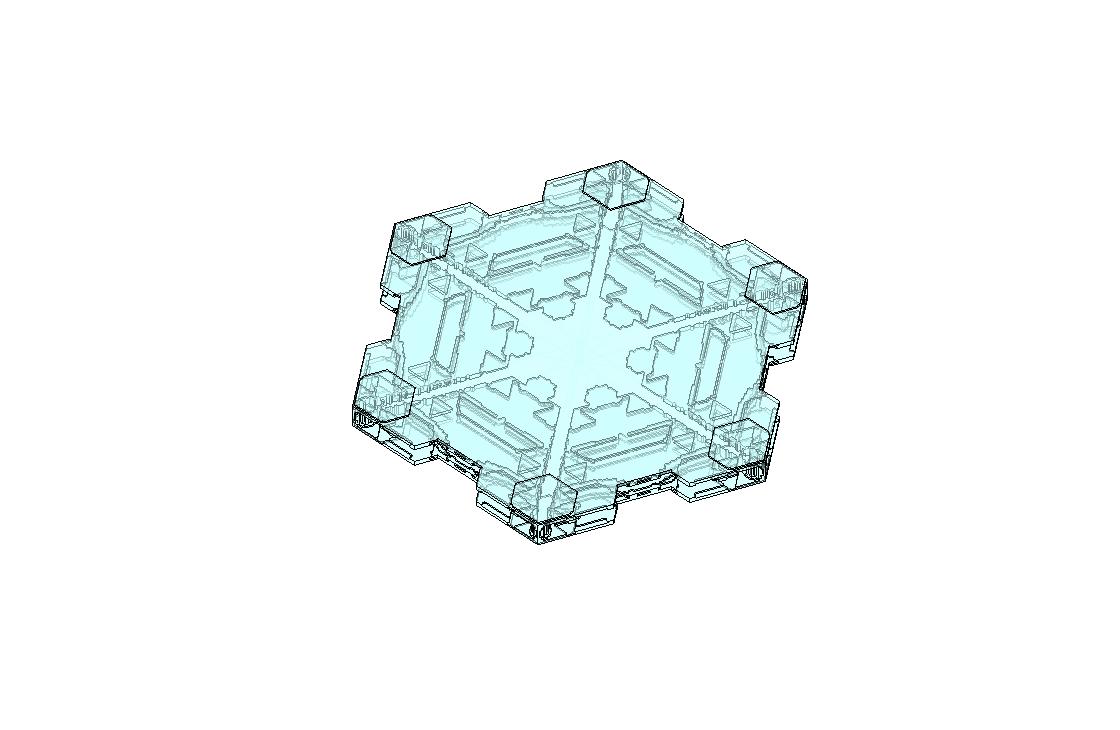}}
{\includegraphics[trim=11cm 5cm 11cm 4cm, clip, height=2.5cm]{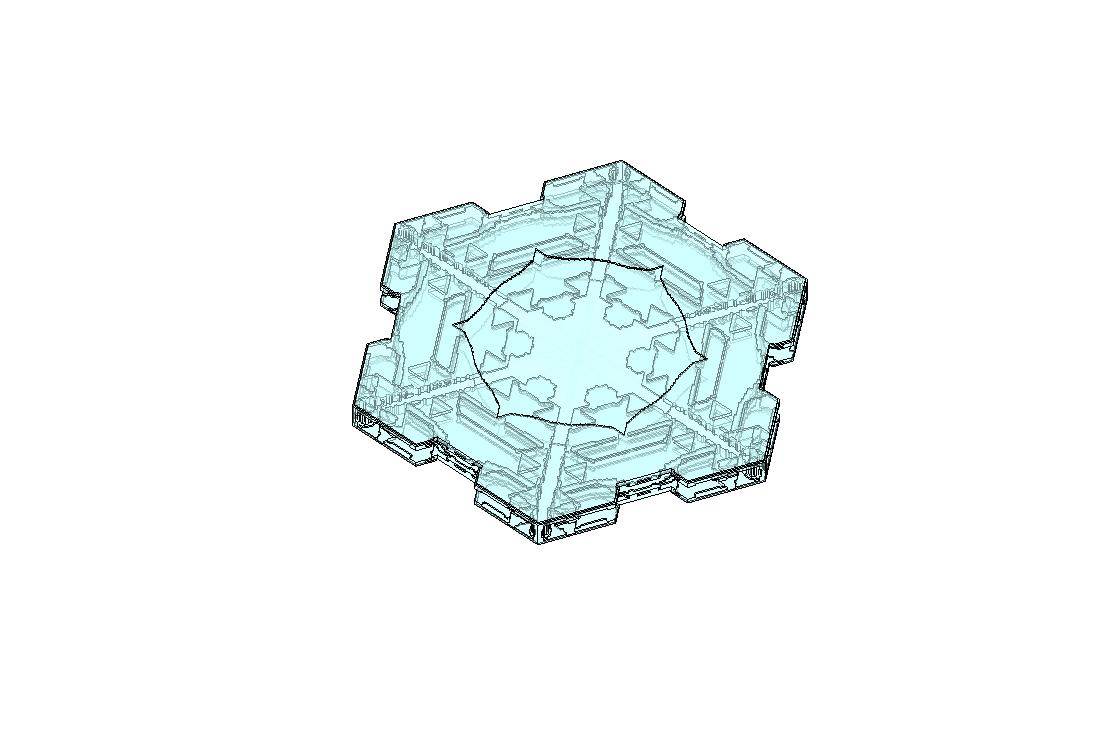}}
{\includegraphics[trim=11cm 5cm 11cm 4cm, clip, height=2.5cm]{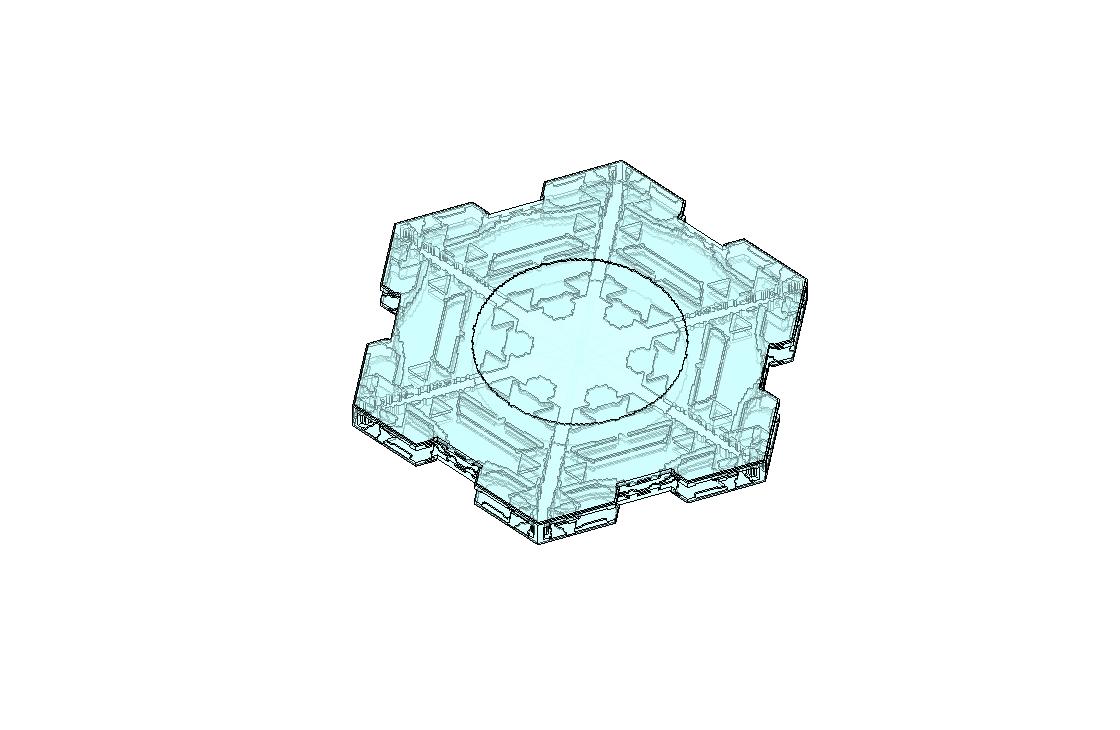}}
{\includegraphics[trim=10cm 5cm 10cm 4cm, clip, height=2.5cm]{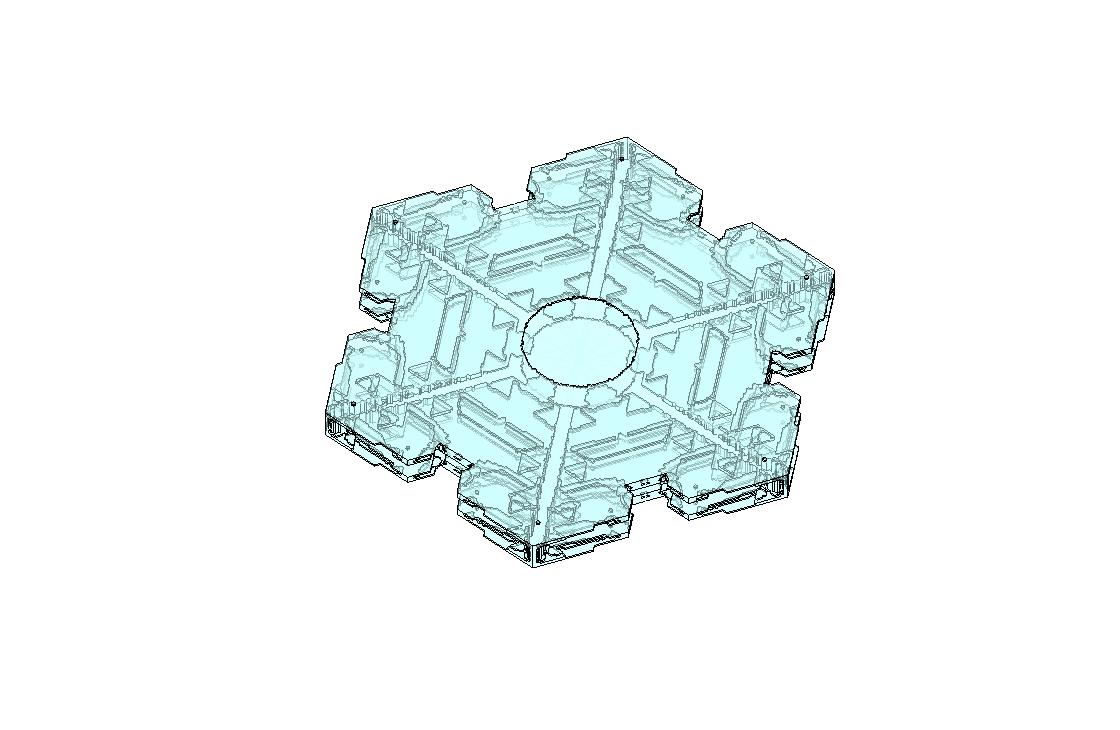}}
 
\end{center}

\begin{center} 
\vskip-0.25cm
{\includegraphics[trim=0cm 0cm 0cm 0cm, clip, height=3cm]{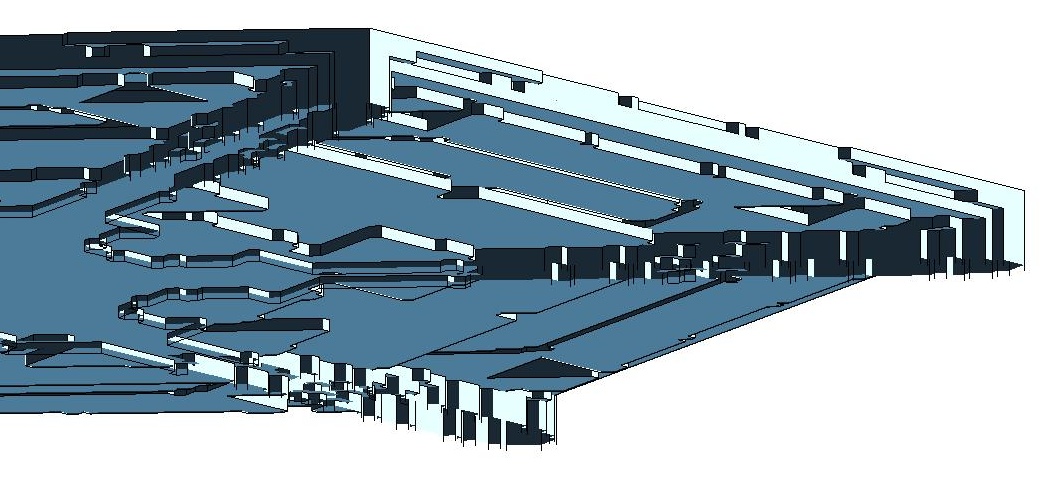}}
\end{center}

\vskip-0.3cm
{\bf Fig.~22.}  The plate of Fig.~21 at $t=19000, 25883, 25900, 25950, 26000, 31671$.
The detail is from the first time, obtained by cutting the crystal along the 
plane $z=0$ and zooming in on the bottom half of the upper portion. 

\vskip1cm

\section{Case study $iv$: the roles of drift and melting}

From some of the electron micrographs at \cite{EMP}, 
it appears possible that the basal facets may have ridges and 
other markings on one side only, while the other side is nearly featureless. 
As far as we are aware, no attempt has been made to ``turn over'' 
these specimens and confirm the asymmetry, 
but \cite{NK,Nel} offer a theoretical explanation. They suggest that the one-sided structure is a consequence 
of early growth and that ridges are actually vestiges of the skeleton of hollow prisms such as Fig.~31 
in Section 11 (see Fig.~3 of \cite{Nel}).  
In fact, it is widely held that the micron-scale prism from which a prototypical snowflake evolves develops 
slight asymmetries in the radii of its two basal facets, and that the larger facet acquires an increasing 
advantage from the feedback effect of diffusion-limited growth. As a result many crystals have a stunted hexagonal 
plate at their center. In \cite{Nak} this effect is described on p.~206 and in sketch 15 of Fig.~369.

Another potential source of asymmetry in the $\bZ$-direction is identified 
in Section 3.5 of \cite{Iwa} and on p.~18 of \cite{TEWF}, based on cloud tunnel experiments 
in the laboratory. Planar snowflakes evidently assume a preferred orientation parallel 
to the ground as they slowly fall, resulting in a small upward drift of the diffusion 
field relative to the crystal.  

We emulate these aspects of asymmetric growth by means of the drift $\phi$ in 
step (1c) of our algorithm and asymmetry of the initial seed as mentioned in Section 3. 
Consider first the snowfake of Fig.~1 and the closely related sectored plate in Fig.~23. 
The former starts from our fundamental prism and never undergoes the sandwich instability, 
but develops ridges on the bottom side and an almost featureless top due to the presence of $\phi = .01$. 
The dynamic parameters of the sectored plate below are identical, 
but growth starts from a mesoscopic prism that is 5 cells high, with radius 7 at the top and 3 at the bottom. 
The idea here is to mimic the situation where the upper basal plate has established an advantage over 
the lower basal plate early in the evolution. As is clear from the side view, in contrast to Fig.~6, 
growth of the lower facet stops completely due to diffusion limitation even though the small drift 
offers a slight advantage in the early stages. (According to \cite{Iwa}, falling snowflakes prefer 
the more aerodynamically stable orientation of 
Fig.~23.) Very many photos of physical snow crystals show evidence of such a stunted simple plate at the center; 
see \cite{Lib6}, pp.~75--76, for further discussion.

\vskip0.3cm
\begin{minipage}[b]{7cm}
\includegraphics[trim=0.7cm 0cm 0.7cm 0cm, clip, height=2.5in]{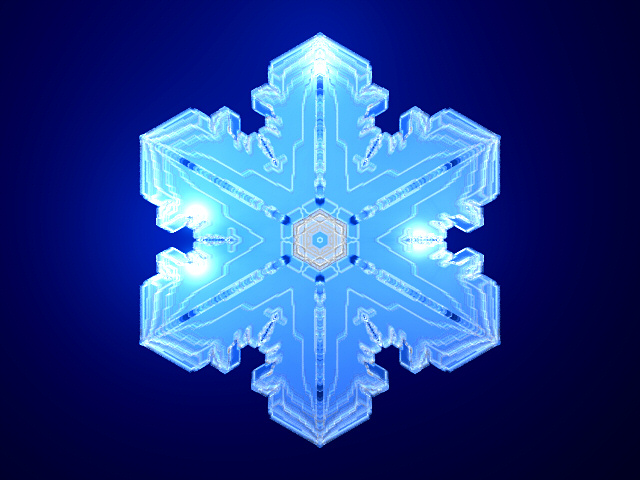}
\end{minipage}
\hskip0.5cm
\begin{minipage}[b][4cm][t]{7cm}
\includegraphics[trim=0cm 0cm 0cm 0cm, clip, height=0.85in]{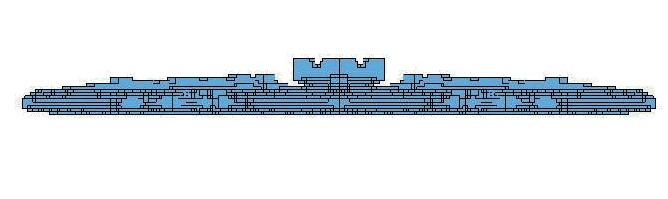}
\end{minipage}

{\bf Fig.~23.} A sectored plate with a stunted double, from the top ({\it left\/}) and side ({\it right\/}). 
\vskip0.4cm

\null\hskip -0.6cm
\begin{minipage}[b]{8cm}
\includegraphics[trim=8cm 1cm 8cm 2cm, clip, height=2.9in]{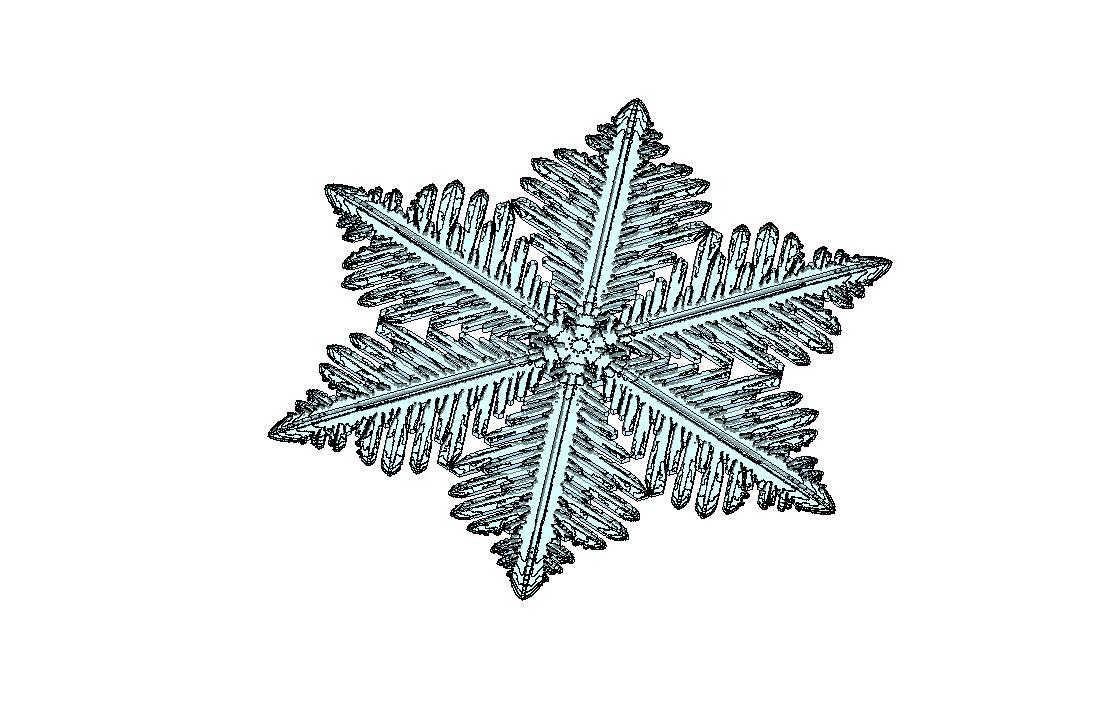}
\end{minipage}
\hskip0.5cm
\begin{minipage}[b][7cm][t]{15cm}
\vskip-0.25cm
\includegraphics[trim=0.7cm 0cm 0.7cm 0cm, clip, height=2.5in]{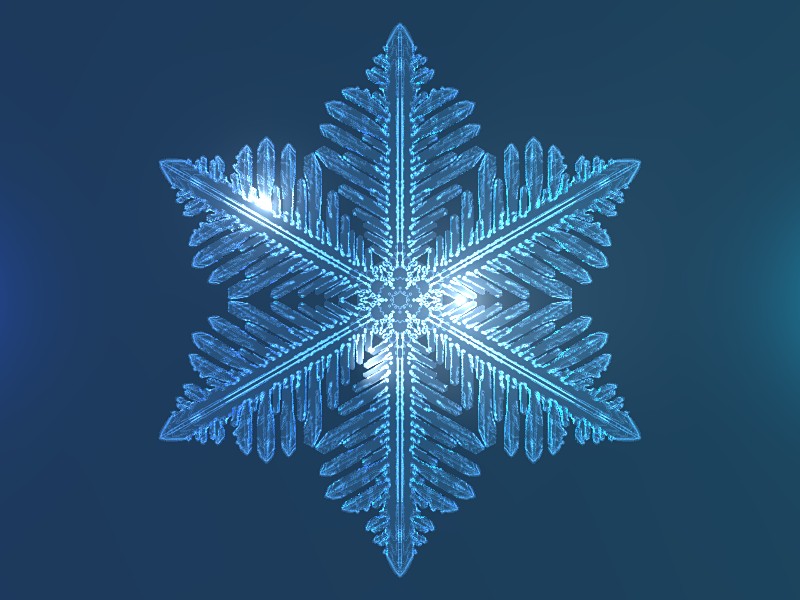}
\end{minipage}
\vskip-0.75cm
{\bf Fig.~24.}  A fern dendrite for $\mu_{10}=\mu_{20}=.005$.
\vskip0.4cm

The remaining examples of this section also start from slightly asymmetric seeds, 
experience a small drift, and have almost all their external markings on one side. 
Our goal is to explore the role of the melting rate, in much the same way we studied density 
dependence in Section 7, by varying $\mu$ in a series of snowfakes with all other parameters held fixed. 
In each instance, the seed has height 3, lower radius 2, and upper radius 1. For the next four crystals, 
$\beta_{01}=3$, $\beta_{10}=\beta_{20}=\beta_{11}=1.4$, $\beta_{30}=\beta_{21}=\beta_{31}=1$, 
$\kappa\equiv .1$, $\phi=.01$, and $\rho=.14$. Moreover $\mu_{01}=.002$, $\mu_{30}=\mu_{11}=\mu_{21}=\mu_{31}=.001$ 
and we vary only the common value of $\mu_{10}=\mu_{20}$. This value governs the speed of tips and --- as 
explained in Section 5 --- has more effect in regions of low density, so an increase inhibits side branching.

Like the sectored plates just discussed, these are relatively rare snowfakes with outside 
ridges on the main arms and most side branches. All our modeling experience suggests that 
crystal tips tend to symmetrize with respect to the $\bT$-direction, managing to avoid the 
sandwich instability only under quite special environmental conditions. We have seen 
little evidence in our simulations for the mechanism of ridge formation proposed in \cite{NK, Nel}, 
so we feel that drift is a more likely explanation of one-sided structures in snowflakes. 

\null\hskip-0.6cm
\begin{minipage}[b]{8cm}
\includegraphics[trim=8cm 1cm 8cm 2cm, clip, height=2.9in]{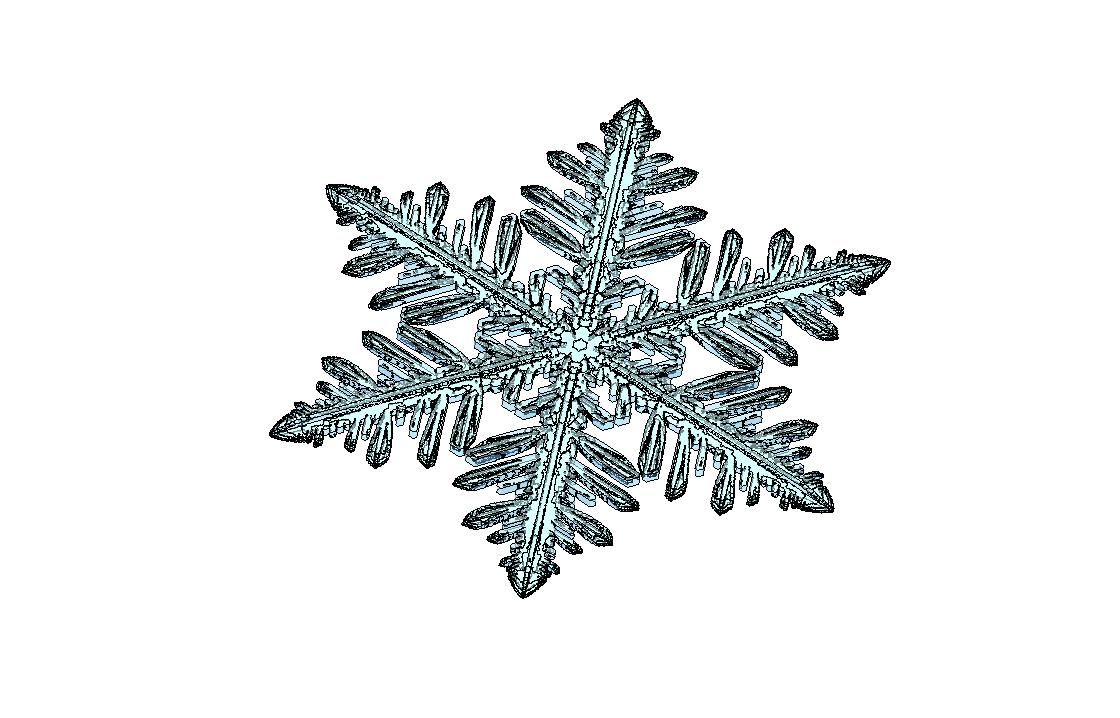}
\end{minipage}
\hskip0.5cm
\begin{minipage}[b][7cm][t]{15cm}
\vskip-0.25cm
\includegraphics[trim=0.7cm 0cm 0.7cm 0cm, clip, height=2.5in]{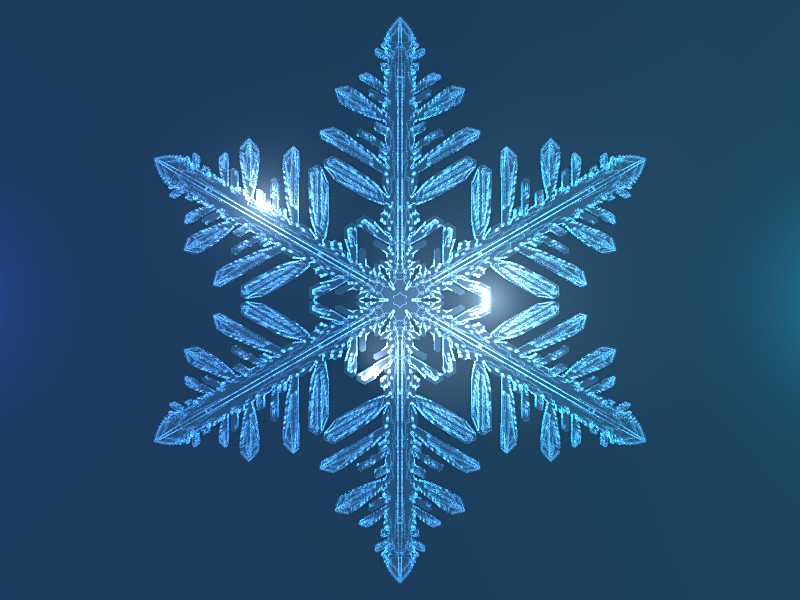}
\end{minipage}
\vskip-0.75cm
{\bf Fig.~25.}  Reduced side branching for $\mu_{10}=\mu_{20}=.008$.
\vskip0.25cm

\null\hskip-0.6cm
\begin{minipage}[b]{8cm}
\includegraphics[trim=8cm 1cm 8cm 2cm, clip, height=2.9in]{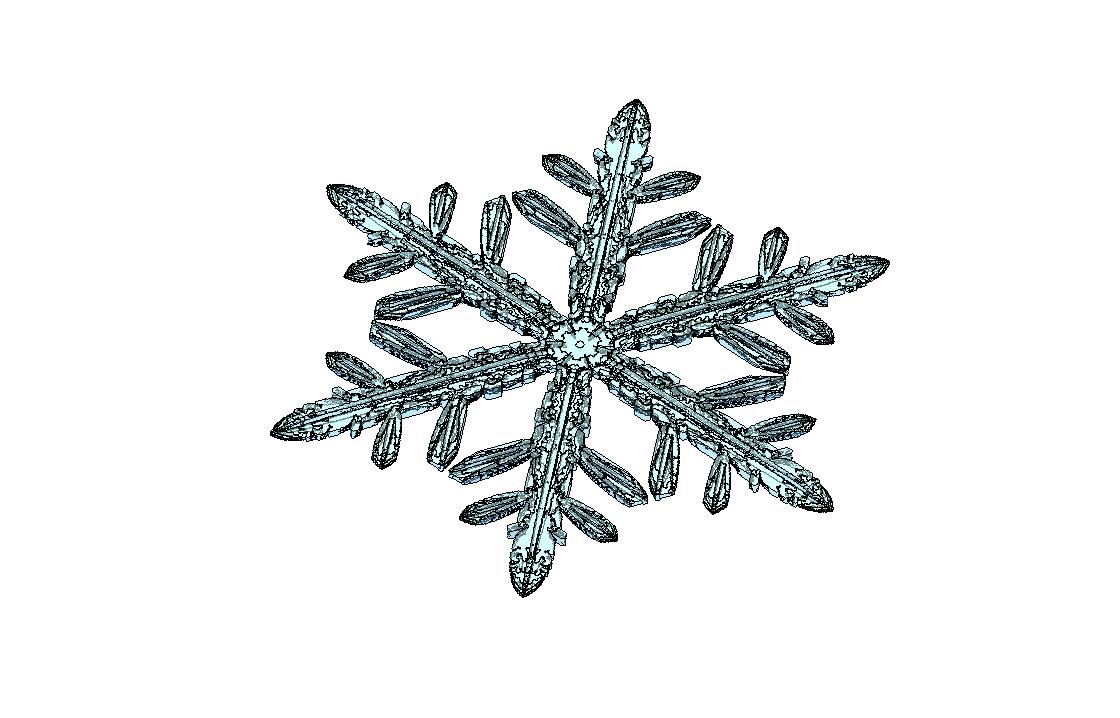}
\end{minipage}
\hskip0.5cm
\begin{minipage}[b][7cm][t]{15cm}
\vskip-0.25cm
\includegraphics[trim=0.7cm 0cm 0.7cm 0cm, clip, height=2.5in]{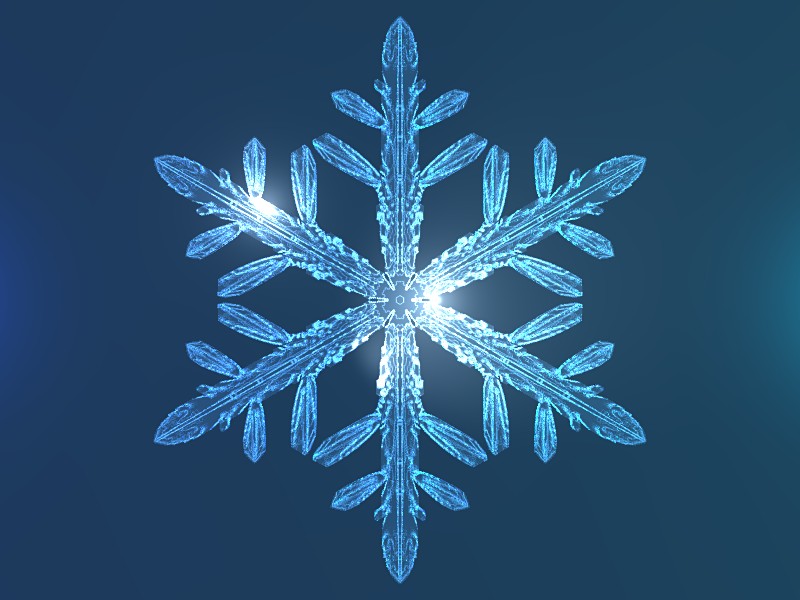}
\end{minipage}
\vskip-0.75cm
{\bf Fig.~26.} Further reduction in the number of side branches for $\mu_{10}=\mu_{20}=.009$. 
\vskip1cm

Starting with the classic fern of Fig.~24, the common prism facet melting threshold  
$\mu_{10}=\mu_{20}$ is gradually increased to twice the original value in Figs.~25--7. 
Stellar dendrites with fewer and fewer side branches result, until the final snowfake 
has only a few short sandwich plates on the sides of each arm.

\null\hskip-0.6cm
\begin{minipage}[b]{8cm}
\includegraphics[trim=8cm 1cm 8cm 2cm, clip, height=2.9in]{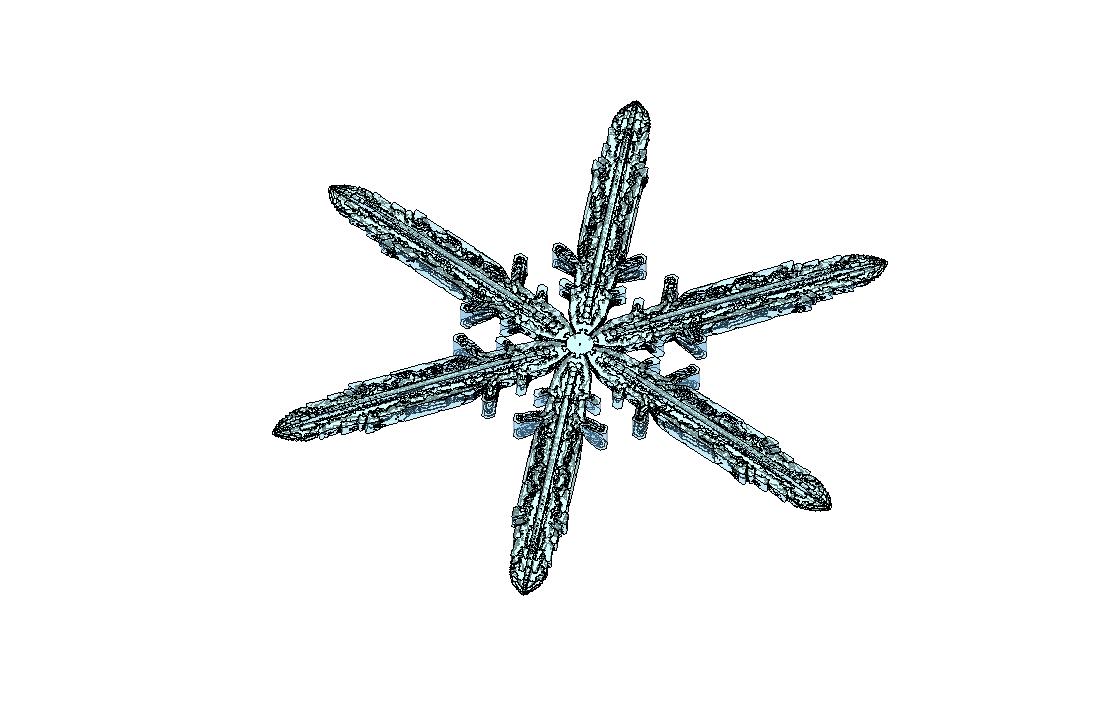}
\end{minipage}
\hskip0.5cm
\begin{minipage}[b][7cm][t]{15cm}
\vskip-0.25cm
\includegraphics[trim=0.7cm 0cm 0.7cm 0cm, clip, height=2.5in]{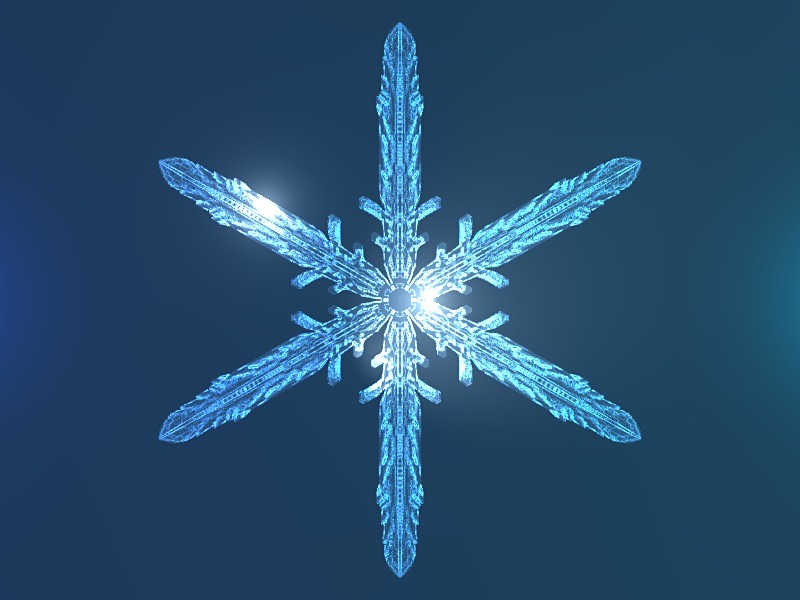}
\end{minipage}
\vskip-0.75cm
{\bf Fig.~27.} When $\mu_{10}=\mu_{20}=.01$, very few side branches remain. 
\vskip1cm

The final example of this section is a classic simple star, a crystal with no side branches 
at all and a characteristic parabolic shape to its tips (cf.~\cite{Lib6}, p.~57 bottom). 
This elegant snowfake required considerable tweaking of parameters; they are:
$\beta_{01}=3.1$, $\beta_{10}=1.05$, $\beta_{20}=1.03$, $\beta_{11}=1.04$,
$\beta_{30}=1.02$, $\beta_{21}=1.01$, $\beta_{31}=1$, $\kappa\equiv .01$,  
$\mu_{01}=\mu_{30}=\mu_{11}=\mu_{21}=\mu_{31}=.01$ , $\mu_{10}=\mu_{20}=.03$, $\phi=.005$, and $\rho=.16$.

\null\hskip-0.6cm
\begin{minipage}[b]{8cm}
\includegraphics[trim=8cm 1cm 8cm 2cm, clip, height=2.9in]{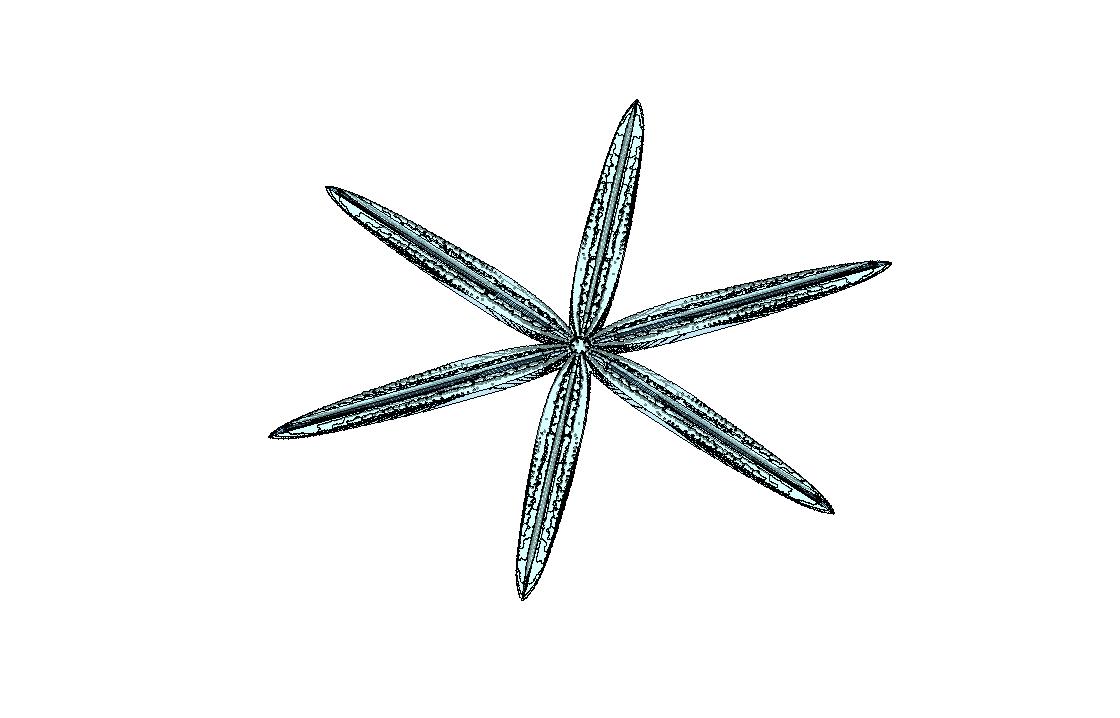}
\end{minipage}
\hskip0.5cm
\begin{minipage}[b][7cm][t]{15cm}
\vskip-0.25cm
\includegraphics[trim=0.7cm 0cm 0.7cm 0cm, clip, height=2.5in]{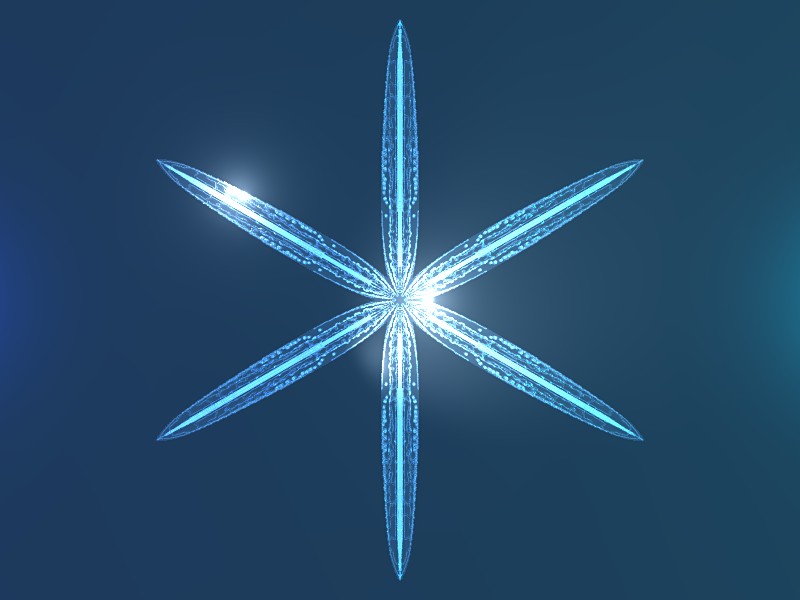}
\end{minipage}
\vskip-0.75cm
{\bf Fig.~28.} A simple star.
\vskip1cm

\vskip0.5cm

\section{Case study $v$ : needles and columns}

Let us now turn to the common but less familiar snow crystals that expand primarily in the $\bZ$-direction. 
As one would expect, these have $\beta_{01}$ small compared to $\beta_{10}$ and $\beta_{20}$, 
but surprisingly small advantage often suffices. We offer three snowfakes that emulate their physical 
counterparts quite well. All start from the canonical seed. Our first example, with a substantial bias 
toward attachment on the basal facets, is a (simple) {\it needle\/}. 
In Fig.~29, $\beta_{01}=2$, $\beta_{10}=\beta_{20}=\beta_{11}=4$,
$\beta_{30}=\beta_{21}=\beta_{31}=1$, $\kappa\equiv .1$, $\mu\equiv .001$, $\phi=0$, and $\rho=.1$. 
This snowfake reproduces structure observed in nature and the laboratory: slender hollow tubes, 
often with cable-like protuberances at the ends (cf.~Fig.~135 of \cite{Nak}, pp.~67--68 of \cite{Lib6}). 

\null\hskip-0.7cm
\begin{minipage}[b]{8cm}
\includegraphics[trim=8cm 1cm 8cm 2cm, clip, height=2.9in]{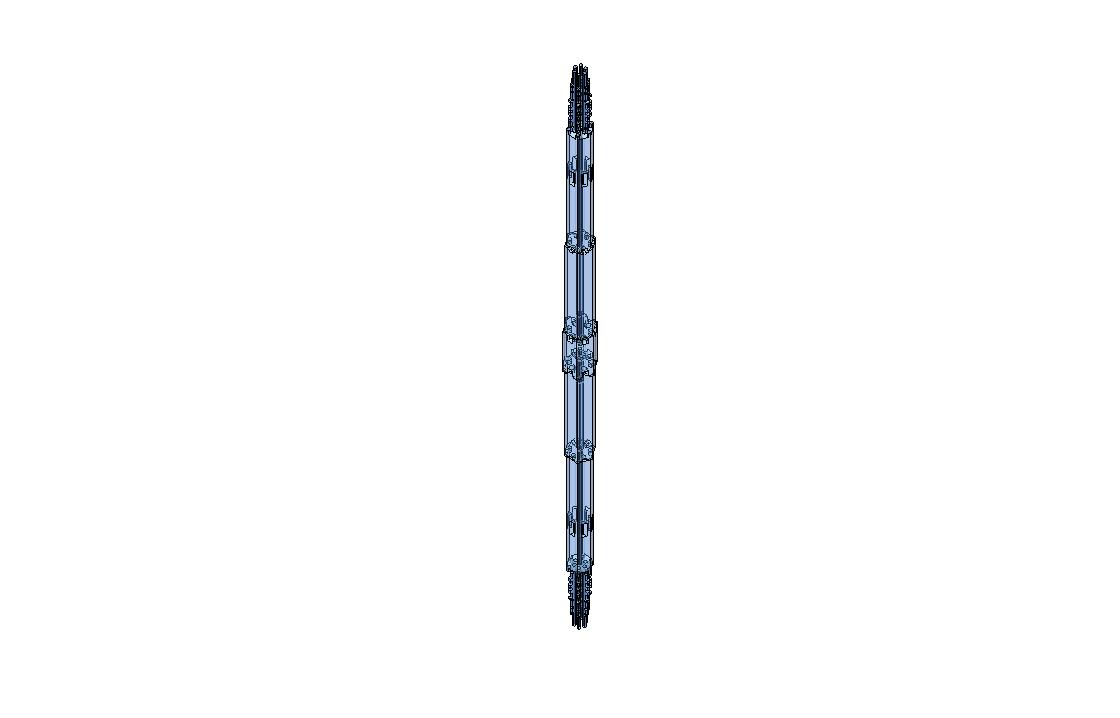}
\end{minipage}
\hskip-1.5cm
\begin{minipage}[b][7cm][t]{15cm}
\vskip-0.25cm
\includegraphics[trim=0cm 0cm 0cm 0cm, clip, height=2.5in]{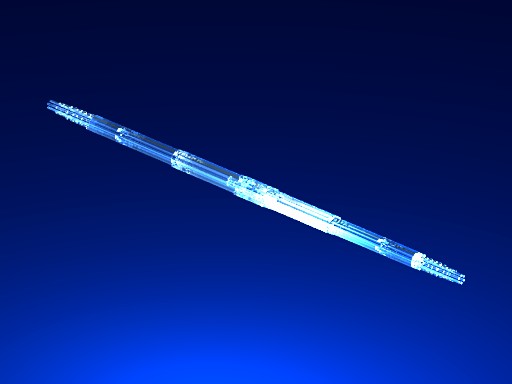}
\end{minipage}
\vskip-0.5cm
{\bf Fig.~29.} A needle.   
\vskip0.6cm

\null\hskip-0.5cm
\begin{minipage}[b]{8cm}
\includegraphics[trim=8cm 1cm 8cm 2cm, clip, height=2.9in]{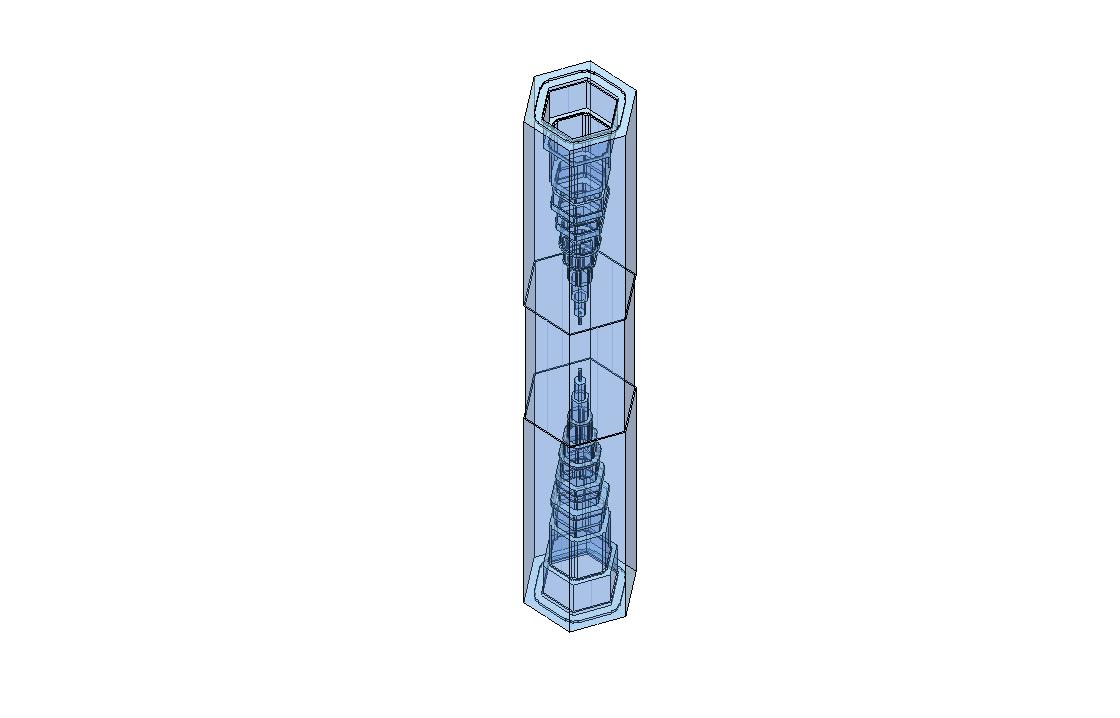}
\end{minipage}
\hskip0.25cm
\begin{minipage}[b][7.5cm][t]{15cm}
\vskip-0.0cm
\includegraphics[trim=0cm 0cm 0cm 0cm, clip, height=2.7in]{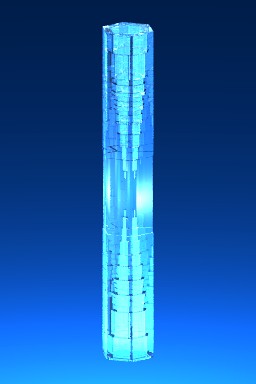}
\end{minipage}
\vskip-0.5cm
{\bf Fig.~30.} A hollow column. 
\vskip1cm

Next, Fig.~30 simulates the common type of snow crystal known as a {\it hollow column\/}. 
Here the bias toward attachment on the basal facets is not as pronounced. The parameter set is:  
$\beta_{01}=1$, $\beta_{10}=\beta_{20}=2$ $\beta_{30}=\beta_{11}=\beta_{21}=.5$ 
$\beta_{31}=1$, $\kappa\equiv .1$, $\mu\equiv .01$, $\phi=0$, and $\rho=.1$.
Evidently, the hole starts developing early on. See pp.~64--66 of \cite{Lib6}) for photos 
of actual hollow columns and a qualitative description of their growth.

The final example of this section is a column whose facets are hollow as well. 
The morphology of 
Fig.~31 occurs when the rates of expansion in the two directions are not very different. 
Photos and a description of this sort of snowflake appear on pp.~35--37 of \cite{Lib6}. 
Here $\beta_{01}=1.5$, $\beta_{10}=\beta_{20}=1.6$ $\beta_{11}=\beta_{30}=\beta_{21}=\beta_{31}=1$, 
$\kappa\equiv .1$, $\mu\equiv .015$, $\phi=0$, and $\rho=.1$.

\null\hskip0.5cm
\begin{minipage}[b]{8cm}
\includegraphics[trim=0cm 0cm 0cm 0cm, clip, height=2.75in]{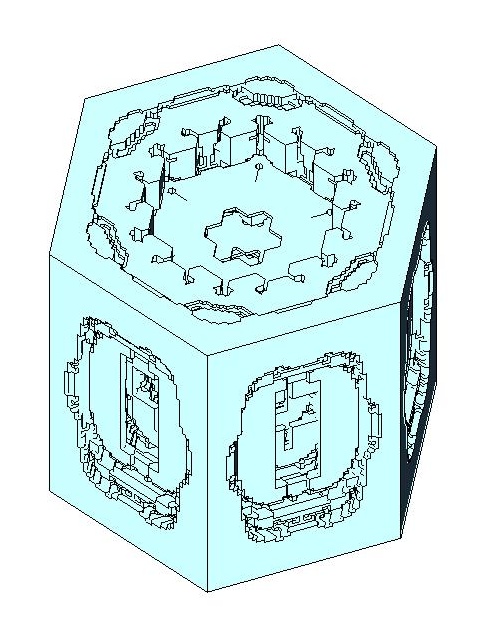}
\end{minipage}
\hskip-.75cm
\begin{minipage}[b][5.7cm][t]{15cm}
\vskip-1cm
\includegraphics[trim=0cm 0cm 0cm 0cm, clip, height=2.5in]{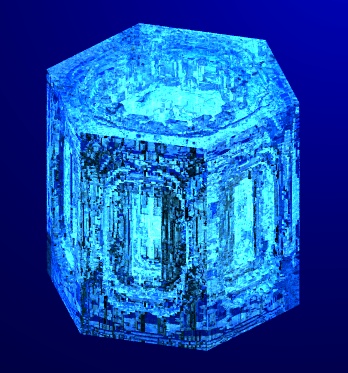}
\end{minipage}
\vskip-0.25cm
{\bf Fig.~31.} A column with hollow prism facets. 
\vskip1cm

\section{Case study $vi$ : change of environment}

In his pioneering research, Nakaya \cite{Nak} reproduced 
several of the most striking types found in nature 
by subjecting the cold chamber in his lab to a precisely controlled schedule 
of temperature and humidity changes, either sudden or gradual. 
Based on such experiments, he argued that {\it plates with dendritic extensions\/}, 
for example, are formed when a snowflake's early growth occurs in the upper atmosphere 
and then it drops to another layer more conducive to branching (\cite{Nak}, p.~16).


In this section we mimic such varying environments by consider the effect of an abrupt change of  
parameters on some of our previous snowfakes. 
Let us begin with two examples of the type cited in the last paragraph: plates with dendritic extensions. 
Both start from a prism that is 3 cells high with radius 2 at the top and 1 at the bottom. 
The first stage for both is a simple plate similar to the snowfake of Fig.~1, but with a delayed branching instability. 
The initial parameters are: $\beta_{01}=3.5$, $\beta_{10}=\beta_{20}=\beta_{11}=2.25$,  
$\beta_{30}=\beta_{21}=\beta_{31}=1$, 
$\kappa\equiv .005$, $\mu\equiv .001$, $\phi=.01$, and $\rho=.12$. 
The first stage runs until time 8000 in the first example, and until time 12500 in the second. 
At that time most parameters remain the same, but in order to promote branching we 
change $\beta_{10}=\beta_{20}=\beta_{11}$ to 1.15 (resp.~1.4) and 
$\mu_{10}=\mu_{20}$ to .006 (resp.~004). The results, once the two snowfakes 
have reached a radius of 200 cells, are shown in Figs.~32--3. 
Predictably, the first example has more branching in its dendritic phase since the 
prism facet attachment threshold is lower. The large image on the cover of {\cite{Lib6}} 
shows a beautiful natural example of this type.

\vskip 0.5cm
\null\hskip-0.6cm
\begin{minipage}[b]{8cm}
\includegraphics[trim=8cm 1cm 8cm 2cm, clip, height=2.9in]{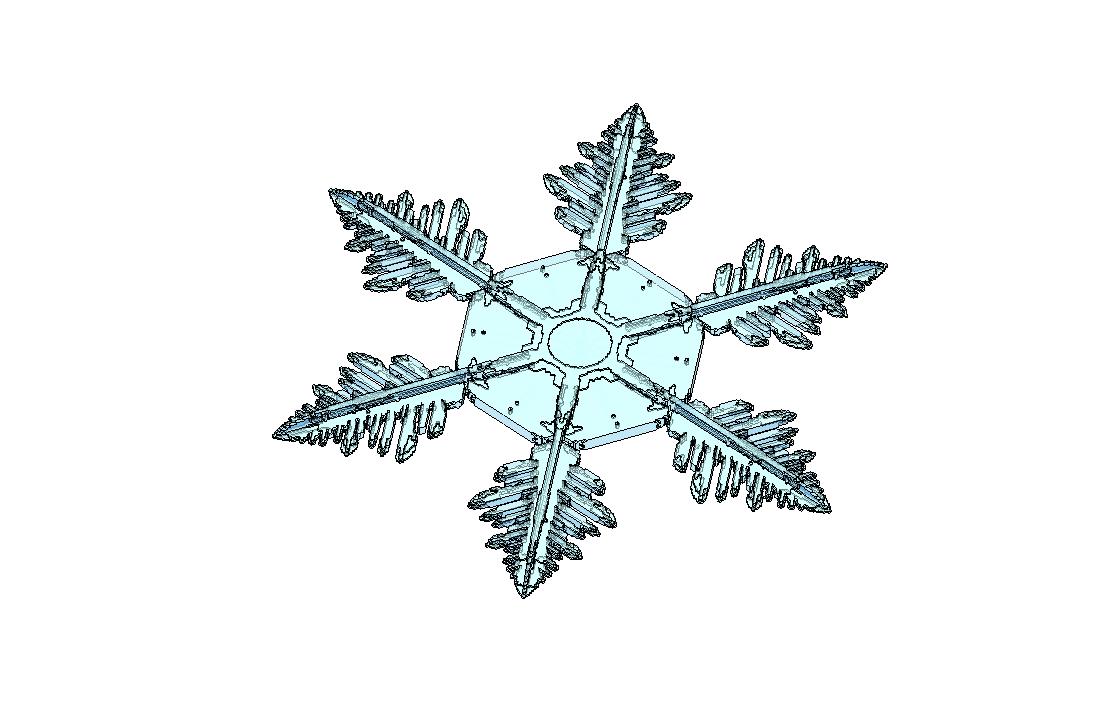}
\end{minipage}
\hskip0.5cm
\begin{minipage}[b][7cm][t]{15cm}
\vskip-0.25cm
\includegraphics[trim=0.7cm 0cm 0.7cm 0cm, clip, height=2.5in]{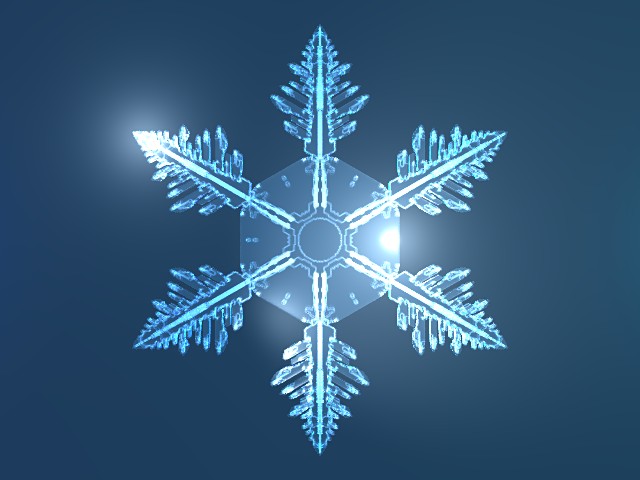}
\end{minipage}
\vskip-0.75cm
{\bf Fig.~32.} A plate with fern extensions.
\vskip1cm

\null\hskip-0.6cm
\begin{minipage}[b]{8cm}
\includegraphics[trim=8cm 1cm 8cm 2cm, clip, height=2.9in]{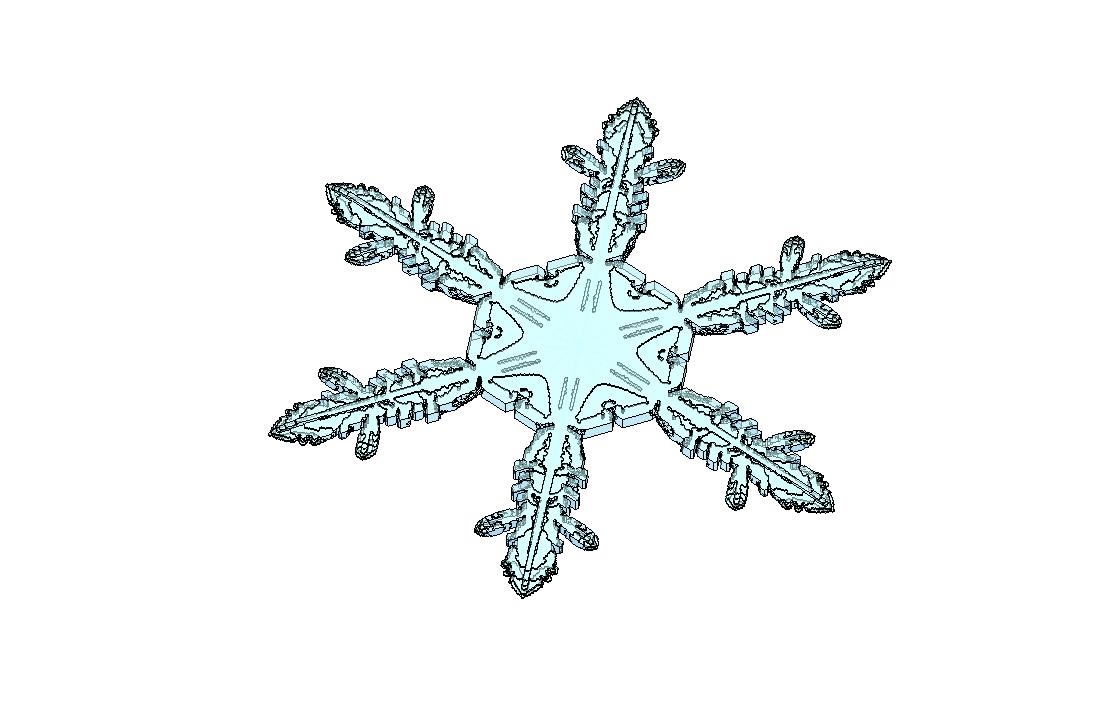}
\end{minipage}
\hskip0.5cm
\begin{minipage}[b][7cm][t]{15cm}
\vskip-0.25cm
\includegraphics[trim=0.7cm 0cm 0.7cm 0cm, clip, height=2.5in]{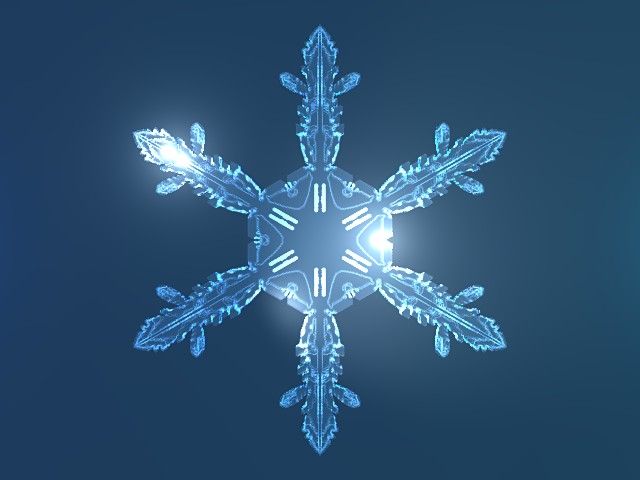}
\end{minipage}
\vskip-0.75cm
{\bf Fig.~33.} A plate with dendrite extensions. 
\vskip1cm

A hybrid evolution at the opposite end of the spectrum is described in 
\cite{Lib6}, pp.~51--53, and many of the most striking snowflakes in \cite{LR} are of this type. 
As presumably in nature, conditions need to be just right for the corresponding snowfake to evolve. 
In this vein, we present three  
snowfakes that begin as stellar dendrites with minimal branching and 
later encounter an environment promoting plates. 
All start from a prism of height 
5 with top radius 6 and bottom radius 2. The first stage runs the simple star dynamics of 
Fig.~28 until time 4000, 3000, or 2000, respectively. Then new parameters for 
the three experiments with higher attachment thresholds are run until time, respectively, 24000, 
20000, and close to 20000. 
Common parameters are: $\beta_{30}=\beta_{31}=1$, $\kappa\equiv .1$, $\rho=.16$. 
In Fig.~34, the remaining parameters are $\beta_{01}=3.0$, $\beta_{10}=\beta_{20}=2.2$, 
$\beta_{11}=2.0$, $\beta_{21}=1.1$, $\mu\equiv .01$, $\phi=.005$. 
Note that in this instance the branches of the star broaden considerably 
after the change of environment, and the tips form sandwich plates.

\vskip 0.25cm
\null\hskip-0.6cm
\begin{minipage}[b]{8cm}
\includegraphics[trim=8cm 1cm 8cm 2cm, clip, height=2.9in]{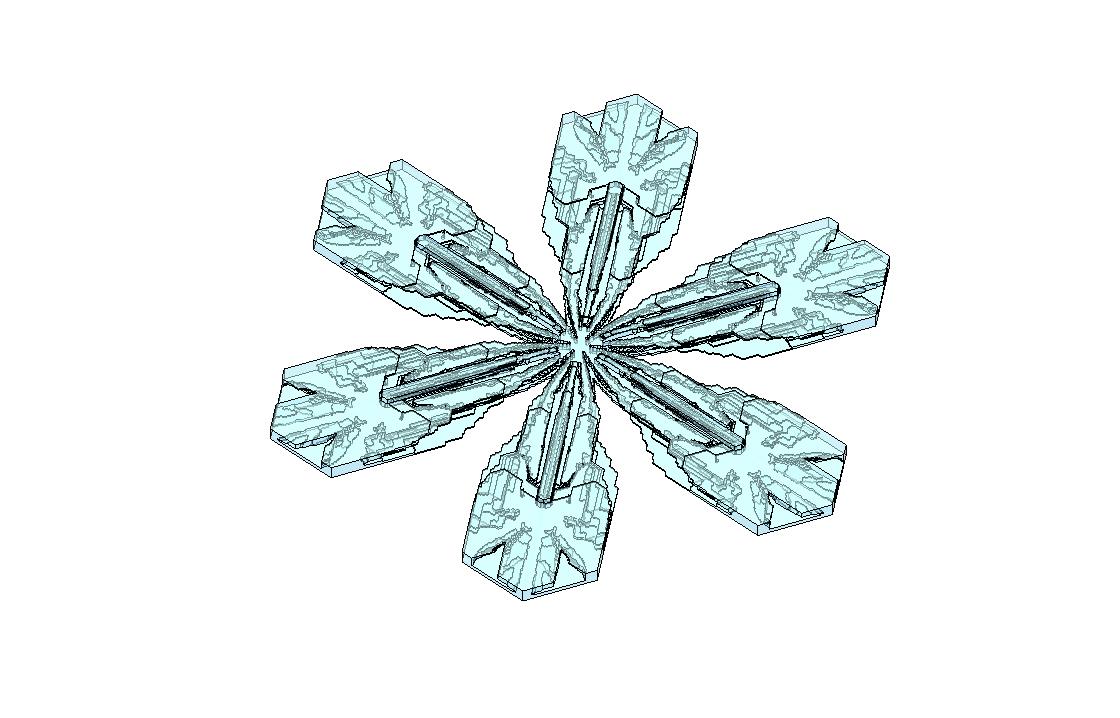}
\end{minipage}
\hskip0.5cm
\begin{minipage}[b][7cm][t]{15cm}
\vskip-0.25cm
\includegraphics[trim=0.7cm 0cm 0.7cm 0cm, clip, height=2.5in]{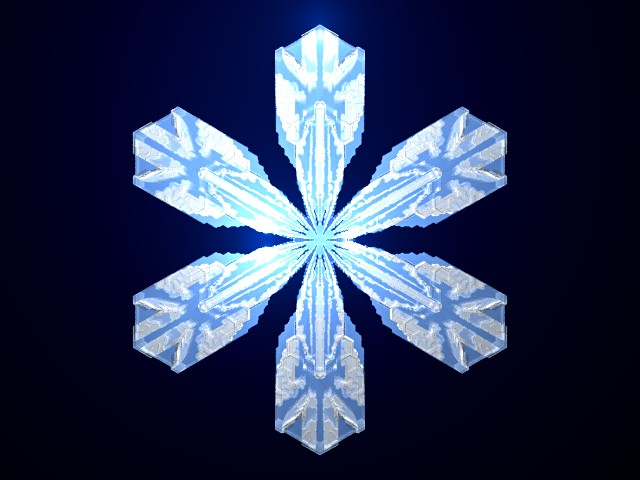}
\end{minipage}
\vskip-0.75cm
{\bf Fig.~34.} A broad-branched stellar crystal with sandwich-plate extensions.
\vskip1cm

\null\hskip-0.6cm
\begin{minipage}[b]{8cm}
\includegraphics[trim=8cm 1cm 8cm 2cm, clip, height=2.9in]{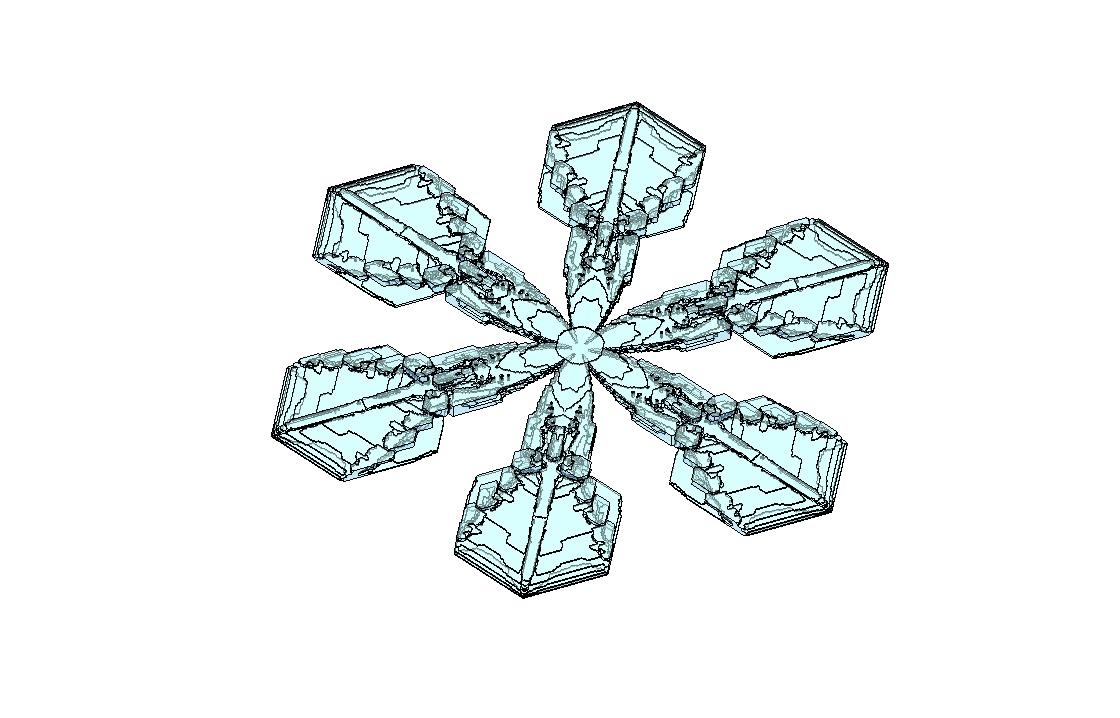}
\end{minipage}
\hskip0.5cm
\begin{minipage}[b][7cm][t]{15cm}
\vskip-0.25cm
\includegraphics[trim=0.7cm 0cm 0.7cm 0cm, clip, height=2.5in]{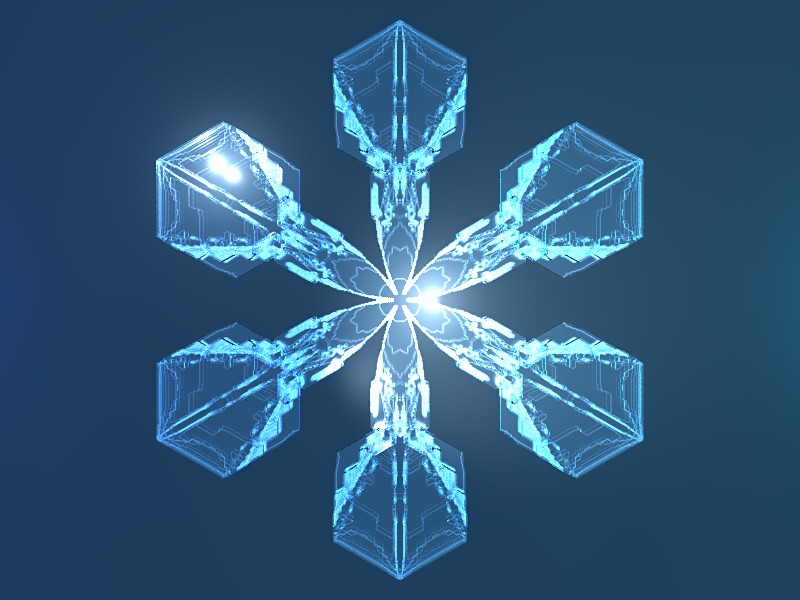}
\end{minipage}
\vskip-0.75cm
{\bf Fig.~35.} A broad-branched stellar crystal with sectored-plate extensions.   
\vskip1cm

By raising the attachment thresholds 
somewhat we avoid the sandwich instability and obtain instead  
the sectored-plate extensions with outside ridges seen in Fig.~35. Here
$\beta_{01}=3.5$, $\beta_{10}=\beta_{20}=2.45$, $\beta_{11}=2.25$, $\beta_{21}=1.1$, 
$\mu_{10}=\mu_{20}=.002$, $\mu= .001$ otherwise, $\phi=.015$. 


Our final broad-branched example interpolates between the previous two. 
The values of $\beta$ are large enough to avoid the sandwich instability, 
but small enough that side branching leads to sectored plate structure of the extensions. 
Here $\beta_{01}=3.0$, $\beta_{10}=\beta_{20}=2.25$, $\beta_{11}=2.05$, $\beta_{21}=1.05$, 
$\mu\equiv .001$, $\phi=.015$. 

\null\hskip-0.6cm
\begin{minipage}[b]{8cm}
\includegraphics[trim=8cm 1cm 8cm 2cm, clip, height=2.9in]{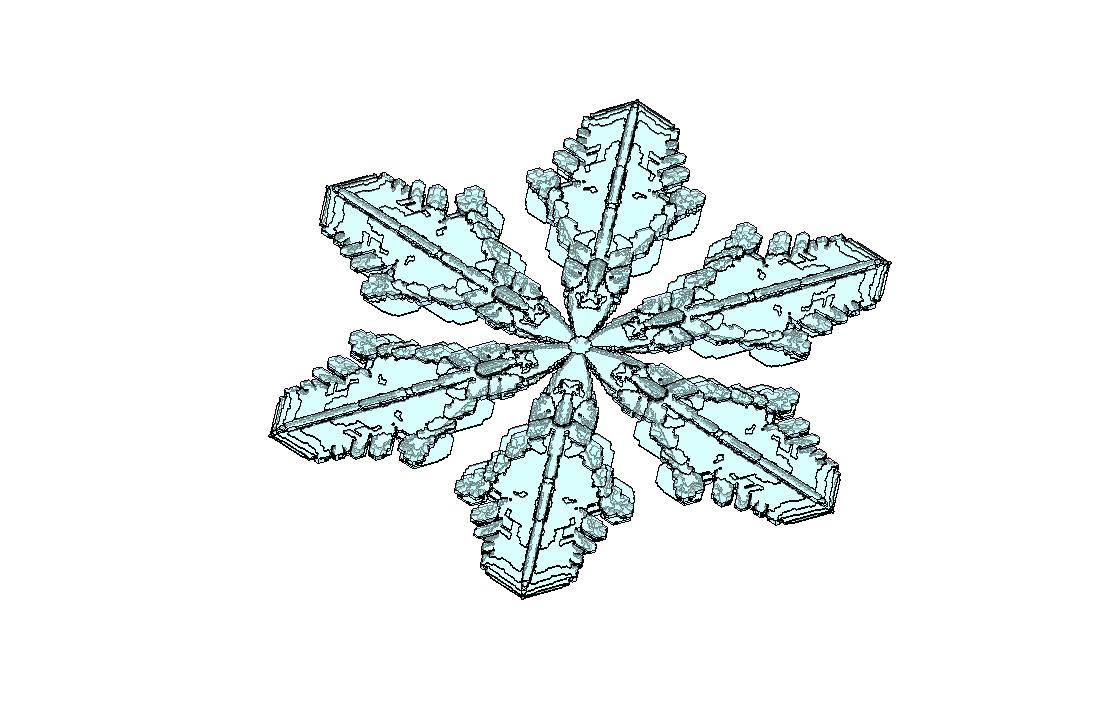}
\end{minipage}
\hskip0.5cm
\begin{minipage}[b][7cm][t]{15cm}
\vskip-0.25cm
\includegraphics[trim=0.7cm 0cm 0.7cm 0cm, clip, height=2.5in]{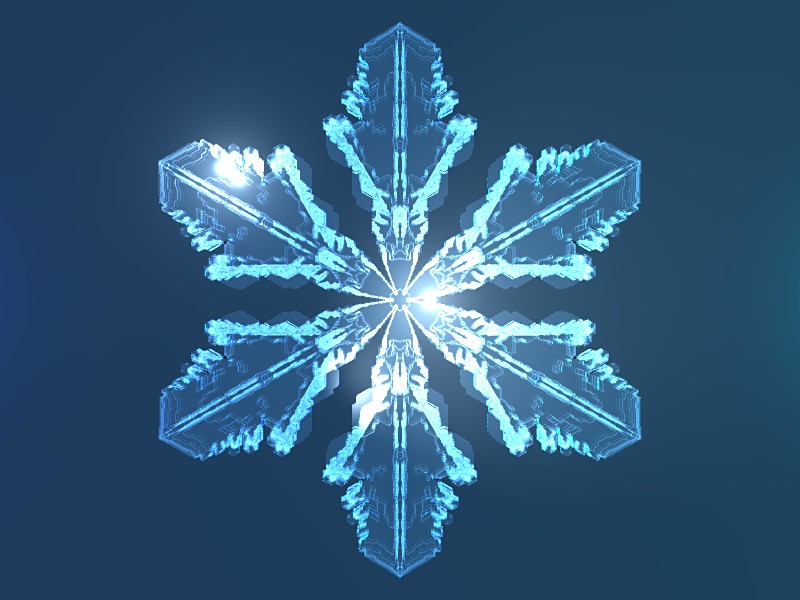}
\end{minipage}
\vskip-0.75cm
{\bf Fig.~36.} Another broad-branched stellar crystal.   
\vskip0.75cm

We conclude this case study with two crystals that combine a 
three-dimensional column and  two-dimensional plates. These are the {\it tsuzumi\/}, or 
{\it capped columns\/}, described on pp.~69--74 of \cite{Lib6}. 
They are thought to arise when crystals are transported to higher and colder 
regions of the atmosphere by a passing storm. Without a preferred orientation, 
it is most reasonable to model these as driftless. Both 
our snowfakes use the canonical seed and evolve with the parameters for 
the hollow column of Fig.~30 until time 20000. Then they run with new parameters that 
promote planar growth, until time  80000 for the first example, 60000 for the second. 
Common values for the two examples are: $\beta_{01}=5$, $\beta_{30}=\beta_{21}=\beta_{31}=1$, 
$\kappa\equiv .1$, $\mu\equiv .001$, $\phi=0$, and $\rho=.1$. 
The difference is the common value $\beta_{10}=\beta_{20}=\beta_{11}$ is $2.4$ in Fig.~37, and $2.1$ in Fig.~38. 
Higher attachment thresholds delay the branching instability in the first capped column 
so the caps are simple plates, as opposed to sectored plates in the second.

The transition period from column to cap in lab tsuzumi is described in some detail by Nakaya (\cite{Nak}, 
p.~221; see also the sketch on p.~222). We remark that 
our snowfake versions evolve in the same way. Namely, for a 
considerable time after the change of environment, outward growth occurs 
almost exclusively along the 18 edges of the hexagonal column. 
This is a diffusion-limited effect similar to the hollowing in Fig.~31. 
Then, rather suddenly, growth in the $\bT$-direction takes over. 

\null\hskip1cm
\begin{minipage}[b]{8cm}
\includegraphics[trim=0cm 0cm 0cm 0cm, clip, height=2.5in]{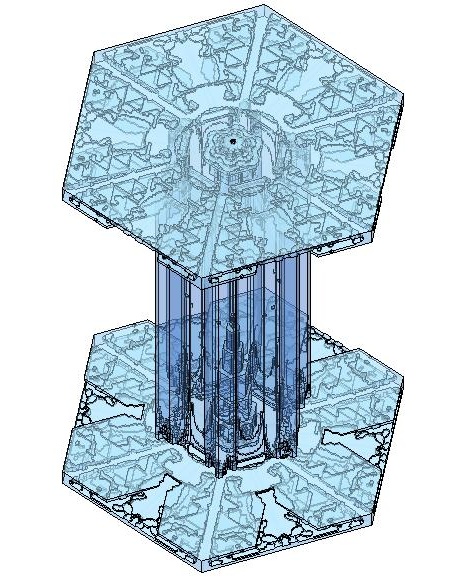}
\end{minipage}
\hskip-1cm
\begin{minipage}[b][5.7cm][t]{15cm}
\vskip-0.8cm
\includegraphics[trim=0cm 0cm 0cm 0cm, clip, height=2.5in]{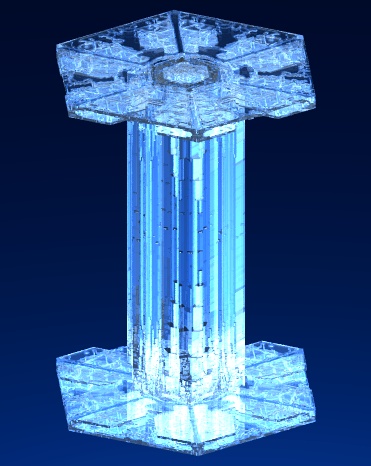}
\end{minipage}
\vskip0cm
{\bf Fig.~37.} A column capped with hexagonal plates. 
\vskip1cm

\null\hskip1cm
\begin{minipage}[b]{8cm}
\includegraphics[trim=0cm 0cm 0cm 0cm, clip, height=2.5in]{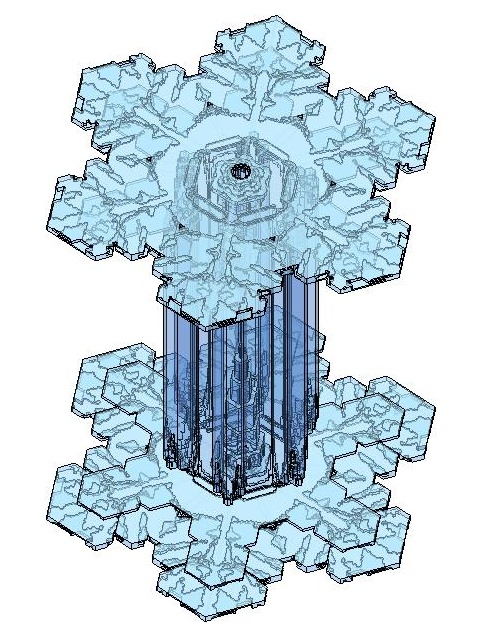}
\end{minipage}
\hskip-1cm
\begin{minipage}[b][5.7cm][t]{15cm}
\vskip-0.8cm
\includegraphics[trim=0cm 0cm 0cm 0cm, clip, height=2.5in]{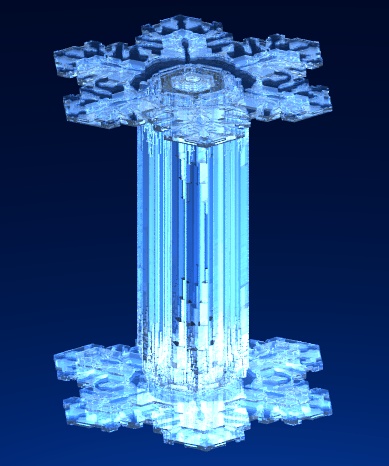}
\end{minipage}
\vskip0cm
{\bf Fig.~38.} A column capped with sectored plates. 
\vskip1cm

\vskip0.5cm

\section{Case study $vii$ : eccentric crystals}

This last section features snowfakes that result from a careful 
search through parameter space and are quite sensitive to any 
change. They are close-to-critical, near the phase boundary 
between dominant growth in the $\bZ$-direction and the $\bT$-direction. 
Consequently they may be rare in nature, though variants of some of 
the forms have been observed, and even represent morphological types in the Magono-Lee 
classification \cite{ML}. All our final examples start from the canonical seed. 

As mentioned in Section 2, starting from a single cell our algorithm has a strong tendency 
to grow rapidly in the $\bZ$-direction due to the immediate onset of a {\it needle instability\/}. 
Even if the initial 
mesoscopic prism is wider in the $\bT$-direction,  
it is still quite common for this instability to arise later on if
the dynamics are close to critical.  
After an initial phase of typical 
planar growth, needles suddenly nucleate at concentric locations scattered over the central plate or arms. 
Fig.~137 of \cite{Nak} shows an excellent example of this type in nature, and our first two examples 
illustrate a similar phenomenon in our model. The conventional explanation for such
hybrid types, called {\it stellar crystals with needles\/} in \cite{ML}, involves a sudden change 
in the environment, but this is one of several cases where our algorithm suggests that homogeneous 
conditions can sometimes produce the same effect.
 
Fig.~39 
has features like a classic planar snowflake 
that has developed {\it rime\/} from attachment of surrounding water droplets. 
In fact these protrusions are potential needle instabilities --- the two symmetric rings 
close to the center and the tips are stunted needles, whereas the intermediate needles have 
successfully nucleated. The parameters of this snowfake are: 
$\beta_{01}=1.58$, 
$\beta_{10}=\beta_{20}=\beta_{11}=1.5$, $\beta_{30}=\beta_{21}=\beta_{31}=1$, 
$\kappa\equiv .1$, $\mu\equiv .006$, $\phi=0$ and $\rho=.1$. 
Partial symmetry of bumps in many natural crystals, 
statistically unlikely to be the result of rime, often indicates vestiges of 
rims and ribs after sublimation, but can also be due to nascent needles, 
as in the middle specimen of Plate 116 in \cite{Nak}. 
Since the locations where needles nucleate are quite sensitive to changes in parameters, 
residual randomness in the mesoscopic dynamics is apt to degrade the symmetry.

\null\hskip-0.4cm
\begin{minipage}[b]{8cm}
\includegraphics[trim=0cm 0cm 0cm 0cm, clip, height=2.3in]{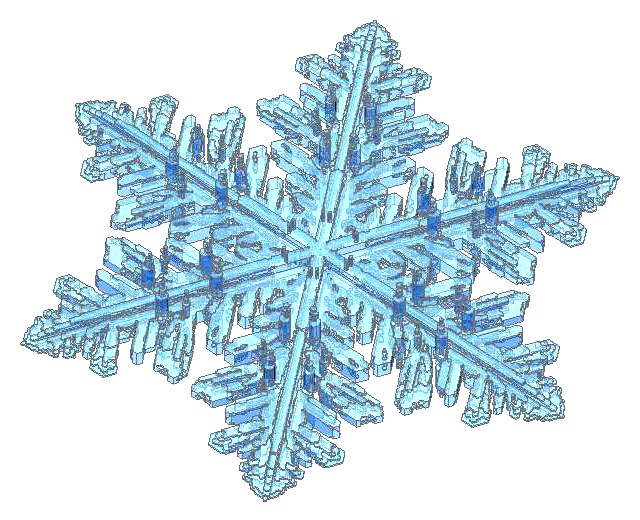}
\end{minipage}
\hskip0.0cm
\begin{minipage}[b][5.7cm][t]{15cm}
\vskip-0.1cm
\includegraphics[trim=0cm 0cm 0cm 0cm, clip, height=2.2in]{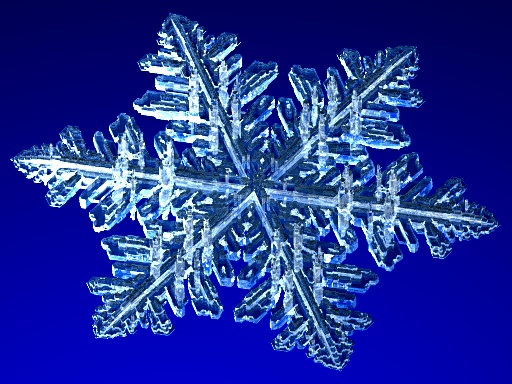}
\end{minipage}
\vskip-0.1cm
{\bf Fig.~39.} A stellar dendrite with stunted and nucleating needles.
\vskip 0.5cm 

The next three examples have 
$\beta\equiv 1$, $\mu\equiv .03$, $\kappa_{10}=\kappa_{20}=.1$, 
$\kappa_{30}=.05$, and $\kappa_{11}=\kappa_{21}=\kappa_{31}=.01$.
The remaining parameters for Fig.~40 are $\kappa_{01}=.11$ and $\rho=.06$. 
This snowfake is a rather extreme instance of a stellar crystal with needles 
in which the planar portion 
is a thick but very narrow simple star. 

\null\hskip-0.5cm
\begin{minipage}[b]{8cm}
\includegraphics[trim=8cm 1cm 8cm 2cm, clip, height=3in]{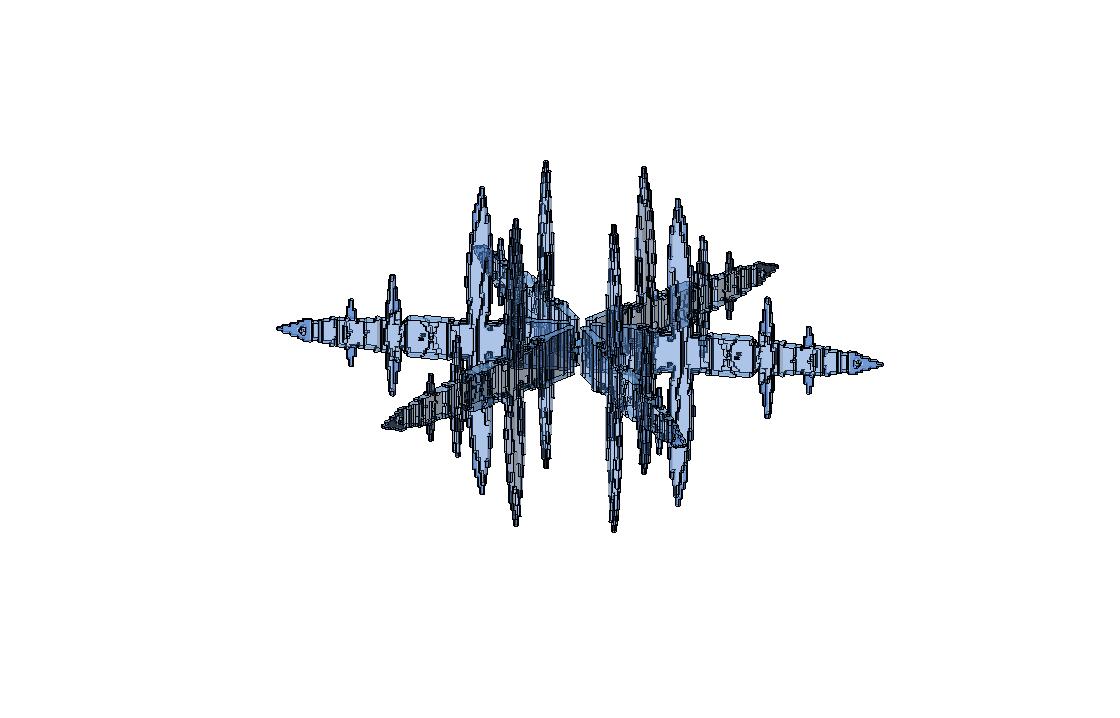}
\end{minipage}
\hskip0.5cm
\begin{minipage}[b][7cm][t]{15cm}
\vskip0.25cm
\includegraphics[trim=0cm 0cm 0cm 0cm, clip, height=2in]{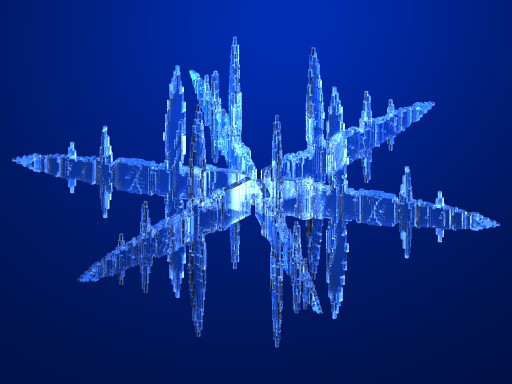}
\end{minipage}
\vskip-1.75cm
{\bf Fig.~40.} A simple star with needles. 
\vskip1cm

Our next two examples seem never to have been seen at all, 
and it is clear why: even if they managed to grow, 
their thin plates would be extremely brittle and susceptible to random 
fluctuations. They are characterized by very small differences 
in the growth rates. After starting as planar crystals, 
they suddenly nucleate thin structures extending into the third dimension. 
In Fig.~41 $\kappa_{01}=.12$ and $\rho=.057$; in Fig.~42 $\kappa_{01}=.116$ and $\rho=.06$. 
For obvious reasons, we call these {\it butterflakes\/}. 
They are idealizations of the {\it stellar crystals with spatial plates\/} 
in \cite{ML}; chaotic snow crystals with thin plates growing every which way are relatively common.

\null\hskip0.5cm
\begin{minipage}[b]{8cm}
\includegraphics[trim=0cm 0cm 0cm 0cm, clip, height=2.4in]{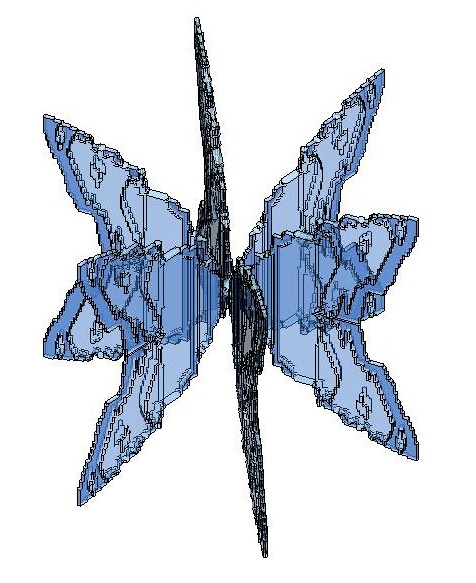}
\end{minipage}
\hskip-0.5cm
\begin{minipage}[b][5.7cm][t]{15cm}
\vskip-0.2cm
\includegraphics[trim=0cm 0cm 0cm 0cm, clip, height=2.2in]{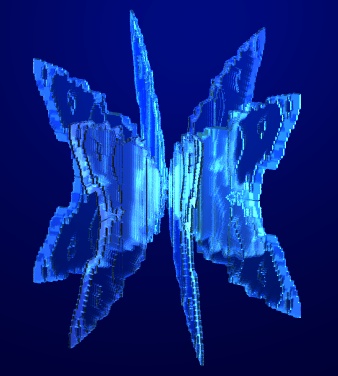}
\end{minipage}
\vskip-0.1cm
{\bf Fig.~41.} A butterflake with wings in the directions of the main arms.   
\vskip-0.4cm

\null\hskip-0.7cm
\begin{minipage}[b]{8cm}
\includegraphics[trim=8cm 1cm 8cm 2cm, clip, height=2.8in]{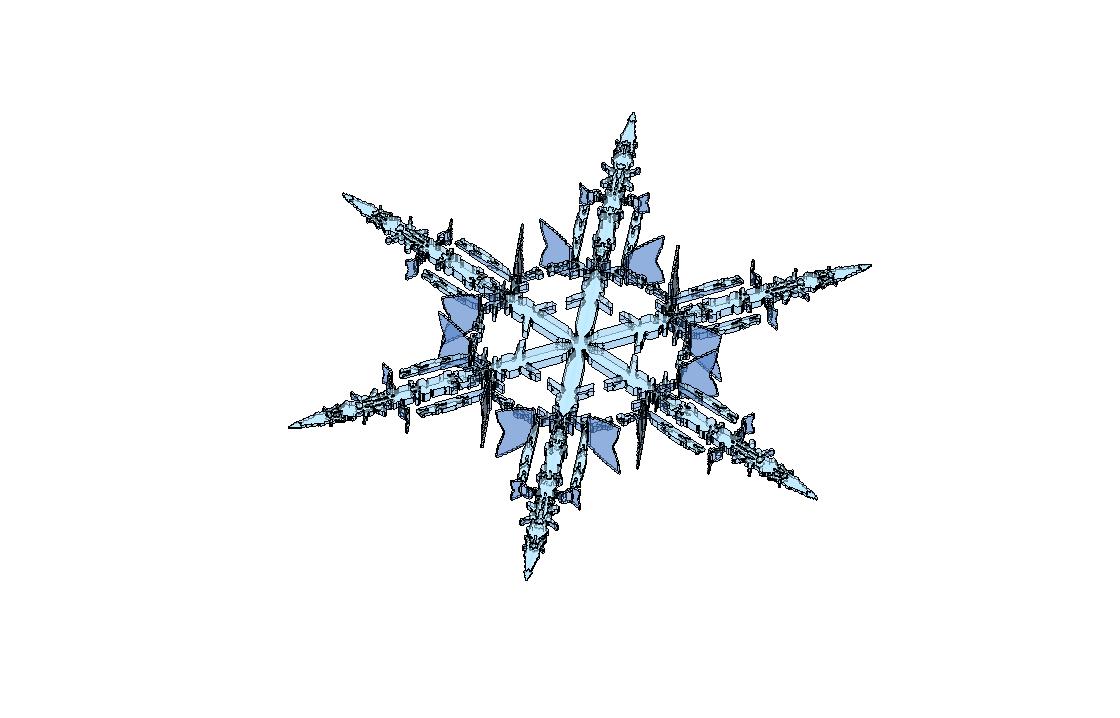}
\end{minipage}
\hskip-0.0cm
\begin{minipage}[b][7cm][t]{15cm}
\vskip 0.5cm
\includegraphics[trim=0cm 0cm 0cm 0cm, clip, height=2.1in]{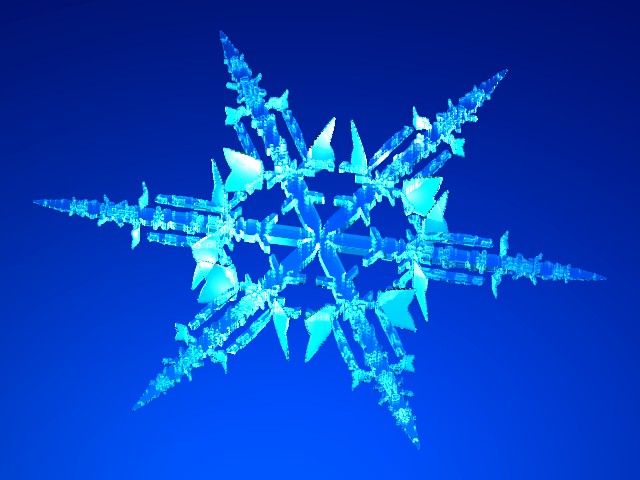}
\end{minipage}
\vskip-1.1cm
{\bf Fig.~42.} A butterflake with side wings. 
\vskip0.5cm

We conclude the paper with a family of five related examples. 
The first is a common sandwich plate (cf.~p.~39, lower right, in \cite{Lib6}) 
with parameter values 
$\beta_{01}=1.41$, $\beta_{10}=\beta_{20}=1.2$ $\beta_{11}=\beta_{30}=\beta_{21}=\beta_{31}=1$, 
$\kappa\equiv .1$, $\mu\equiv .025$, $\phi=0$, and $\rho=.09$.

\null\hskip-0.7cm
\begin{minipage}[b]{8cm}
\includegraphics[trim=8cm 1cm 8cm 2cm, clip, height=2.7in]{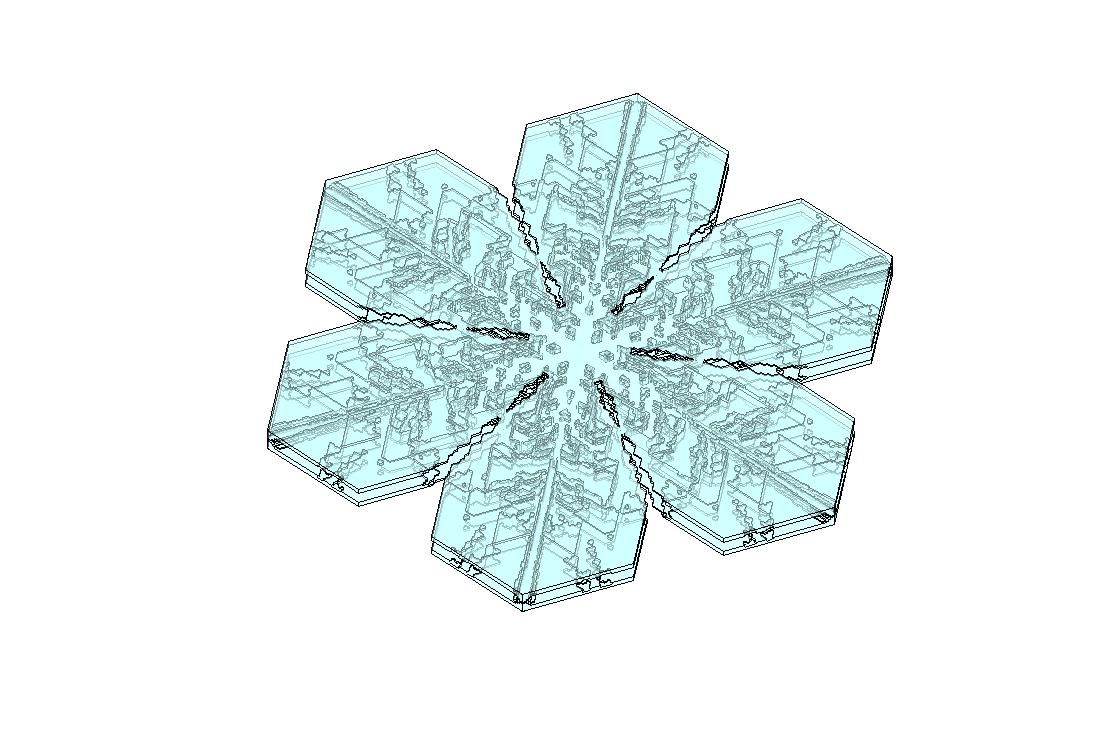}
\end{minipage}
\hskip0.3cm
\begin{minipage}[b][7cm][t]{15cm}
\vskip0.45cm
\includegraphics[trim=0.7cm 0cm 0.7cm 0cm, clip, height=2.2in]{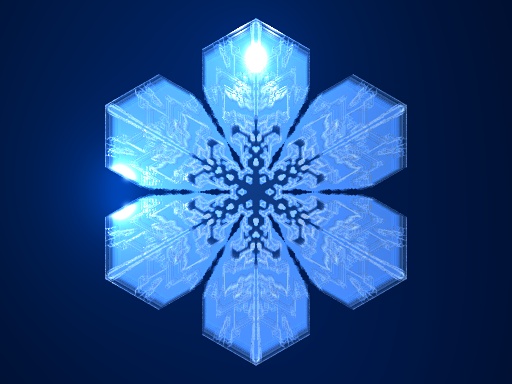}
\end{minipage}
\vskip-0.9cm
{\bf Fig.~43.} A sandwich plate with broad branches. 
\vskip0.5cm

The remaining four are minor perturbations, which nevertheless look 
quite different. Namely, even though their model parameters are 
constant over time, they undergo ``exploding tips'' quite similar 
to crystals such as the one in Fig.~35 that results from inhomogeneous environmental conditions. 
The principle behind all four variants is the same: 
eventually, the growing tip thickens and slows down considerably. 
Usually this happens close to the beginning of the evolution 
(as, in fact, occurred in the dynamics leading to Fig.~43), so the snowfake is unremarkable. 
But with some experimentation we find cases when the onset of the sandwich instability 
is delayed and the final picture can be quite dramatic. The 
complex inner patterns are the result of extraordinarily intricate dynamics. 
Parameter values that differ from those of Fig.~43 are given in the captions. 
\vskip-0.5cm

\null\hskip-0.7cm
\begin{minipage}[b]{8cm}
\includegraphics[trim=8cm 1cm 8cm 2cm, clip, height=2.7in]{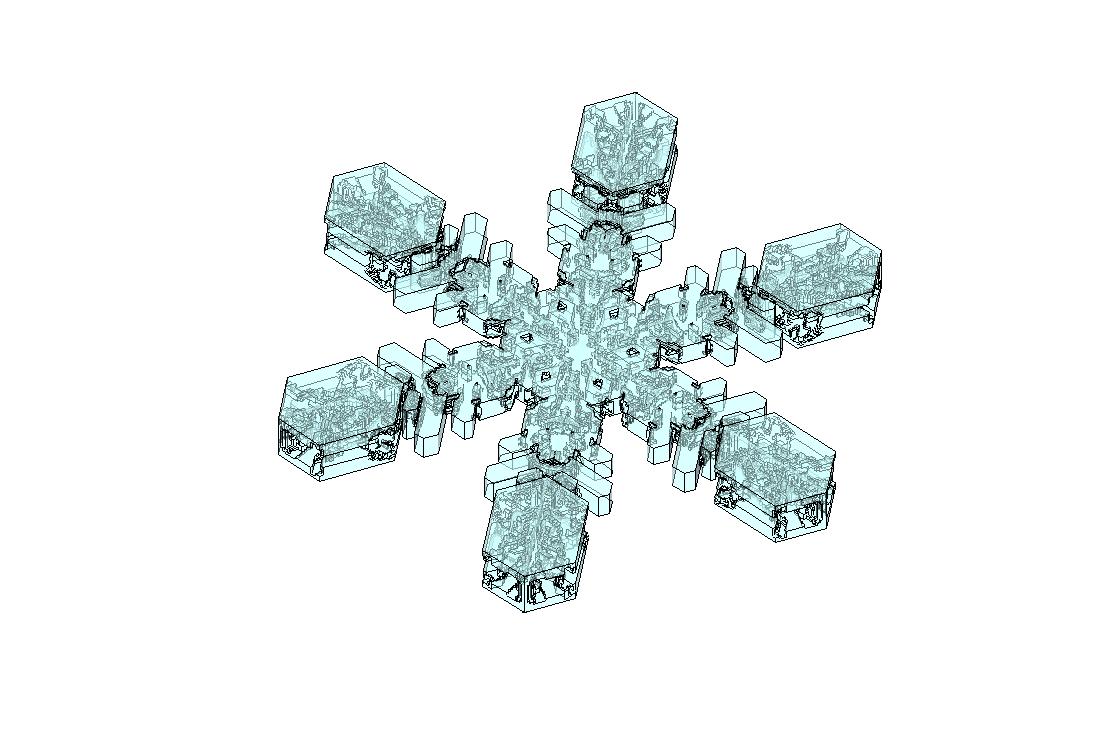}
\end{minipage}
\hskip0.3cm
\begin{minipage}[b][7cm][t]{15cm}
\vskip0.45cm
\includegraphics[trim=0.7cm 0cm 0.7cm 0cm, clip, height=2.2in]{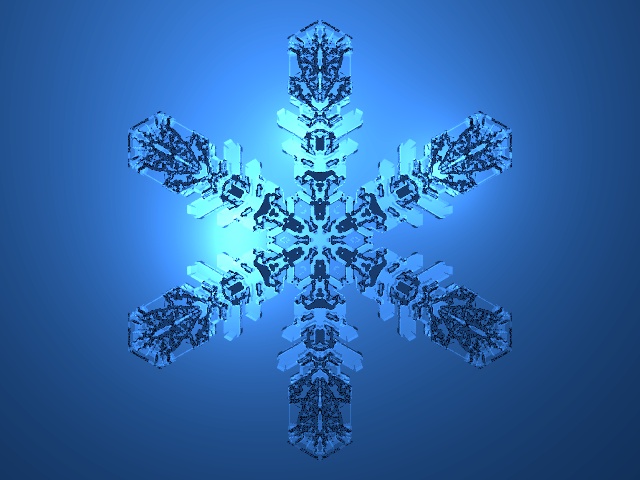}
\end{minipage}
\vskip-0.9cm
{\bf Fig.~44.} Perturbed parameters: $\beta_{01}=1.25$, $\rho=.091$. 
\vskip.05cm

\null\hskip-0.7cm
\begin{minipage}[b]{8cm}
\includegraphics[trim=8cm 1cm 8cm 2cm, clip, height=2.7in]{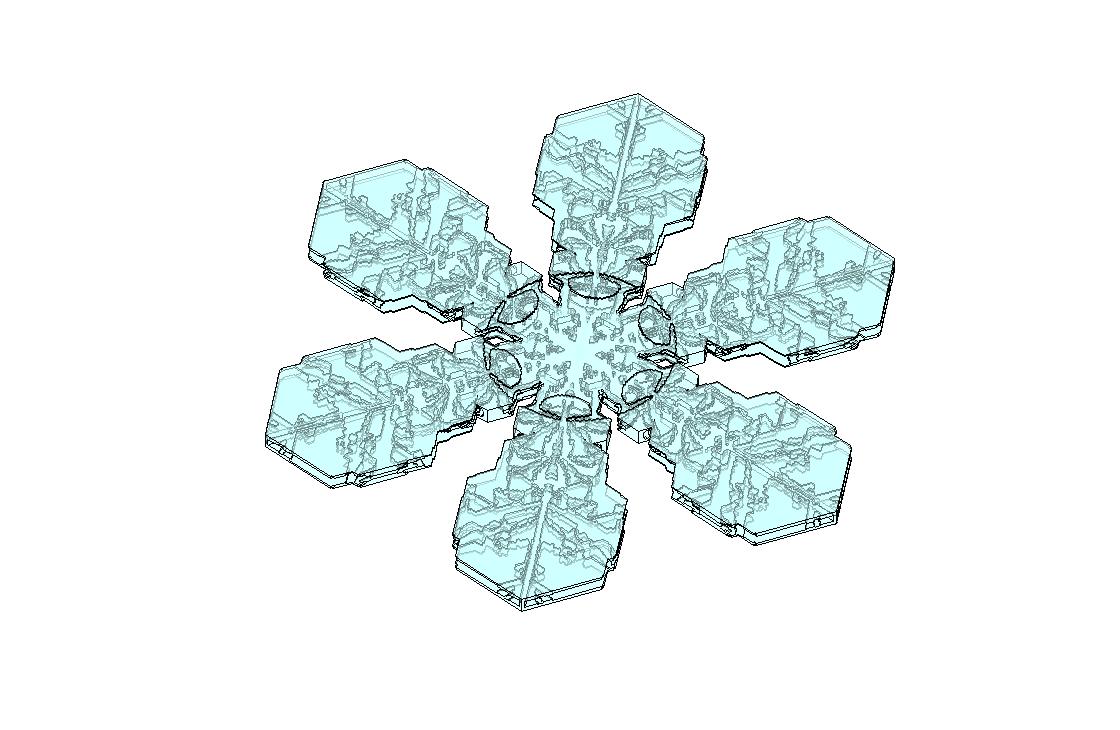}
\end{minipage}
\hskip0.3cm
\begin{minipage}[b][7cm][t]{15cm}
\vskip0.45cm
\includegraphics[trim=0.7cm 0cm 0.7cm 0cm, clip, height=2.2in]{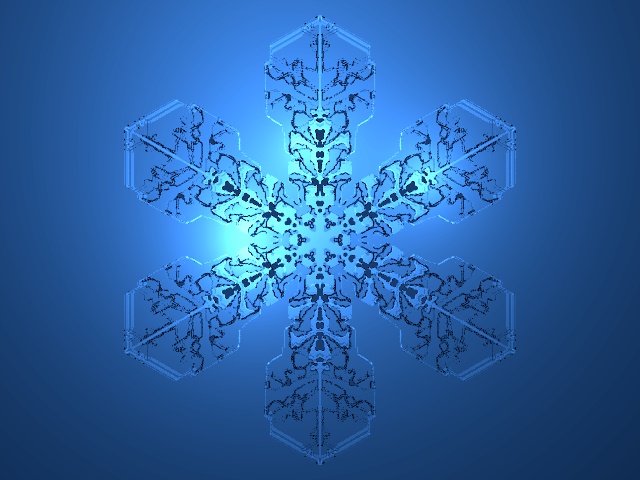}
\end{minipage}
\vskip-0.9cm
{\bf Fig.~45.} Perturbed parameter: $\beta_{01}=1.5$. 
\vskip.05cm

\null\hskip-0.7cm
\begin{minipage}[b]{8cm}
\includegraphics[trim=8cm 1cm 8cm 2cm, clip, height=2.7in]{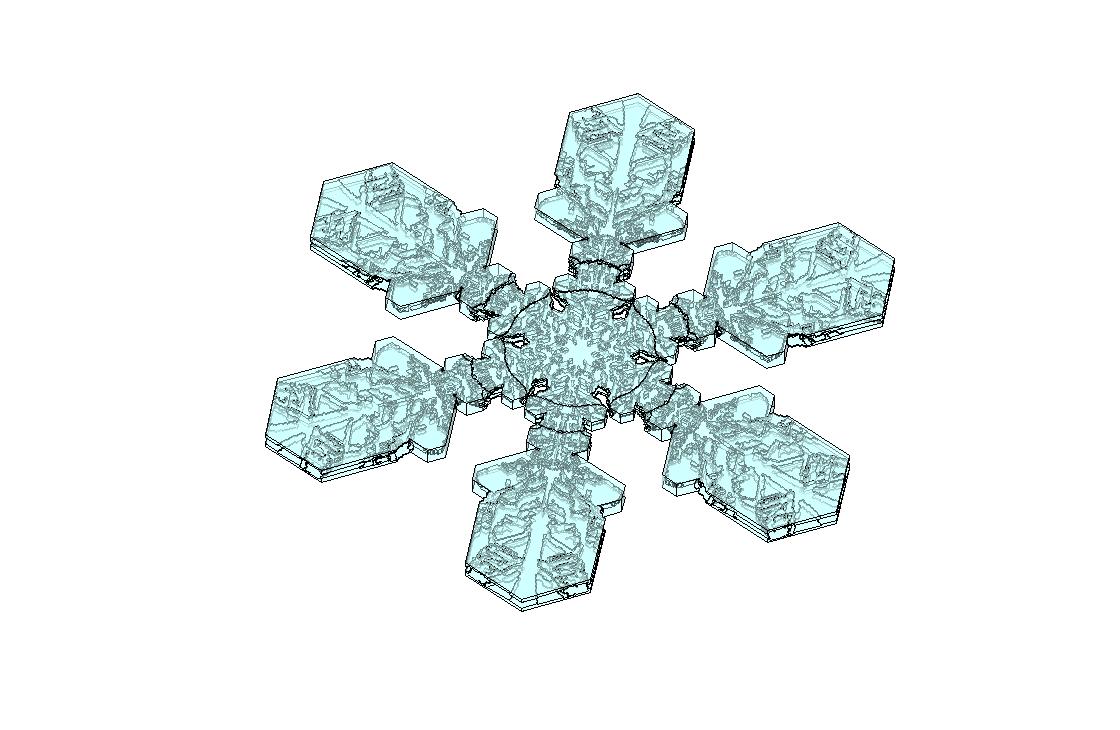}
\end{minipage}
\hskip0.3cm
\begin{minipage}[b][7cm][t]{15cm}
\vskip0.45cm
\includegraphics[trim=0.7cm 0cm 0.7cm 0cm, clip, height=2.2in]{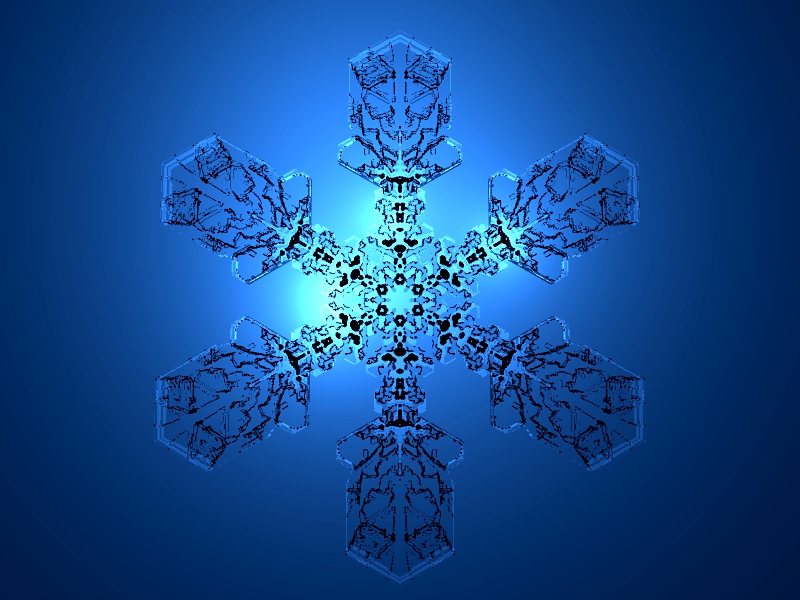}
\end{minipage}
\vskip-0.9cm
{\bf Fig.~46.} Perturbed parameter: $\beta_{01}=1.19$. 
\vskip-.25cm

\null\hskip-0.7cm
\begin{minipage}[b]{8cm}
\includegraphics[trim=8cm 1cm 8cm 2cm, clip, height=2.7in]{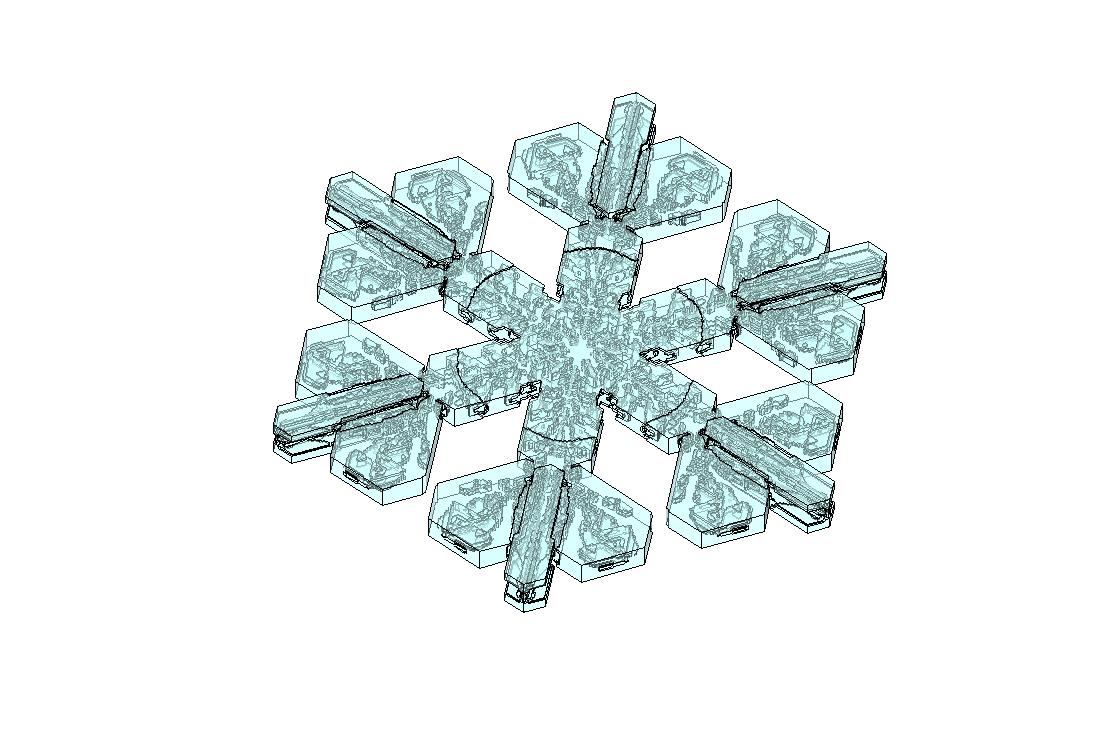}
\end{minipage}
\hskip0.3cm
\begin{minipage}[b][7cm][t]{15cm}
\vskip0.45cm
\includegraphics[trim=0.7cm 0cm 0.7cm 0cm, clip, height=2.2in]{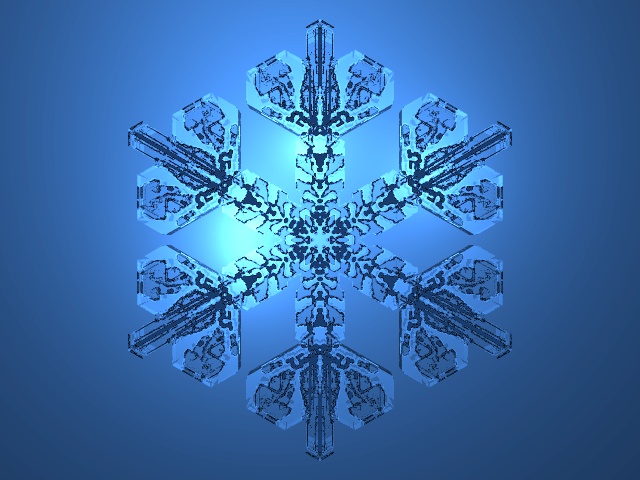}
\end{minipage}
\vskip-0.9cm
{\bf Fig.~47.} Perturbed parameter: $\beta_{01}=1.25$. 

\end{document}